\documentclass[sigconf,9pt]{acmart}

\usepackage[english]{babel}
\usepackage{blindtext}
\usepackage{tikz}
\usepackage{amsmath}
\usepackage{enumitem}
\setlist{noitemsep,topsep=0pt,parsep=0pt,partopsep=0pt,leftmargin=0.5cm}
\usepackage{xspace}
\usepackage{color}
\usepackage{outlines}
\usepackage{multirow}
\usepackage{multicol}
\usepackage{tabularx}
\usepackage{balance}
\usepackage{graphicx}
\usepackage{subfig}
\usepackage{caption}

\renewcommand\footnotetextcopyrightpermission[1]{} % removes footnote with conference info
\setcopyright{none}
\newcommand{\eg}{\emph{e.g.,} }

\newcommand{\ie}{{\em i.e.}, }

\newcommand{\vs}{{\em vs.} }
\newcommand{\etc}{{\em etc.}\xspace}
\newcommand{\Fig}[1]{Fig.~\ref{fig:#1}} % remove xspace, causes too many issues
\newcommand{\Tab}[1]{Table~\ref{tab:#1}\xspace}
\newcommand{\Sec}[1]{\S\ref{sec:#1}\xspace}
\newcommand{\App}[1]{Appendix~\ref{app:#1}\xspace}

\newcommand{\nop}[1]{}

\newcommand{\mycaptionsize}{\small}

\newenvironment{parafont}{\fontfamily{ptm}\selectfont}{}
\newcommand{\Para}[1]{\vspace{4pt}\noindent\begin{parafont}\fontsize{9.0}{10.8}\textbf{\textit{#1}}\end{parafont}}
\newcommand{\ct}{\small \tt}

\usepackage{listings}
\newcommand{\ngs}[1]{}
\newcommand{\as}[1]{}
\newcommand{\qwx}[1]{}

\newcommand{\TheCompiler}{{\sc K2}\xspace}
\newcommand{\MinThroughputGain}{{\sc 0}\xspace}
\newcommand{\MaxThroughputGain}{{\sc 4.75\%}\xspace}
\newcommand{\MinLatencyReduction}{{\sc 1.36\%}\xspace}
\newcommand{\MaxLatencyReduction}{{\sc 55.03\%}\xspace}
\newcommand{\NumBenchmarks}{19\xspace}
\newcommand{\AverageInstrCountReduction}{13.95\%\xspace}
\newcommand{\MinEqChkTimeReduction}{2\xspace}
\newcommand{\MaxEqChkTimeReduction}{7\xspace}
\newcommand{\AvgEqChkTimeReduction}{6\xspace}

\begin{document}

\title[Synthesizing Safe and Efficient Kernel Extensions for Packet
  Processing]{Synthesizing Safe and Efficient Kernel Extensions\\for
  Packet Processing}
\newcommand{\aut}[2]{#1\texorpdfstring{$^{#2}$}{(#2)}}
\author{\aut{Qiongwen Xu}{1}, \aut{Michael D. Wong}{2}, \aut{Tanvi
    Wagle}{1}, \aut{Srinivas Narayana}{1}, \aut{Anirudh Sivaraman}{3}}
\affiliation{\aut{}{1}\textit{Rutgers University}\quad
             \aut{}{2}\textit{Princeton University}\quad
             \aut{}{3}\textit{New York University}\\
             \country{k2\_compiler@email.rutgers.edu}}
\newcommand{\noexpauthors}{Qiongwen Xu, Michael D. Wong, Tanvi
  Wagle, Srinivas Narayana, and Anirudh Sivaraman}
\renewcommand{\shortauthors}{Qiongwen Xu et al.}

\begin{abstract}

Extended Berkeley Packet Filter (BPF) has emerged as a powerful method to extend
packet-processing functionality in the Linux operating system.
BPF allows users to write code in high-level languages (like C or
Rust) and execute them at specific hooks in the kernel, such as the
network device driver.
To ensure safe execution of a user-developed BPF program in kernel
context, Linux uses an in-kernel static checker. The checker allows a
program to execute only if it can prove that the program is
crash-free, always accesses memory within safe bounds, and avoids
leaking kernel data.

BPF programming is not easy. One, even modest-sized BPF programs are
deemed too large to analyze and rejected by the kernel checker. Two, the
kernel checker may incorrectly determine that a BPF program exhibits
unsafe behaviors.  Three, even small performance optimizations to BPF
code (\eg 5\% gains) must be meticulously hand-crafted by expert
developers.
Traditional optimizing compilers for BPF are often
inadequate since the kernel checker's safety constraints are
incompatible with rule-based optimizations.

We present \TheCompiler, a program-synthesis-based compiler that automatically
optimizes BPF bytecode with formal correctness and safety guarantees.
\TheCompiler produces code with 6--26\% reduced size,
\MinLatencyReduction--\MaxLatencyReduction lower average
packet-processing latency, and
\MinThroughputGain--\MaxThroughputGain higher throughput
(packets per second per core) relative to the best {\ct
  clang}-compiled program, across benchmarks drawn from
Cilium, Facebook, and the Linux kernel.
\TheCompiler incorporates several domain-specific techniques
to make synthesis practical by accelerating
equivalence-checking of BPF programs by \AvgEqChkTimeReduction 
orders of magnitude.
\end{abstract}

\maketitle

\section{Introduction}
\label{sec:introduction}

The CPU efficiency of processing packets at servers is of paramount importance,
given the increasing volumes of data from large-scale applications, the
stagnation of Moore's law, the monetization of CPU cores in cloud computing, and
the stringent throughput and latency requirements of high-end applications. The
networking community has responded with several efforts, including innovations
in operating system packet-processing~\cite{tso, gro,
  scaling-linux-networking-stack, napi-linuxconf01}, user-space
stacks~\cite{netmap, dpdk, af-xdp, snap-sigcomm19, andromeda-nsdi18, ix,
  arrakis}, and programmable NIC offloads~\cite{azurenic-nsdi18,
  netronome-hotchips-talk, fungible-dpu, mellanox-bluefield,
  mellanox-bluefield2, marvell-octeon-tx}.

Recently, extended Berkeley Packet Filter (BPF) has emerged as a popular method
to achieve flexible and high-speed packet processing on the Linux operating
system. With roots in packet filtering in the early
90s~\cite{bpf-packet-capture-usenix93}, BPF has since evolved into a
general-purpose in-kernel virtual machine~\cite{xdp-conext18,
  bpf-universal-kernel-vm} with an expressive 64-bit RISC instruction set. BPF
code\footnote{In this paper, we use the term BPF throughout to denote the
  extended version of BPF, rather than ``classic'' BPF used to write packet
  filters.} has been widely deployed in production systems---implementing load
balancing~\cite{katran-facebook-talk, kubeproxy-replacement}, DDoS
protection~\cite{cloudflare-l4drop}, container policy
enforcement~\cite{cilium-google-kubernetes-engine}, application-level
proxying~\cite{calico-ebpf-dataplane}, intrusion-detection~\cite{suricata-xdp},
and low-level system monitoring~\cite{brendan-gregg-reinvent-talk-2019,
  datadog-file-integrity-monitoring}. Every packet sent to
Facebook~\cite{katran-facebook-talk} and CloudFlare~\cite{cloudflare-l4drop} is
processed by BPF software.

BPF enables users to extend the functionality of the operating system
without developing kernel software~\cite{bpf-microkernel}. The user
writes code in a high-level language (\eg C, Rust), and uses a standard compiler toolchain (\eg
Clang-9) to produce BPF bytecode. The operating system leverages
an in-kernel {\em static checker}, which analyzes the BPF bytecode to
determine if it is safe to run in kernel context. Specifically, the checker attempts to
prove that the program terminates, does not crash, does not access
kernel memory beyond safe permitted bounds, and does not leak
privileged kernel data. If the program is proved safe by the
kernel checker, it is downloaded into kernel memory and run
without any additional run-time checks. Otherwise, the
program is rejected. BPF programs can
be executed within several performance-critical parts of the
packet-processing stack~\cite{bpf-program-types}, like the
network device driver~\cite{xdp-conext18}, traffic
control~\cite{bpf-traffic-control}, congestion
control~\cite{bpf-tcp-cc-struct-ops}, and socket
filters~\cite{bpf-packet-capture-usenix93}.

BPF is unique in the combination of flexibility, safety, and
performance it enables for packet processing.  Unlike a kernel module
that may potentially crash the kernel or corrupt kernel memory, a BPF
program accepted by the kernel checker is guaranteed not to misbehave,
assuming that the checker and the BPF run-time are bug-free. Unlike
kernel-bypass stacks, BPF does not pin CPU cores, and retains the
user-kernel privilege separation and standard management tools (\eg
{\ct tcpdump}) available on operating systems~\cite{xdp-conext18,
  kopi-hotos21}.

Despite the promises of BPF, it is not easy to develop high-quality
packet-processing code in BPF today. We outline three challenges.

\Para{Challenge 1: Performance.} Optimizing the performance of BPF
code today is tantamount to optimizing assembly code. Userspace
profiling tools do not exist. Optimization support in compilers is
inadequate. For the benchmarks we tested, the standard compilation
toolchain (based on Clang-9) produced identical code under
optimization flags {\ct -O2} and {\ct -O3}, missing opportunities
available to optimize the {\ct -O2} code (\Sec{evaluation}).  Anecdotally, it is known
that even expert developers have to put in painstaking work to improve
performance of BPF code by small margins~\cite{accelerate-with-af-xdp,
  kubernetes-load-balancing-with-xdp,
  netronome-smaller-programs-greater-performance}. Yet, small
improvements are worthwhile: reducing even a few clock cycles per
packet is crucial to meeting the line rate at high speeds given
the limited budget of CPU clock cycles available to process each
packet~\cite{xdp-conext18,
  cpu-savings-xdp-load-balancer-kubeproxy-replacement-cilium,
  lowlatency-kth-thesis19}. Further, cutting the CPU usage
of networking decreases interference to workloads
co-hosted on the same machine~\cite{cpu-cost-of-networking,
  iron-nsdi18}.

\Para{Challenge 2: Size.} Running BPF programs beyond a
modest size poses challenges. The kernel checker limits the
complexity\footnote{Older kernels (prior to v5.2) rejected
programs with more than 4096 BPF bytecode instructions. On
modern kernels, this limit is still applicable to
non-privileged BPF program
types~\cite{bpf-program-size-limit} such as socket filters
and container-level packet
filters~\cite{bpf-nonprivileged-program-types}.  Since
kernel v5.2, there is a limit of 1
million~\cite{bpf-increase-complexity-limit,
  bpf-program-size-limit} on the number of instructions
examined by the checker's static analysis, which is a form
of symbolic execution~\cite{symbolic-exec-clarke76} with
pruning heuristics. Unfortunately, the number of examined
instructions explodes quickly with branching in the program,
resulting in many programs even smaller than 4096
instructions long being rejected due to this limit~\cite{
  cilium-complexity-issue-2, cilium-complexity-issue-3,
  cilium-complexity-issue-4, cilium-complexity-issue-5,
  cilium-complexity-issue-6}.} of the programs that it deems
acceptable~\cite{bpf-kernel-documentation, bpf-design-qa} to
keep the time it takes to load user programs small. In
practice, programs with even a few thousand instructions end
up being rejected~\cite{cilium-complexity-issues-list}.
%% due to the requirement that the
%% kernel checker must complete its safety checks fast, to
%% enable users to load BPF programs quickly. Hence, the kernel
%% checker rejects programs exceeding fixed limits on the
%% number and complexity of BPF bytecode instructions in the
%% program.\footnote{Historically (before kernel v5.2),
%% programs with more than 4096 instructions were rejected by
%% the kernel checker. Since kernel v5.2, there is a limit of 1
%% million~\cite{bpf-increase-complexity-limit,
%%   bpf-program-size-limit} on the number of instructions
%% examined by the checker's static analysis---a form of
%% symbolic execution~\cite{symbolic-exec-clarke76} with
%% pruning heuristics---which explodes quickly with branching
%% in the program. In practice, many programs smaller than 4096
%% instructions are still rejected by this new
%% limit~\cite{cilium-complexity-issues-list,
%%   cilium-complexity-issue-2, cilium-complexity-issue-3,
%%   cilium-complexity-issue-4, cilium-complexity-issue-5,
%%   cilium-complexity-issue-6}. Also, the old limit is still
%% applicable~\cite{bpf-program-size-limit} to older stable
%% kernels and non-privileged BPF program types such as socket
%% filters and container-level packet
%% filters~\cite{bpf-nonprivileged-program-types}.}
%
Further, hardware platforms supporting BPF offload are very
sensitive to program size, given their limited amount of
fast memory to hold the
program~\cite{netronome-smaller-programs-greater-performance}.
Compiler support for code compaction is deficient: for most
of our benchmarks, we found that {\ct clang -Os} produces
code of the same size as {\ct clang -O2}. The only recourse
for developers under size pressure is to refactor their
program~\cite{fristonio-bpf-size-issue,
  documentation-on-bpf-complexity-1,
  documentation-on-bpf-complexity-2}.

\Para{Challenge 3: Safety.} It is difficult to get even small programs past the kernel checker. The
checker's static analysis is incomplete and imprecise: it rejects many programs
which have semantically-equivalent rewrites that can be accepted
(\Sec{safety}). This makes it tricky to develop compilers that produce runnable
BPF bytecode. The developers of Clang's BPF backend work specifically towards
producing instruction sequences that the kernel checker will accept, \eg
\cite{clang-verifier-1, clang-verifier-2, clang-verifier-3, clang-verifier-4,
  clang-verifier-5}.
Producing checker-acceptable code is a major challenge in designing a BPF
backend to the {\ct gcc} compiler~\cite{compiling-to-bpf-with-gcc, bpf-in-gcc}.

Fundamentally, generating optimized, compact, and safe BPF code is challenging
due to the incompatibility between checker-enforced safety restrictions and
rule-based optimizations (\Sec{phase-ordering-problem}). We call this the {\em
  phase-ordering} problem in BPF compilation: producing safe, checker-acceptable
code precludes many traditional rule-based optimizations. Conversely, applying
optimizations produces code that the kernel checker rejects.

\Para{A synthesis-based compiler.} We present \TheCompiler, a compiler
which uses {\em program synthesis} to automatically generate safe,
compact, and performant BPF bytecode, starting from
unoptimized bytecode.  Program synthesis is the
task of searching for a program that
meets a given specification~\cite{sygus-cacm18}. An example of a
specification is that the outputs of the synthesized program must
match that of a source program on all inputs. Synthesis works by
searching through the space of programs, typically guided by
user-provided restrictions on the structure of the synthesized
program. For example, the synthesizer may search for programs that fit
a user-defined grammar~\cite{sketch-pldi05, sketch-asplos06}, use
smaller library components~\cite{oracle-component-synthesis-icse10,
  loopfree-pldi11}, or use a low-level instruction
set~\cite{binary-translation-osdi08, lens-asplos16, souper17,
  dataflow-pruning-oopsla20}.

While traditional compilers are designed to emit ``reasonable'' code within a
small time budget, synthesis-based compilers can produce high-quality code by
searching the space of programs more extensively over a longer time period. We
believe that the longer compilation time is worthwhile for BPF
programs, given their prevalence in deployed systems, their sensitivity to
performance, the difficulty of achieving even small performance gains, and their
portability across machines and
architectures~\cite{bpf-portability-core, btf-kernel-documentation}.

\TheCompiler makes three contributions.

\Para{Contribution 1: Stochastic synthesis for BPF
  (\Sec{stochastic-search}).}  \TheCompiler adapts stochastic
synthesis~\cite{stoke-asplos13, conditionally-correct-superopt-pldi15,
  loopsuperopt-pldi17} to the domain of the BPF instruction set. At a
high level, the algorithm runs a Markov chain to search for programs
with smaller values of a cost function that incorporates
correctness, safety, and performance. A new candidate program is
synthesized probabilistically using one of several 
rewrite rules that modify the current state (program) of the Markov chain.
The Markov chain transitions to the new state (synthesized program)
with a probability proportional to the reduction in the cost relative
to the current program. We show how we set up \TheCompiler to optimize
programs with diverse cost functions under safety constraints. We have
incorporated several domain-specific rewrites to
accelerate the search. At the end of the search, \TheCompiler produces
multiple optimized versions of the same input program.

\Para{Contribution 2: Techniques to equivalence-check BPF programs
  (\Sec{formalization},
  \Sec{optimizing-equivalence-checking}).}
\TheCompiler synthesizes programs that are formally
shown to be equivalent to the original program. 
To perform equivalence-checking, we formalize the input-output behavior of BPF
programs in first-order
logic (\Sec{formalization}). Our formalization includes the arithmetic
and logic instructions of BPF handled by earlier treatments of
BPF~\cite{untrusted-extensions-pldi19, serval-sosp19, jitsynth-cav20,
  jitterbug-osdi20}, and goes beyond prior work by incorporating
aliased memory access (using pointers) as well as BPF maps and helper
functions (\Sec{background}).
Equivalence-checking occurs within the inner loop
of synthesis, and it must be efficient
for synthesis to remain practical.
We present several
domain-specific techniques that
reduce the time required to check the input-output equivalence of two
BPF programs by five orders of magnitude
(\Sec{optimizing-equivalence-checking}). Consequently,
\TheCompiler can optimize real-world BPF
code used in production systems. 

\Para{Contribution 3: Techniques to check the safety of BPF programs
  (\Sec{safety}).} At each step of stochastic search, \TheCompiler
evaluates the safety of the candidate program. \TheCompiler
incorporates safety checks over the program's control flow and memory
accesses, as well as several kernel-checker-specific constraints. To
implement these checks, \TheCompiler employs static analysis and
discharges first-order-logic queries written over the candidate
program.

\TheCompiler resolves the phase-ordering problem of BPF compilation by
considering both performance and safety of candidate programs at each
step of the search.  While \TheCompiler's safety checks have
significant overlap with those of the kernel checker, the two sets of
checks are distinct, as the kernel checker is a complex body of code
that is under active development~\cite{bpf-mailing-list}. It is
possible, though unlikely, that \TheCompiler deems a program safe but
the kernel checker rejects it. To guarantee that \TheCompiler's
outputs are acceptable to the kernel checker, \TheCompiler has a
post-processing pass where it loads each of its best output programs
into the kernel and weeds out any that fail the
kernel checker. While the existence of this pass may appear to
bring back the phase-ordering problem, it is merely a fail-safe: as of
this writing, all of \TheCompiler's output programs resulting from the
search already pass the kernel checker.

\TheCompiler can consume BPF object files emitted by {\ct clang} and
produce an optimized, drop-in replacement. We present an evaluation of
the compiler across \NumBenchmarks programs drawn from the Linux kernel,
Cilium, and Facebook. Relative to the best {\ct clang}-compiled
variant (among {\ct -O2/-O3/-Os}), \TheCompiler can reduce
the size of BPF programs
by between 6--26\%, reduce average latency by \MinLatencyReduction--\MaxLatencyReduction,
and improve throughput (measured in packets
per second per core) by
\MinThroughputGain--\MaxThroughputGain. This is in
comparison to a state of the art
where significant effort is required from expert developers to produce
5--10\% performance gains~\cite{accelerate-with-af-xdp,
  katran-facebook-talk}.

\TheCompiler is an existence proof that domain-specific
application of program synthesis techniques is a viable
approach to automatically optimizing performance-critical
packet-processing code. We call upon the community to
explore such technology to alleviate the developer burden of
improving performance in other contexts like user-space
networking and programmable NICs. \TheCompiler's source
code, including all of our experimental scripts, is
available at \url{https://k2.cs.rutgers.edu/}.

\section{Background and Motivation}
\label{sec:background}

\subsection{Extended Berkeley Packet Filter (BPF)}

BPF is a general-purpose in-kernel virtual machine and instruction
set~\cite{bpf-universal-kernel-vm} that enables users to write operating system
extensions for Linux~\cite{bpf-microkernel}. A standard compiler (\eg Clang-9)
can be used to turn C/Rust programs into BPF {\em bytecode}, whose format is
independent of the underlying hardware architecture.

BPF programs are event-driven. BPF bytecode can be attached to specific events within the operating
system, such as the arrival of a packet at the network device
driver~\cite{xdp-conext18}, packet enqueue within Linux traffic
control~\cite{tc-bpf-man}, congestion control
processing~\cite{bpf-tcp-cc-struct-ops}, and socket system
call invocations~\cite{bpf-program-types}.

Stateful functionality is supported using {\em BPF helper functions.}  Helper
functions are implemented as part of the kernel and can be called by the BPF
program with appropriate parameters. For example, there are helper functions
that provide access to persistent key-value storage known as a {\em map}.  The
map-related helper functions include lookup, update, and delete. The arguments
to the map helpers include pointers to memory and file descriptors that uniquely
identify the maps. The list and functionality of helpers in the kernel are
steadily increasing; there are over 100 helpers in the latest kernel as of
this writing~\cite{bpf-helper-calls}.

The BPF instruction set follows a 64-bit RISC architecture. Each program has
access to eleven 64-bit registers, a program stack of size 512 bytes (referred to
by the stack pointer register {\ct r10}), and access to the memory containing
program inputs (such as packets) and some kernel data
structures (\eg socket buffers). The BPF instruction set
includes 32 and 64-bit arithmetic and logic
operations, signed and unsigned operations, and pointer-based load and store
instructions. BPF programs can be executed efficiently by
leveraging just-in-time (JIT) compilation to popular architectures like {\ct
  x86\_64} and ARM. BPF is not intended to be a Turing-complete language; it does not support
executing unbounded loops. User-provided BPF programs are run directly in kernel context. To ensure that it
is safe to do so, Linux leverages an in-kernel static checker.

\subsection{Phase Ordering in BPF Compilers}
\label{sec:phase-ordering-problem}

We illustrate why it is challenging to optimize BPF bytecode while
simultaneously satisfying the safety constraints enforced by the kernel
checker. These examples emerged from our experimentation with the checker in
kernel v5.4. In the programs below, we use {\ct r0 ... r9} for general-purpose BPF
registers. {\ct r10} holds the stack pointer.

\Para{Example 1. Invalid strength reduction.} The sequence
\begin{lstlisting}
  bpf_mov rY 0       // rY = 0
  bpf_stx rX rY      // *rX = rY
\end{lstlisting}
for some registers {\ct rX} $\neq$ {\ct rY} can usually be optimized to
the simpler single instruction
\begin{lstlisting}
  bpf_st_imm rX 0    // *rX = 0
\end{lstlisting}
However, the kernel checker mandates that a pointer into the program's
``context memory''~\cite{ctx-store-restriction-verifier} cannot be used to
store an immediate value. If {\ct rX} were such a pointer,
the program would be rejected.

\Para{Example 2. Invalid coalescing of memory accesses.} Consider the
instruction sequence
\begin{lstlisting}
  bpf_st_imm8 rX off1 0     // *(u8*)(rX + off1) = 0
  bpf_st_imm8 rX off2 0     // *(u8*)(rX + off2) = 0
\end{lstlisting}
where {\ct rX} is a safely-accessible memory address, and {\ct off1} and {\ct
  off2} are offsets such that {\ct off2} = {\ct off1} + 1. Usually, two such
1-byte writes can be combined into one 2-byte write:
\begin{lstlisting}
  bpf_st_imm16 rX off1 0    // *(u16*)(rX + off1) = 0
\end{lstlisting}
However, the kernel checker mandates that a store into the stack must be aligned
to the corresponding write
size~\cite{stack-aligned-restriction-verifier}. If {\ct rX} is {\ct r10}, the
stack pointer, and {\ct off1} is not 2-byte aligned, the checker will reject
the rewritten program.

In general, applying optimizations that pass the checker's constraints requires
compilers to be aware of the specific restrictions that impact each
optimization. The checker has numerous
restrictions~\cite{bpf-verifier-source-code, bpf-verifier-selftests}, making it
tedious to consider the cross-product of optimizations and safety conditions.

\subsection{\TheCompiler: A Program-Synthesis-Based Compiler}
\label{sec:compiler-overview}

We present \TheCompiler, a compiler that leverages {\em
  program synthesis} to consider correctness, performance,
and safety of programs together rather than piecemeal, to
resolve the phase-ordering problem between efficiency and
safety in BPF optimization.

Program synthesis is the combinatorial search problem of
finding a program that satisfies a given
specification. \App{program-synthesis-background} overviews
program synthesis approaches in the literature.
Given a sequence of instructions in the BPF bytecode format, we are interested
in synthesizing an alternative sequence of BPF instructions that satisfies the
specification that: (i) the synthesized program is equivalent to the source
program in its input-output behavior, (ii) the synthesized program is safe, and
(iii) the synthesized program is more efficient than the source program. The
precise definitions of efficiency and safety will be discussed in
\Sec{stochastic-search} and \Sec{safety}.

\begin{figure}
    \centering
    \includegraphics[width=0.3\textwidth]{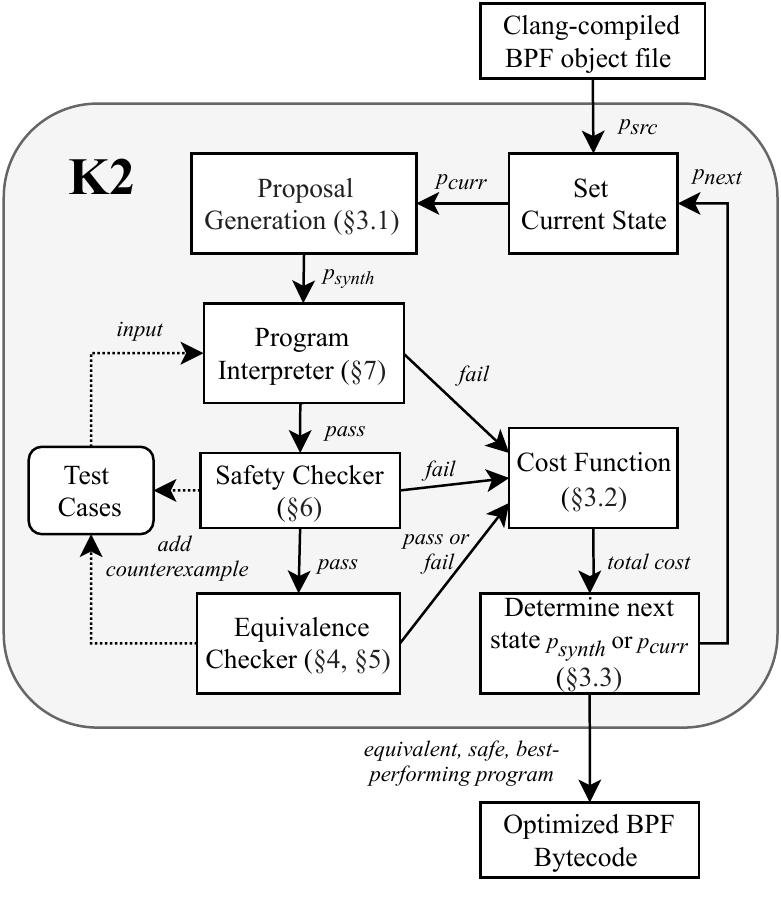}
    \caption{\mycaptionsize An overview of the \TheCompiler
      compiler. Solid arrows represent the flow of
      control. Dotted arrows represent the flow of data.}
    \label{fig:compiler-overview}
\end{figure}

\Fig{compiler-overview} presents an overview of
\TheCompiler, which synthesizes programs satisfying the
specification above.  \TheCompiler consumes Clang-compiled BPF bytecode,
and implements the {\em stochastic search} procedure described in \Sec{stochastic-search}. The
search process synthesizes {\em proposals,} which are candidate rewrites of the
bytecode. The proposal is evaluated against a suite of automatically-generated
test cases to quickly prune programs which are not equivalent
 to the source
program, or unsafe. If the proposal passes all tests, \TheCompiler uses formal
equivalence-checking (\Sec{formalization},
\Sec{optimizing-equivalence-checking}) and formal safety-checking (\Sec{safety})
to determine the value of a {\em cost function} over the proposal. 
The cost combines
correctness, safety, and performance characteristics, 
and is used to guide the search process towards better programs.
Formal equivalence-checking and safety-checking may generate {\em
  counterexamples,} \ie inputs where the proposal's output differs from that of
the original bytecode, or the proposal exhibits unsafe behaviors. These tests
are added to the test suite, to enable quick pruning of similar
programs in the future.
We
describe aspects of the compiler's implementation, including the BPF program
interpreter we developed in \Sec{implementation}.

\section{Stochastic Optimization of BPF}
\label{sec:stochastic-search}

The \TheCompiler compiler translates programs from BPF bytecode to BPF
bytecode. \TheCompiler uses the stochastic optimization framework, introduced in
STOKE~\cite{stoke-asplos13}, which applies a Markov Chain Monte Carlo (MCMC)
sampling approach to optimize a cost function over the space of programs.

At a high level, MCMC is a method to sample states from a probability
distribution over states. When we apply MCMC to program optimization, the state
is a program of a fixed size. A well-known MCMC sampler, the Metropolis-Hastings
(MH) algorithm~\cite{mcmc-in-practice-text96}, works as
follows. From an initial state, at
each step, the algorithm proposes a new state to transition to, using transition
probabilities between states (\Sec{proposal-generation}). The algorithm computes
a cost function over the proposal (\Sec{search-objective-function}) and
determines whether to {\em accept} or {\em reject} the proposed new state
(\Sec{proposal-acceptance}) based on the cost.  If the proposal is accepted, the
proposed state becomes the new state of the Markov chain. If not, the current
state is the new state of the Markov chain. In the asymptotic limit, under mild
conditions on the transition probabilities~\cite{mcmc-in-practice-text96}, the
set of all accepted states form a representative sample of the steady-state
probability distribution.

\Para{Why stochastic synthesis?} Among the program synthesis approaches in the literature
(\App{program-synthesis-background}), \TheCompiler adopts
stochastic search primarily because it can optimize complex
cost functions, \eg the number of cache misses during
program execution, with complex constraints, \ie
safety. MCMC uses a standard transformation to turn very
general cost functions (\Sec{search-objective-function})
into steady-state probability distributions, enabling it to
perform optimization by sampling from the corresponding
distribution~\cite{mcmc-in-practice-text96, stoke-asplos13,
  ethanfetaya-mcmc-lecture-notes}.

\subsection{Proposal Generation}
\label{sec:proposal-generation}

The Markov chain starts by setting its initial state to
$p_{src}$, the input program. Starting from any current state $p_{curr}$, we
generate a candidate rewrite, \ie a proposal $p_{synth}$,
using one of the rules below, chosen randomly with fixed
probabilities $prob_{(.)}$:
\begin{enumerate}
\item \textbf{Replace an instruction ($prob_{ir}$)}: at random, choose
  an instruction from $p_{curr}$, and modify both its opcode
  and operands. For example, change
  {\ct bpf\_add r1 4} to {\ct bpf\_mov r4 r2}.
\item \textbf{Replace an operand ($prob_{or}$)}: at random, choose an
  instruction and replace one of its operands with another value of the same
  type. For example, change {\ct bpf\_add r1 4} to {\ct bpf\_add r1 10}.
\item \textbf{Replace by NOP ($prob_{nr}$)}: at random, choose an
  instruction and replace it with a {\ct nop}, effectively
  reducing the number of instructions in the program.
\item \textbf{Exchange memory type 1 ($prob_{me1}$)}: at
  random, choose an instruction, and if it is a memory-based
  instruction (\ie a load or a store), sample a new width
  for the memory operation and a new immediate or register
  operand. The instruction's memory address operand (\ie
  address base and offset) as well as its type (load \vs
  store) are unchanged.  For example, change {\ct r1 =
    *(u16*)(r2 - 4)} to  {\ct r3 = *(u32*)(r2 - 4).}
\item \textbf{Exchange memory type 2 ($prob_{me2}$):} at random,
  choose an instruction, and if it is a memory-based
  instruction, sample a new width for the memory
  operation. All other instruction operands are
  unchanged. For example, change  {\ct r1 = *(u16*)(r2 - 4)}
  to {\ct r1 = *(u32*)(r2 - 4).}
\item \textbf{Replace contiguous instructions
  ($prob_{cir}$)}: at random, choose up to $k$ contiguous
  instructions (we pick $k=2$) and replace all of them with
  new instructions.
\end{enumerate}

These rewrite rules define the transition probabilities of
the Markov chain, which we denote by $tr(p_{curr}
\rightarrow p_{synth})$. We use the probabilities
$prob_{(\cdot)}$ shown in \Tab{parameter-settings-sets}
(\App{parameter-set}). In our experience, any probabilities
that allow the Markov chain to move ``freely'' through the
space of programs suffice to find programs better than the
input. 

\Para{Non-orthogonality of rewrite rules.} The rewrite rules
above are not mutually exclusive in the program
modifications they affect. For example, replacement by NOP
(rule 3) is just a specific version of the more general
instruction replacement (rule 1). Given enough time, a
small set of general rules is sufficient to explore the
space of programs. However, the existence of more specific
rules accelerates the convergence of the Markov chain to
better programs.

\Para{Domain-specificity of rewrite rules.} STOKE and its
variants~\cite{stoke-asplos13, loopsuperopt-pldi17} proposed
variants of the rewrite rules (1--3) above. Rules (4), (5),
and (6) are domain-specific rules that \TheCompiler uses to
accelerate the search for better BPF programs. Rules (4) and
(5) help identify memory-based code optimizations
(\Sec{optimizations-discovered-by-k2}). Rule (6) captures
one-shot replacements of multiple instructions, \eg
replacing a register addition followed by a store into a
single memory-add instruction.  These domain-specific rules
improve both the quality of the
resulting programs and the time to find better
programs (\Sec{evaluation}, \App{parameter-set}). 

\subsection{Cost Function}
\label{sec:search-objective-function}

We compute a cost function over each candidate program.  The cost function
$f(p)$ contains three components: an error cost, a performance cost,
and a safety cost. 

\Para{Error cost.} The error cost function $err(p)$ is 0 if and only if the
program $p$ produces the same output as the source program $p_{src}$ on all
inputs. We would like a function that provides a smooth measure of the
correctness of program $p$ with respect to the source program $p_{src}$, to
guide the search towards ``increasingly correct'' programs. Similar to STOKE, we
incorporate test cases as well as formal equivalence checking
(\Sec{formalization} \& \Sec{optimizing-equivalence-checking}) to compute an
error cost.  Using a set of tests $T$ and executing $p_{synth}$ on each test
$t \in T$, we set
\begin{equation}
\label{eq:error-cost-function}
  \begin{aligned}
err(p)\ := & \ c \cdot \sum_{t \in T}
\mathit{diff}(o_{p_{synth}(t)}, o_{p_{src}(t)})\ + \\
& \mathit{unequal}\ \cdot\ \mathit{num\_tests}
\end{aligned}
\end{equation}

where:

\begin{itemize}
    \item $o_{p_{synth}(t)}$ and $o_{p_{src}(t)}$ are the outputs of the proposal and
      the source program on test case $t$, 
    \item {$\mathit{diff}(x, y)$} is a measure of the distance between two
      values. We consider two variants: (i) $\mathit{diff}_{pop}(x, y) :=
      \mathit{popcount}(x \oplus y)$ is the number of bits that differ between
      $x$ and $y$, and (ii) $\mathit{diff}_{abs}(x, y) := \mathit{abs}(x - y)$,
      which represents the absolute value of the numerical difference between
      $x$ and $y$. Relative to STOKE, which only considers $\mathit{popcount}$
      as the semantic distance between values, we also find that many
      packet-processing programs require numeric correctness (\eg counters),
      captured via $\mathit{diff}_{abs}(.)$.
    \item $c$ is a normalizing constant denoting the weight of each test
      case. STOKE adds the full error cost for each test case, setting $c =
      c_{full} = 1$. We also explore a second variant, $c_{avg} = 1/|T|$, where
      $|T|$ is the number of test cases, to normalize the contributions of the
      many test cases we require to prune complex, ``almost correct'' BPF
      programs.
    \item $\mathit{unequal}$ is 0 if the first-order-logic formalization of the
      two BPF programs (\Sec{formalization}) finds that the programs are
      equivalent,
      else it is 1. We only run
      equivalence-checking if all test cases pass, since it
      is time-consuming. If any test case fails, we set
      $\mathit{unequal}$ to 1.
    \item $\mathit{num\_tests}$ includes two variants: (i) the number of
      test cases on which $p$ produced incorrect outputs, and (ii) the number of
      test cases on which $p$ produced {\em correct}
      outputs. STOKE uses only the first variant.
      We consider the second variant to distinguish a program that is
      equivalent to the source program from one that satisfies all the test
      cases but is not equivalent.
\end{itemize}

Considering all variants from equation (\ref{eq:error-cost-function}), there are
8 error cost functions. We run MCMC with each cost function in parallel and
return the best-performing programs among all of them.

\Para{Performance cost.} We use two kinds of performance costs corresponding to
different scenarios, namely optimizing for program {\em size} and program {\em
  performance.}

The function $perf_{inst}(p_{synth})$ (instruction count) is the number of extra
instructions in $p_{synth}$ relative to $p_{src}$.

The function $perf_{lat}(p_{synth})$ is an estimate of the additional latency of
executing program $p_{synth}$ relative to $p_{src}$.
Unfortunately, executing a candidate BPF program $p_{synth}$ to directly
measure its latency is unviable, since the kernel checker
will reject most candidate programs.
Instead,  we profile every instruction 
of the BPF instruction set by executing each opcode millions of
times on a lightly loaded system, and determining an average execution time
$\mathit{exec}(i)$ for each opcode $i$. The performance cost function is
the difference
of the sum of all the opcode latencies, \ie $perf_{lat}(p_{synth}) := \sum_{i_{synth} \in p_{synth}}
\mathit{exec}(i_{synth}) - \sum_{i_{src} \in p_{src}}
\mathit{exec}(i_{src})$.

\nop{
  While there is significant correlation between the latency estimated through the
method above and the actual execution latency of the programs, the performance
cost can be improved in at least two ways. First, packet-processing
programs exhibit many memory accesses, which show significant variability in
latency due to the processor's cache hierarchy. In general, estimating execution
latency for memory-access-heavy programs is a challenging
problem~\cite{stoke-asplos13, lowlatency-kth-thesis19,
  nfv-perf-prediction-sigcomm20}. Second, we consider the sum of the estimated
latencies of {\em all} instructions in the program, disregarding the actual set
of instructions that are executed on any given input. Modeling the performance
of specific code paths accurately, \eg to optimize precisely the ``fast path''
of the packet-processing program, is a topic we leave for future work. 
}

\Para{Safety cost.} To our knowledge, \TheCompiler is the first synthesizing
compiler to incorporate generic safety constraints in first-order logic into
synthesis. The safety properties considered by \TheCompiler are
described in \Sec{safety}. Our approach to dealing with unsafe programs is
simple: once a program $p_{synth}$ is deemed unsafe, we set $safe(p_{synth})$ to
a large value $ERR\_MAX$, leaving just a small probability for it to be accepted
into the Markov chain. We set $safe(p_{synth}) = 0$ for safe programs.  We do
not simply reject unsafe programs because the path from the current program to a
more performant and safe program in the Markov chain may pass through an
unsafe program (for some intuition on why, see
Fig. 4 in \cite{stoke-asplos13}). We leave
formulating smooth cost functions to guide the search
through progressively ``safer'' programs to future work.

\nop{
\eg to rewrite programs rejected by the Linux BPF static checker into a
semantically-equivalent instruction sequence that is accepted by the checker.
More generally, incorporating safety as a first-class consideration into program
synthesis is a way to break the phase ordering problem, \ie enabling several
optimizations while being safe, which is important for plugin compilers like
those of BPF and  WebAssembly, which need to isolate plugin faults from the rest
of the system.\as{Reworded. Please check previous sentence.}
}

The final cost function we use is $\alpha * err(p_{synth}) + \beta *
perf(p_{synth}) + \gamma * safe(p_{synth})$. We run parallel Markov chains with
different $(\alpha, \beta, \gamma)$ and return the programs
with the least performance costs.

\subsection{Proposal Acceptance}
\label{sec:proposal-acceptance}

To determine whether a candidate proposal should be used as the next state of
the Markov chain, the cost $f(p_{synth})$ is
turned into the probability of $p_{synth}$ in the
steady-state distribution, as follows~\cite{mcmc-in-practice-text96}:
\begin{equation}
  \pi(p_{synth}) = e^{-\beta \cdot f(p_{synth})}/Z
\end{equation}
where $Z = \sum_{p} e^{-\beta \cdot f(p)}$. The Metropolis-Hastings algorithm
computes an {\em acceptance probability} for $p_{synth}$ as follows:
\begin{equation}
  \label{eq:acceptance-probability}
\alpha = \mathit{min}\left(1, \frac{\pi(p_{synth}) \cdot tr(p_{synth} \rightarrow p_{curr})}{\pi(p_{curr}) \cdot tr(p_{curr} \rightarrow p_{synth})}\right)
\end{equation}

With probability $\alpha$, the next state of the Markov
chain is set to $p_{synth}$, else the next state is just
$p_{curr}$.
Here, the $tr(.)$ are the transition probabilities between
programs (\Sec{proposal-generation}). Intuitively,
$p_{synth}$ is always accepted if its cost is lower than
that of $p_{curr}$. Otherwise, $p_{synth}$ is
accepted with a probability that decreases with the increase in the cost of 
$p_{synth}$ relative to $p_{curr}$.
%
%% Note that the constant $Z$ is not required to compute
%% equation (\ref{eq:acceptance-probability}).

\TheCompiler repeats the process in \Sec{proposal-generation},
\Sec{search-objective-function}, and \Sec{proposal-acceptance} from the new
state of the Markov chain, looping until a timeout.

\section{Checking the Equivalence of BPF Programs}
\label{sec:formalization}
\label{sec:program-equivalence-checking}

\TheCompiler synthesizes output programs that are formally
shown to be equivalent to the input program.
To do this, we first formalize the input-output behavior of
the two programs in first-order
logic, using the theory of bit
vectors~\cite{decision-procedures-book}.
We identify the input and output registers of the two
programs based on the kernel hook they attach
to~\cite{bpf-kernel-documentation}.
Then, we dispatch the logic query below to a solver:
\begin{lstlisting}
inputs to program 1 == inputs to program 2
(* $\land$ *) input-output behavior of program 1 
(* $\land$ *) input-output behavior of program 2
(* $\Rightarrow$ *) outputs of program 1 != outputs of program 2
\end{lstlisting}

If the formula is satisfiable, there is a common input that causes the outputs
of the two programs to differ, which is added to the test suite
(\Sec{stochastic-search}). If the formula is unsatisfiable, the two programs
are equivalent in terms of input-output behaviors. 

The rest of this section describes how we obtain the input-output behavior of a
single program in first-order logic. Our formalization handles arithmetic and
logic instructions (\Sec{arithmetic-instruction-formalization}), memory access
instructions (\Sec{memory-instruction-formalization}), and BPF maps and other
helper functions (\Sec{map-formalization}). We have checked the soundness of our
formalization using a test suite that compares the outputs produced by the logic
formulas against the result of executing the instructions
with given inputs.

\Para{Preliminaries.}
\label{sec:preliminaries-equivalence-checking}
We begin by reordering the instructions in the program so that all control flow only moves
forward. This is possible to do when a BPF program
does not contain any loops. Then, we convert the entire program into
static-single-assignment (SSA) form~\cite{compiler-implementation-appel04,
  ssa-toplas91}. The result after SSA conversion is a sequence of BPF bytecode
instructions where (i) each assignment to a register uses a
fresh label with a version number, \eg
{\ct bpf\_mov r0 1; bpf\_mov r0 2} is turned into {\ct bpf\_mov r0\_v1 1;
  bpf\_mov r0\_v2 2}, and (ii) each statement is associated with a well-defined
path condition~\cite{static-analysis-book}. For example, in the instruction
sequence,
\begin{lstlisting}
  bpf_jeq r1 0 1   // if r1 != 0:
  bpf_mov r2 1     //    r2 = 1
\end{lstlisting}
the second instruction is associated with the path condition {\ct r1!=0.} 

At the highest level, we construct first-order formulas corresponding to each
instruction, and conjoin them, \ie through the logical conjunction operator
$\land$, to produce a final formula that represents the input-output
relationship of the entire program. Now we discuss how \TheCompiler formalizes
each kind of instruction.

\subsection{Arithmetic And Logic Instructions}
\label{sec:arithmetic-instruction-formalization}

To model register-based arithmetic and logic instructions, we represent each
version of each register using a 64-bit-wide bit vector data type. The action of
each instruction is formalized by representing its impact on all the registers
involved. Our formalization handles both 32-bit and 64-bit opcodes, as well as
signed and unsigned interpretations of the data.

As an example, consider the 32-bit arithmetic instruction {\ct bpf\_add32 dst
  src} (opcode {\ct 0x04}) which has the action of taking the least significant
32 bits of the registers {\ct dst} and {\ct src}, adding them, and writing back
a (possibly truncated) 32-bit result into {\ct dst}, zeroing out the most
significant 32 bits of the {\ct dst} register.

Suppose we are given a single instruction {\ct bpf\_add32 dst\_x src\_y} (after
SSA) where {\ct x} and {\ct y} represent the version numbers of {\ct dst} and
{\ct src}, respectively. Suppose the result is stored in {\ct dst\_z}. This
instruction results in the formula
\begin{lstlisting}
  (tmp == (dst_x.extract(31, 0) +
    src_y.extract(31, 0) ) ) (* $\land$ *)
  (dst_z == concat(
    bv32(0), tmp.extract(31, 0) ) )
\end{lstlisting}
where {\ct tmp} is a fresh variable to hold the intermediate result of the
32-bit addition of {\ct dst\_x} and {\ct src\_y}, {\ct extract(a, b)} represents
the effect of picking up bits a...b of a given bit vector, {\ct concat(x, y)}
represents the bit vector produced by concatenating the two bit vectors x and y,
with x occupying the higher bits of significance in the result, and {\ct
  bv32(0)} is a 32-bit bit vector representing 0.

Similarly, we have constructed semantic representations of all 64-bit and 32-bit
arithmetic and logic instructions~\cite{bpf-instruction-set}. 

\subsection{Memory Access Instructions}
\label{sec:memory-instruction-formalization}

BPF supports memory load and store instructions of varying
sizes~\cite{bpf-instruction-set} using pointers. We encode memory operations
directly in the theory of bit vectors to produce an efficient encoding in a
single first-order theory. We show how this encoding occurs in three steps.

\Para{Step 1: Handling loads without any stores.} Suppose a BPF program contains
no stores to memory, and only load instructions, \ie {\ct bpf\_ld rX rY}. To
keep the descriptions simple, from here on we will use the notation {\ct
  rX = *rY} to represent the instruction above.

The key challenge in encoding loads is handling {\em aliasing}, \ie different
pointers {\ct rY} might point to the same memory region, and hence the different
{\ct rX} must have the same value.

Suppose the $i^{th}$ load instruction encountered in the program reads from
memory address {\ct rY\_i} and loads into register {\ct rX\_i}. Then for the
$i^{th}$ load, we conjoin the formula \begin{lstlisting}
  $\underset{j < i}{\land}$ (rY_j == rY_i (* $\Rightarrow$ *) rX_j == rX_i)
\end{lstlisting}

Formulating this formula requires maintaining all the previous loads in the
program that might affect a given load instruction. To achieve this,
\TheCompiler maintains a {\em memory read table} for the program: the source and
destination of each load is added to this table in the order of appearance in
the post-SSA instruction sequence. \TheCompiler handles partial overlaps in
loaded addresses by expanding multi-byte loads into multiple single-byte loads.

\Para{Step 2: Handling stores and loads in straight-line programs.} Stores
complicate the formula above due to the fact that a load of the form {\ct rX =
  *rY} must capture the {\em latest write} to the memory pointed to by {\ct
  rY}. For example, in the instruction sequence {\ct rX\_1 = *rY; *rY = 4; rX\_2 =
  *rY}, the first and second load from {\ct rY} may return different values to
be stored in {\ct rX\_1} and {\ct rX\_2}.

Suppose the program contains no branches. Then, the latest write to a memory
address can be captured by the most recent store instruction (in order of
encountered SSA instructions), if any, that writes to the
same address.
\TheCompiler maintains a {\em memory write table}, which records the memory
address and stored variable corresponding to each store in the
program. Suppose $k$ stores of the form {\ct *rY\_i = rX\_i} ({\ct i} from {\ct
  1 $\cdots$ k}) have been encountered in the program before the load
instruction {\ct rX\_l = *rY\_l}. Then, the load is encoded by the formula
\begin{lstlisting}
  (* $\underset{j: j \leq k}{\land}$ *) (* $\underset{i: j < i \leq k}{\land}$ *) ! (rY_i == rY_l)
           (* $\land$ *)    rY_j  == rY_l
           (* $\Rightarrow$ *)    rX_j  == rX_l 
\end{lstlisting}
The formula $\underset{i: j < i \leq k}{\land}$ {\ct ! (rY\_i == rY\_l)}
asserts that the address loaded isn't any of the addresses
from stores that are more recent than store $j$. Hence, if
{\ct rY\_j == rY\_l}, the loaded value must come from store $j$.

\nop{The above condition only captures loads whose values
  were written in prior stores. } Informally, the overall
formula for a load instruction takes the form: if the
address was touched by a prior store, use the value from
that store, otherwise use the aliasing clauses from the
``loads-only'' case in step (1) above.\footnote{It is
possible for a load to occur without a prior store \eg when
an instruction reads from input packet memory.} Together,
step (1) and step (2) complete \TheCompiler's encoding of
memory accesses for straight-line programs.

\Para{Step 3: Handling control flow.} We construct a single formula per
instruction including control flow akin to bounded model
checking~\cite{bounded-model-checking-biere99}. Our key insight to generalize
the encoding from step (2) above is to additionally check whether the path
condition of the load instruction is implied by the path condition of the prior
store:

\begin{lstlisting}
  (* $\underset{j: j \leq k}{\land}$ *) (* $\underset{i: j < i \leq k}{\land}$ *) ! (rY_i == rY_l (* $\land$ *) pc_i (* $\Rightarrow$ *) pc_l)
           (* $\land$ *)    (rY_j  == rY_l (* $\land$ *) pc_j (* $\Rightarrow$ *) pc_l)
           (* $\Rightarrow$ *)    rX_j  == rX_l 
\end{lstlisting}

Note that the path conditions {\ct pc\_j} of each load or
store $j$ are already computed by \TheCompiler during the
preliminary SSA pass.

\subsection{BPF Maps and Helper Functions}
\label{sec:map-formalization}

BPF helpers (\Sec{background}) provide special functionality
in a program, including stateful operation. Due to space
constraints, we only briefly discuss our formalization of
BPF maps---the most frequently used helpers---in this
subsection. A more detailed treatment of maps and other
helpers is available in \App{map-formalization}.

\Para{Maps.} BPF maps are similar to memory, in that they
can be read and written using a key (rather than an
address). However, two features make BPF maps very different
from memory.

First, the inputs to lookup, update, or delete a map entry in BPF's map API are
all pointers to memory holding a key or value. This results in {\em two levels
  of aliasing}: distinct pointers may point to the same location in
memory (like regular pointer aliasing); additionally, distinct locations in
memory may hold the same key, which must result in the same value upon a map
look-up. Intuitively, we handle these two levels by keeping two pairs of tables
for read and write operations. The first pair of read/write tables tracks the
contents of the addresses corresponding to the key and value
pointers, as in \Sec{memory-instruction-formalization}. The
second pair of read/write tables tracks the updates to the
value pointers corresponding to the map's actual keys.

Second, keys in a map can be deleted, unlike addresses in
memory. Our encoding treats a deletion as a update of the
value pointer to {\ct 0} (a null pointer) for the
corresponding key, so that any subsequent lookup returns
null, mimicking the BPF lookup function semantics.

\Para{Other helper functions.} Each helper function
considered for optimization ideally should be formalized
using its specific semantics. We have added formalizations
for helpers used to obtain random numbers, access the
current Unix timestamp, adjust memory headroom in a packet
buffer, and get the ID of the processor on which the program
is running.

The list of BPF helpers currently numbers in the hundreds
and is growing~\cite{bpf-helper-calls,
  bpf-calling-kernel-functions}. For most helpers, it is
possible to model the function application as a call to an
{\em uninterpreted function}
$f(.)$~\cite{calculus-of-computation-book}: the only
governing condition on the input-output behavior of the
function is that calling it with the same inputs will
produce the same outputs, \ie $x == y \Rightarrow f(x) ==
f(y)$. (Stateful functions include the state as part of the
inputs.)  While such modeling is general, it limits the
scope of optimization across function calls, since it is
impossible to prove equivalence of code involving
uninterpreted functions without requiring that the sequence
of function calls and the inputs to each function call must
be exactly the same in the input and the output programs.

\section{Fast Equivalence Checking}
\label{sec:optimizing-equivalence-checking}

The formulas generated in
\Sec{arithmetic-instruction-formalization}--\Sec{map-formalization} are in
principle sufficient to verify the equivalence of all BPF programs we have
tested. However, the corresponding verification task is too slow
(\Sec{evaluation}).
Equivalence-checking time grows quickly with the number of branches, the number
of memory accesses, and the number of distinct maps looked up in the BPF
programs. Equivalence checking is in the inner loop of synthesis
(\Fig{compiler-overview}): large verification times render synthesis
impractical.

We have developed several optimizations that accelerate
equivalence-checking times by \AvgEqChkTimeReduction orders of magnitude on
average over the programs we tested. This section summarizes
the key ideas; more details are available in
\App{optimizing-equivalence-checking}. Several optimizations
leverage lightweight static analysis that is only feasible
due to the restrictions in the BPF instruction set.

The time to solve a logic formula is often reduced significantly by assigning
specific values to, \ie {\em concretizing,} formula terms whose value is
otherwise unconstrained, \ie
symbolic~\cite{bounded-model-checking-biere99, symbolic-exec-clarke76,
  symbolic-exec-king76, finding-code-oopsla18,
  fixing-code-vmcai20}. Our first
three optimizations are of this kind.

\Para{I. Memory type concretization.} All pointers to memory in BPF programs
have well-defined provenance, \ie it is possible to develop a static analysis to
soundly and completely track the {\em type} of memory (stack, packet, \etc) that
each pointer references. This allows \TheCompiler to maintain separate read and
write tables (\Sec{memory-instruction-formalization}) for each memory
type. Consequently, the size of aliasing-related formulas reduces from
$O((\sum_{t} N_{t})^2)$ to $O(\sum_{t} N_{t}^2)$, where
$N_{t}$ refers to the number of accesses to memory of a specific type $t$.

\Para{II. Map type concretization.} Similar to memory-type concretization, a
simple static analysis can soundly and completely determine the map that is used
for a specific lookup or update instruction. This has the effect of breaking map
accesses across several maps in the two-level map tables
(\Sec{map-formalization}) into separate map-specific two-level tables.

\Para{III. Memory offset concretization.} Many packet-processing programs perform
reads and writes into memory at offsets that can be determined at compile time,
for example, specific packet header fields. We developed a ``best-effort''
static analysis to soundly determine if a pointer holds a reference to a
compile-time-known offset into a memory region.  If such a constant offset is
determined, a formula like {\ct rY\_i == rY\_l} (appearing in
\Sec{memory-instruction-formalization}) can be simplified to {\ct constant ==
  rY\_l}, or even {\ct constant1 == constant2}. The latter doesn't even require
a solver to be evaluated, and can result in several cascading clause
simplifications. In the limit, if all offsets can be concretely determined, this
optimization has the effect of modeling the entire memory as
if it is a set of named registers. If we cannot statically determine concrete offsets, we fall
back to the symbolic formulas described in
\Sec{memory-instruction-formalization}.

\Para{IV. Modular verification.} \TheCompiler scales to large programs by
synthesizing and verifying instruction sequences of smaller length within
localized ``windows'' in the program, and then combining the results across the
windows. Hence, \TheCompiler pares down the verification task to
correspond to the size of the window rather than that of the
full program.
Effectively, this would turn \TheCompiler into a peephole
optimizer~\cite{peephole-cacm65}. However, traditional peephole optimizers
necessitate that the rewrites must apply in {\em any} program context, rejecting
many strong optimizations that could work conditionally
within a specific part of the program (\eg {\ct r1 *= r3}
may be changed into {\ct r1 $< <$= 2} if the value of {\ct r3}
is known to be 2).  To discover strong
optimizations but keep equivalence-checking fast, we
develop window-based formulas that use
stronger preconditions and weaker postconditions than
peephole optimizers. \TheCompiler leverages variable {\em
  liveness} (as in prior work~\cite{peephole-asplos06}) as
well as {\em concrete values} of the live variables, both of
which are inferred through static analysis:

\begin{lstlisting}
variables live into window 1
== variables live into window 2
(* $\land$ *) inferred concrete valuations of variables
(* $\land$ *) input-output behavior of window 1 
(* $\land$ *) input-output behavior of window 2
(* $\Rightarrow$ *) variables live out of window 1
    != variables live out of window 2
\end{lstlisting}

\Para{V. Caching.}  We cache the outcomes of equivalence-checking a candidate
program to quickly determine if a structurally-similar program was checked
earlier. This has the effect of reducing the number of times we call the solver.
We canonicalize the program by removing dead code before checking the cache.

\section{Safety of BPF programs}
\label{sec:safety}

\TheCompiler ensures that the programs returned by the compiler are
\textit{safe}, which requires proving specific control-flow and
memory-access safety properties about the output
programs, described below.

\TheCompiler's safety checks are implemented using static
analysis and first-order logic queries over the candidate
programs generated at each step of the stochastic search
(\Sec{stochastic-search}). By considering safety with
optimization at each step, \TheCompiler resolves the
phase-ordering problem (\Sec{phase-ordering-problem}) that
hampers traditional optimizing compilers for BPF.

\TheCompiler's safety checks are distinct from those of the
kernel checker, though there is a significant overlap
between them.  We developed safety-checking directly within
\TheCompiler, eschewing the alternative approach of invoking
the kernel checker on a candidate program at each step of
search, for two reasons.
First, in addition to reporting that a program is unsafe,
\TheCompiler's safety queries also return
a {\em safety counterexample}, \ie an input that
causes the program to exhibit unsafe behaviors. The
counterexample can be added to the test suite
(\Fig{compiler-overview}) to prune unsafe programs by
executing them in the
interpreter, rather than using an expensive kernel checker
(system) call. This has the overall effect of speeding up
the search loop.
Second, the kernel checker is a complex piece of software
that is evolving constantly. We believe
that, over the long term, a logic-based declarative encoding
of the safety intent will make it easier to understand and
maintain the compiler's safety constraints.

\TheCompiler guarantees that the output programs returned to
the user will pass the kernel checker. \TheCompiler
achieves this using a post-processing pass: outputs
from \TheCompiler's search loop which fail the kernel checker
are removed before presenting them to the user. As of this
writing, all the outputs from \TheCompiler's search already
pass the kernel checker without being filtered by this
post-processing.

Now we discuss \TheCompiler-enforced safety properties in detail.

\Para{Control flow safety.} The structure of BPF jump
instructions~\cite{bpf-instruction-set} allows the set of possible jump
targets in the program to be determined at compile time.
Hence, \TheCompiler constructs the complete control flow graph over basic blocks
at compile time~\cite{compiler-implementation-appel04}. Programs synthesized by
\TheCompiler satisfy the following safety properties:
\begin{enumerate}
\item There are no unreachable basic blocks. 
\item The program is loop-free (\ie no ``back-edges'' in the control flow),
    and hence, terminates. \TheCompiler ensures this during
    proposal generation (\Sec{proposal-generation}) by only producing jump offsets taking
    control flow ``forward'' in a topologically-sorted list of basic blocks.
  \item The program has no out-of-bounds jumps. \TheCompiler ensures this by
    only synthesizing jump targets that are within the program's valid
    set of instructions.
\end{enumerate}

The rest of the safety checks below are implemented using first-order
logic queries. Logic queries provide safety counterexamples, which
also allow \TheCompiler to prune an unsafe program using the
interpreter rather than an expensive solver query down the road.  To
our knowledge, \TheCompiler is the first to leverage counterexamples
for both correctness and safety during synthesis.

\Para{Memory accesses within bounds.} \TheCompiler ensures that programs it
synthesizes only access operating system memory within the bounds they are
allowed to. The access bounds for each type of memory are known ahead of
time. For example, the size of the program stack is fixed to 512
bytes~\cite{bpf-kernel-documentation}; packet inputs are provided with metadata
on the start and end addresses; and BPF map values have a pre-defined fixed size
based on the known attributes of the map.

\TheCompiler leverages a sound and complete static analysis to determine the
type of memory that a load or store instruction uses. Then, \TheCompiler
formulates a first-order query to determine if there are any program inputs that
cause the access to violate the known safe bounds of that memory. \TheCompiler
considers both the offset and the size of the access, and models the types of
pointers returned from BPF kernel helper functions very precisely. For example,
the instruction sequence corresponding to {\ct r0 = bpf\_map\_lookup(...); r1 =
  *r0; } will produce a safety counterexample for the case when the lookup
returns a NULL pointer. However, {\ct r0 = bpf\_map\_lookup(...); if (r0 != 0)
  \{ r1 = *r0; \} } is considered safe, since the path condition ensures a valid
value for {\ct r0}.

\Para{Memory-specific safety considerations.}  The BPF kernel checker explicitly
requires that a stack memory address cannot be read by a BPF program before that
address is written to~\cite{bpf-kernel-documentation}. The same rule applies to
registers which are not program inputs. This restriction is distinct
from placing safe bounds on an address that is read, since an address that is
considered unsafe to read at one moment, \ie before a write, is considered safe
to read after the write. \TheCompiler leverages the memory write table
(\Sec{memory-instruction-formalization}) to formulate a first-order query that
checks for semantically-safe loads from the stack under all program
inputs. Further, the stack pointer register {\ct r10} is read-only;
\TheCompiler's proposal generation
avoids sampling {\ct r10} as an instruction operand whenever that operand might
be modified by the instruction
(\Sec{proposal-generation}).

\Para{Access alignment.} The kernel checker enforces that memory
loads and stores of a certain size happening to specific memory types (\eg the
stack) must happen to addresses aligned to that size. That is, an address $a$
with an $N$-byte load or store must be such that $a\ (mod\ N) == 0$. For
example, the two instructions {\ct bpf\_stxw} and {\ct bpf\_stxdw} will require two
different alignments, up to 32 bits and up to 64 bits, respectively.

Somewhat surprisingly, all of the safety properties above can be decided with
sound and complete procedures due to the simplicity of the BPF instruction set. 

\Para{Modeling checker-specific constraints.} We encode several other specific
properties enforced by the kernel checker. These checks
can distinguish semantically-equivalent code sequences that meet with different
verdicts (accept versus reject) in the Linux checker. 
We added these checks ``on-demand'', as we encountered programs from
\TheCompiler that failed to load. A selection of
kernel-checker-specific safety properties we encoded include:
\begin{enumerate}
\item Certain classes of instructions, such as ALU32, NEG64, OR64, \etc are
  disallowed on pointer memory;
\item storing an immediate value into a pointer of a specific type
  (PTR\_TO\_CTX~\cite{bpf-verifier-source-code}) is disallowed;
\item Registers {\ct r1 $\cdots$ r5} are clobbered and unreadable after a helper
  function call~\cite{bpf-kernel-documentation};
\item aliasing pointers with offsets relative to the base address of a
  (permitted) memory region is considered unsafe.
\end{enumerate}
Our encoding of kernel checker safety properties is incomplete; we believe it
will be necessary to keep adding to these checks over time as the kernel checker
evolves. A distinct advantage of a synthesis-based compiler is that such checks
can be encoded once and considered across {\em all possible} optimizations,
rather than encoded piecemeal for each optimization as in a rule-based compiler.

\section{Implementation}
\label{sec:implementation}
\label{sec:interpreter}

We summarize some key points about the implementation of \TheCompiler here.
More details are available in \App{implementation}.

\TheCompiler is implemented in 24500 lines of C++ code and C code, including
proposal generation, program interpretation, first-order logic
formalization, optimizations to equivalence-checking, and safety
considerations. \TheCompiler consumes BPF bytecode compiled by
{\ct clang} and produces an optimized, drop-in replacement. The interpreter and
the verification-condition-generator of \TheCompiler can work with multiple BPF
hooks~\cite{bpf-program-types}, fixing the inputs and outputs
appropriately for testing and equivalence-checking.  \TheCompiler uses {\ct
  Z3}~\cite{z3} as its internal logic solver for discharging
equivalence-checking and safety queries.

\TheCompiler includes a high-performance BPF interpreter
that runs BPF bytecode using an optimized jumptable
implementation similar to the kernel's internal BPF
interpreter~\cite{bpf-kernel-interpreter-source}. We encoded
a declarative specification of the semantics of most
arithmetic and logic instructions in BPF using C
preprocessor directives.  This enabled us to auto-generate
code for both \TheCompiler's interpreter and verification
formula generator from the same specification of the BPF
instruction set, akin to solver-aided
languages~\cite{rosette-onward13, serval-sosp19}.

\section{Evaluation}
\label{sec:evaluation}

In this section, we answer the following questions:

\begin{figure}
    \centering
    \includegraphics[width=0.30\textwidth]{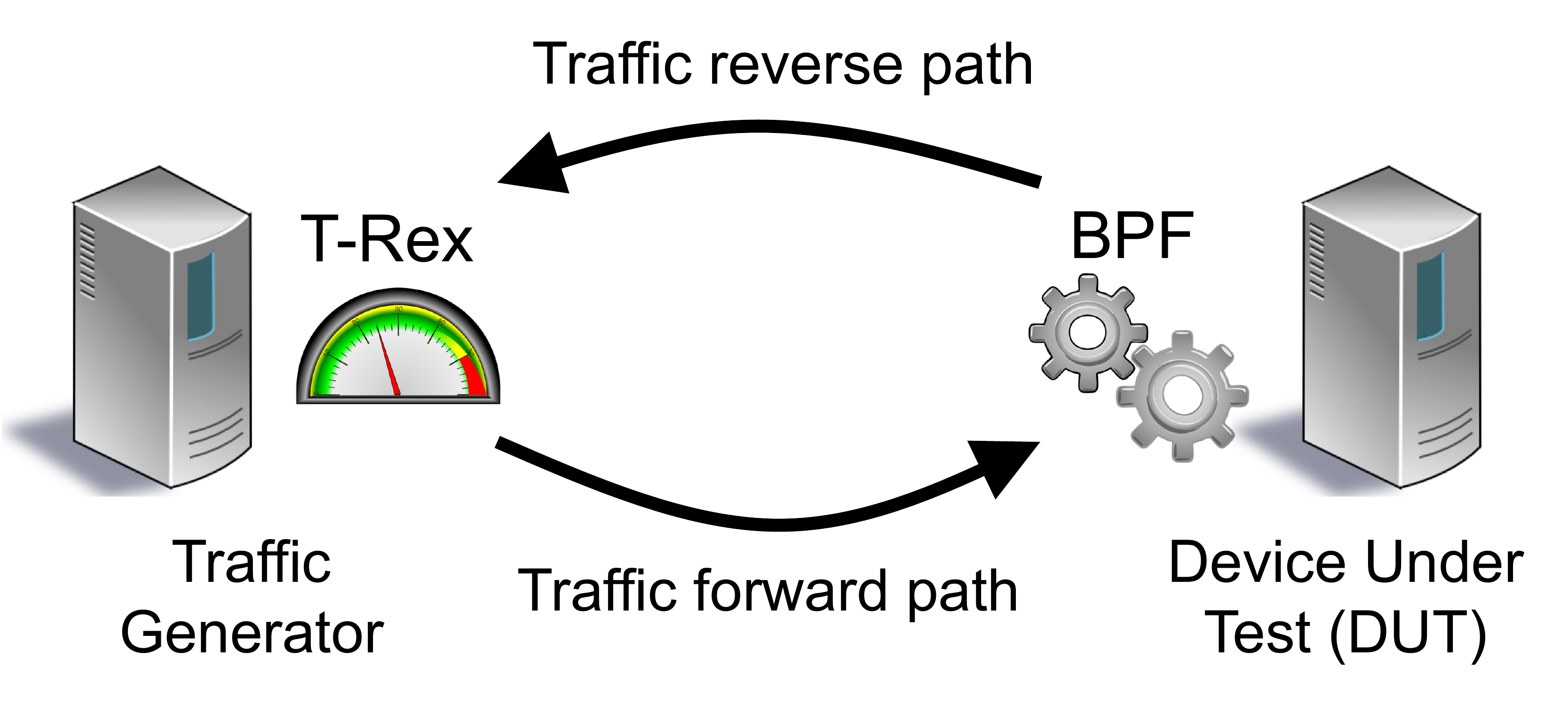}
    \caption{{\mycaptionsize Evaluation setup to measure the throughput and
        latency benefits of \TheCompiler
        (\Sec{throughput-latency-improvements}).}}
    \label{fig:throughput-latency-measurement-setup}
\end{figure}

\label{sec:quality-of-synthesized-programs}

\noindent(1) How compact are \TheCompiler-synthesized programs?\\
\noindent(2) How beneficial is \TheCompiler to
packet latency and throughput?\\
\noindent(3) Does \TheCompiler produce safe, kernel-checker-acceptable programs?\\
\noindent(4) How useful are the optimizations to equivalence
checking (\Sec{optimizing-equivalence-checking})?\\
\noindent(5) How effective are \TheCompiler's search
parameters to find good programs?\\
\noindent(6) How beneficial are \TheCompiler's
domain-specific rules (\Sec{proposal-generation})?

For questions (1) and (2), we compare \TheCompiler-synthesized
results with the best program produced by {\ct clang}
(across {\ct -O1/O2/O3/Os.}).

First, we describe how \TheCompiler is set up to produce the program we compare
against {\ct clang}. We choose the desired performance goal, which is either to
reduce the instruction count or program latency. Then, we set off multiple runs
of \TheCompiler in parallel, starting from the output of
{\ct clang -O2}, and run them until a timeout. Each run uses a different parameter setting
for its Markov chain (\Sec{search-objective-function}). In
particular, we explore the 16 parameter settings described in
\App{parameter-set}. Among these parallel runs, we choose the top-$k$
best-performing programs which are safe and equivalent to the source program
across all the Markov chains. We set $k=1$ for the instruction count performance
goal and $k=5$ for the latency goal. Since the latency-based cost function
used inside \TheCompiler is just an estimate of performance
(\Sec{search-objective-function}), we {\em measure} the average
throughput/latency performance of the top-$k$ programs and return the best
program.

We obtain our benchmark programs from diverse sources, including the Linux
kernel's BPF samples, recent academic literature~\cite{hxdp-osdi20}, and
programs used in production from Cilium and Facebook. We have considered \NumBenchmarks BPF
programs in all, which attach to the network device
driver (XDP), transport-level sockets, and system calls.

\begin{table*}[!htbp]
\centering
\begin{tabular}{|l|c|c|c|c|c|c|c|c|c|}

\hline
\multicolumn{1}{|c|}{ \multirow{2}*{Benchmark} } & \multicolumn{2}{c|}{Number of basic blocks} & \multicolumn{5}{c|}{Number of instructions} & \multicolumn{2}{c|}{When smallest prog. is found}\\
\cline{2-10}
\multicolumn{1}{|c|}{} & All & Longest path & \texttt{-O1} & \texttt{-O2/-O3} & \texttt{-Os} & \TheCompiler & Compression & Time (sec) & Iterations \\\hline
(1) xdp\_exception & 5 & 5 & 18 & 18 & 18 & 16 & 11.11\% & 79 & 372,399 \\\hline
(2) xdp\_redirect\_err & 5 & 5 & 19 & 18 & 18 & 16 & 11.11\% & 10 & 889 \\\hline
(3) xdp\_devmap\_xmit & 6 & 6 & 36 & 36 & 36 & 29 & 19.44\% & 1,201 & 659,903 \\\hline
(4) xdp\_cpumap\_kthread & 5 & 5 & 24 & 24 & 24 & 18 & 25.00\% & 1,170 & 628,354 \\\hline
(5) xdp\_cpumap\_enqueue & 4 & 3 & 26 & 26 & 26 & 21 & 19.23\% & 1,848 & 300,438 \\\hline
(6) sys\_enter\_open & 13 & 6 & 24 & 24 & 24 & 20 & 16.67\% & 519 & 834,179 \\\hline
(7) socket/0 & 13 & 6 & 32 & 29 & 29 & 27 & 6.90\% & 6 & 914 \\\hline
(8) socket/1 & 20 & 17 & 35 & 32 & 32 & 30 & 6.25\% & 9 & 3,455 \\\hline
(9) xdp\_router\_ipv4 & 5 & 5 & 139 & 111 & 111 & 99 & 10.81\% & 898 & 354,154 \\\hline
(10) xdp\_redirect & 18 & 16 & 45 & 43 & 43 & 35 & 18.60\% & 523 & 228,101 \\\hline
(11) xdp1\_kern/xdp1 & 5 & 4 & 72 & 61 & 61 & 56 & 8.20\% & 472 & 739,416 \\\hline
(12) xdp2\_kern/xdp1 & 15 & 11 & 93 & 78 & 78 & 71 & 8.97\% & 157 & 100,811 \\\hline
(13) xdp\_fwd & 19 & 14 & 170 & 155 & 155 & 128 & 17.42\% & 6,137 & 2,851,203 \\\hline
(14) xdp\_pktcntr & 4 & 3 & 22 & 22 & 22 & 19 & 13.64\% & 288 & 614,569 \\\hline
(15) xdp\_fw & 24 & 22 & 85 & 72 & 72 & 65 & 9.72\% & 826 & 342,009 \\\hline
(16) xdp\_map\_access & 6 & 6 & 30 & 30 & 30 & 26 & 13.33\% & 27 & 69,628 \\\hline
(17) from-network & 21 & 18 & 43 & 39 & 39 & 29 & 25.64\% & 6,871 & 4,312,839 \\\hline
(18) recvmsg4 & 4 & 4 & 98 & 94 & 94 & 81 & 13.83\% & 3,350 & 904,934 \\\hline
(19) xdp-balancer & 247 & 96 & DNL & 1,811 & 1,771 & 1,607 & 9.26\% & 167,428 & 10,251,406  \\\hline
Avg. of all benchmarks & & & & & & & 13.95\% & 10,096 & 1,240,505 \\\hline
\end{tabular}
\caption{\mycaptionsize \TheCompiler's improvements in
  program compactness across benchmarks from the Linux
  kernel (1--13), Facebook (14, 19), hXDP~\cite{hxdp-osdi20}
  (15, 16), and Cilium (17, 18). ``DNL'' means that the
  program variant did not load as it was rejected by the
  kernel checker.}
\label{tab:eval-instruction-count}
\end{table*}

\Para{Program Compactness.}
\label{sec:instruction-count-reduction}
\Tab{eval-instruction-count} reports the number of
instructions in \TheCompiler-optimized programs relative to
those of {\ct clang -O1/-O2/-O3/-Os} ({\ct -O2} and {\ct
  -O3} are always identical). We show the compression
achieved, the overall compile time, the time to uncover the
smallest program for each benchmark, as well as some metrics
on the complexity of the program being optimized, such as
the number of total basic blocks and the length of the
longest code path (measured in basic blocks). In all cases,
\TheCompiler manages to compress the program beyond the best
known {\ct clang} variant, by a fraction that ranges from
6--26\%, with a mean improvement of
\AverageInstrCountReduction. The average time to find the
best program\footnote{This average excludes the largest
benchmark {\ct xdp-balancer}, which is an outlier.} is
about 22 minutes; often, the best program
can be found much sooner.

Notably, \TheCompiler can handle programs with more than 100
instructions, something that even state-of-the-art
synthesizing compilers find
challenging~\cite{lens-asplos16}. The time to reach the best
program displays significant variability. Programs with more instructions
take longer to compress by the same relative
amount. However, we do not find any significant
relationship between optimization time and the number of
basic blocks. Some examples of optimizations are in
\Sec{optimizations-discovered-by-k2}.

\Para{Latency and throughput improvements.}
\label{sec:throughput-latency-improvements}
We measure the improvements in packet-processing throughput
and latency obtained by optimizing programs with
\TheCompiler. (Improvements in the compiler's {\em
  estimated} performance are presented in
\App{estimated-runtime-gains}.)

We use two server-class machines on
CloudLab~\cite{cloudlab-atc19} to set up a high-speed
traffic generator (T-Rex~\cite{trex-trafficgen}) and a
device-under-test (DUT). Our setup is visualized in
\Fig{throughput-latency-measurement-setup}. The DUT runs a
subset of our benchmark BPF programs that attach to the
network device driver using the XDP
hook~\cite{xdp-conext18}. The servers house 10-core Intel
Broadwell (E5-2640v4) 2.4 GHz processors with a PCIe 3.0 bus
and 64 GB of memory. The servers are equipped with Mellanox
ConnectX-4 25G adapters. Test traffic moves from the traffic
generator to the DUT and back to the traffic generator to
form a loop, in the spirit of the benchmarking methodology
outlined in RFC 2544~\cite{rfc2544}, allowing us to measure
both the packet-processing throughput and the round-trip
latency to forward via the DUT. Within the CloudLab network,
the two machines connect over a Mellanox switch.
%% We use the T-Rex high-speed traffic
%% generator~\cite{trex-trafficgen} as both the source and the
%% sink of the traffic from the DUT (except when the DUT runs a
%% program which drops all packets. In this configuration,
%% T-Rex can provide high-fidelity throughput and latency
%% measurements since it observes traffic sent to and received
%% from the DUT in one place.

We tuned the DUT following instructions from the XDP benchmarking configurations
described in \cite{xdp-conext18}.  Specifically, we set up Linux Receive-Side Scaling
(RSS)~\cite{scaling-linux-networking-stack}, IRQ affinities for NIC receive
queues~\cite{smp-irq-affinity}, PCIe descriptor compression,
the maximum MTU for the
Mellanox driver to support BPF, and the RX descriptor ring
size for the NIC.
%% The only loads other than packet-processing on the
%% machine were our SSH connections to monitor the
%% experiments.
Our configurations and benchmarking scripts are publicly
available from the project web page~\cite{project-web-page}.

We report program throughput as the {\em maximum loss-free
  forwarding rate} (MLFFR~\cite{rfc2544}) of a single core.
This is measured by increasing the offered load from the
traffic generator slowly and recording the load beyond which the
packet loss rate rises sharply.  We measure throughput in
millions of packets per second (Mpps) at 64-byte packet
size.  We use the minimum packet size since network-related
CPU usage is proportional to packets per second rather than
bytes per second, and XDP programs can easily saturate 100 Gbit/s
on a single core with larger packet
sizes~\cite{xdp-conext18}. Since latency varies with the
load offered by the traffic generator, we report the
latencies of the program variants at four separate offered
loads: (i) low (load smaller than the throughput of the slowest
variant), (ii) medium (load equal to the throughput of the
slowest variant), (iii) high (load equal to the throughput of the
fastest variant), and (iv) saturating (load higher than the
throughput of all known variants). We average
the results of 3 trials, with each result obtained after
waiting 60 seconds or until the numbers stabilize.

\TheCompiler's measured improvements in throughput and
latency over the best {\ct clang}-compiled variant of the
same program are summarized in \Tab{eval-throughput-mlffr}
and \Tab{eval-latency-mlffr}. \TheCompiler provides small
improvements in throughput ranging from
\MinThroughputGain--\MaxThroughputGain, while \TheCompiler's
latency benefits range from
\MinLatencyReduction--\MaxLatencyReduction.  These benefits
arise from target-specific optimizations with the latency
cost function. (\App{throughput-latency-profile} shows
detailed pictures of packet latency at varying loads.)  More
work remains before fully attaining the potential benefits
of synthesis (\Sec{discussion}).

% new version
 \begin{table}[!htbp]
 \centering
 \begin{tabular}{|l|c|c|c|c|}\hline
  Benchmark & \texttt{-O1}  & \texttt{-O2/-O3}  & \TheCompiler  & Gain  \\\hline
  xdp2  & 8.855 & 9.547 & 9.748 & 2.11\% \\\hline
  xdp\_router\_ipv4 & 1.496 & 1.496 & 1.496 & 0.00\% \\\hline
  xdp\_fwd  & 4.886 & 4.984 & 5.072 & 1.77\% \\\hline
  xdp1  & 16.837  & 16.85 & 17.65 & 4.75\% \\\hline
  xdp\_map\_access  & 14.679 & 14.678 &  15.074 & 2.70\% \\\hline
  xdp-balancer & DNL & 3.292 & 3.389 & 2.94\% \\\hline
 \end{tabular}
 \caption{\mycaptionsize Throughput reported as the maximum
   loss-free forwarding rate (MLFFR) in millions of packets
   per second per core (\Sec{evaluation}).}
 \label{tab:eval-throughput-mlffr}
 \end{table}

 \begin{table*}[!htbp]
 \scriptsize
 \centering
 \begin{tabular}{|l|c|c|c|c|c|c|c|c|c|c|c|c|c|c|c|c|}\hline
  Benchmark & Low  & \texttt{clang}  & \TheCompiler  & Reduction
            & Medium  & \texttt{clang}  & \TheCompiler  & Reduction
            & High  & \texttt{clang}  & \TheCompiler  & Reduction
            & Saturating  & \texttt{clang}  & \TheCompiler  & Reduction  \\\hline
  xdp2 & 9 & 29.148 & 25.676 & 11.91\% 
       & 9.5 & 51.157 & 30.237 & 40.89\% 
       & 9.7 & 89.523 & 40.259 & 55.03\% & 10.3 
       & 103.872 & 97.754 & 5.89\% \\\hline
  xdp\_router\_ipv4 & 1 & 63.323 & 59.834 & 5.51\% 
                    & 1.5 & 84.450 & 76.929 & 8.91\% 
                    & 1.5 & 84.450 & 76.929 & 8.91\% 
                    & 1.8 & 619.291 & 610.119 & 1.48\% \\\hline
  xdp\_fwd & 4.4 & 32.272 & 30.358 & 5.93\% 
           & 5 & 87.291 & 71.645 & 17.92\% 
           & 5 & 87.291 & 71.645 & 17.92\% 
           & 5.2 & 192.936 & 188.199 & 2.46\% \\\hline
  xdp-balancer & 3 & 38.650 & 37.152 & 3.88\% 
               & 3.3 & 73.319 & 55.741 & 23.97\% 
               & 3.4 & 237.701 & 119.497 & 49.73\% 
               & 3.7 & 296.405 & 292.376 & 1.36\% \\\hline
 \end{tabular}
 \caption{\mycaptionsize Average latencies (in microseconds)
   of the best {\ct clang} and \TheCompiler variants at
   different offered loads (in millions of packets per
   second). We consider 4 offered loads: low (smaller than
   the slowest throughput of {\ct clang} or \TheCompiler),
   medium (the slowest throughput among {\ct clang} and
   \TheCompiler), high (the highest throughput among {\ct
     clang} and \TheCompiler), and saturating (higher than
   the fastest throughput of {\ct clang} or \TheCompiler).}
 \label{tab:eval-latency-mlffr}
 \end{table*}

\nop{
% old version
 \begin{table*}[!htbp]
 \centering
 \begin{tabular}{|l|c|c|c|c|c|c|c|c|c|c|c|}
 \hline
 \multicolumn{1}{|c|}{ \multirow{2}*{Benchmark} }& \multicolumn{4}{c|}{Avg. Throughput} &\multicolumn{4}{c|}{Avg. Latency} & \multicolumn{3}{c|}{Max. Latency}\\
 \cline{2-11}
 \multicolumn{1}{|c|}{} & \texttt{-O1} & \texttt{-O2/-O3} & \TheCompiler &  Gain &  \texttt{-O1} & \texttt{-O2/-O3} & \TheCompiler & \texttt{Reduction} & \texttt{-O1} & \texttt{-O2/-O3} & \TheCompiler \\
 \hline
 xdp1 & 6.46 &  6.89 & 6.87 & -0.03\% & N/A &	N/A & N/A &	 N/A & N/A & N/A & N/A \\
 \hline
 xdp2 & 5.60 & 5.55 & 5.87 & 4.82\% & 	45.00 & 45.00 & 33.00 & 13.00 &	81.00 & 71.00 & 83.00  \\
 \hline
 xdp\_map\_access & 16.67 & 16.54 & 17.00 & 2.03\% &	N/A &  N/A &N/A & N/A & N/A & N/A & N/A \\
 \hline
 xdp\_router\_ipv4 & 0.68 & 0.70 & 0.71 & 1.15\% & 293.00 & 258.00 & 173.00 & 85.00 &	498.00 & 389.00 & 329.00  \\
 \hline
 xdp\_fwd & 1.44 & 1.42 & 1.52 &	 5.55\% & 148.00 & 135.00 &	70.00 & 65.00 & 215.00 &  217.00 & 198.00  \\
 \hline
 xdp\_fw & 1.48 & 1.48 &	1.53 & 3.38\% & 143.00 &	134.00 & 103.00 & 31.00 & 193.00 & 196.00 & 194.00 \\
 \hline
 \end{tabular}
 \caption{Throughput (Mpps) and Latency ($\mu$s)}
 \label{tab:eval-throughput-latency}
 \end{table*}
}

\Para{Safety of synthesized programs.}
\label{sec:safety-evaluation}
We loaded the XDP program outputs produced by \TheCompiler
into the kernel. All 38 out of the 38 programs found by
\TheCompiler's search were successfully accepted by the
kernel checker, even without \TheCompiler's safety post-processing
(\Sec{safety}). \Tab{eval-safety} in \App{additional-eval}
lists the programs we loaded into the kernel.

\begin{table*}[!htbp]
\footnotesize
\centering
\begin{tabular}{|l|c|c|c|c|c|c|c|c|c|c|}
\hline
\multicolumn{2}{|c|}{Benchmark} & \multicolumn{1}{c|}{I, II, III, IV} &\multicolumn{2}{c|}{I, II, III} & \multicolumn{2}{c|}{I, II} &\multicolumn{2}{c|}{I} & \multicolumn{2}{c|}{None} \\
% \cline{1-11}
\hline
% \multicolumn{1}{|c|}{} &
 name & \#inst & time ($\mu$s) &  time ($\mu$s) & slowdown & time ($\mu$s)& slowdown & time ($\mu$s) & slowdown & time ($\mu$s) & slowdown \\\hline
(1) xdp\_exception  & 18  & 25,969  & 465,113 & 18$\times$ & 5,111,940 & 197$\times$ & 2,160,340 & 83$\times$ & 21,119,700  & 813$\times$  \\\hline
(2) xdp\_redirect\_err  & 18  & 30,591  & 855,942 & 28$\times$ & 3,795,580 & 124$\times$ & 1,574,910 & 51$\times$ & 11,105,500  & 363$\times$  \\\hline
(3) xdp\_devmap\_xmit & 36  & 48,129  & 42,887,200  & 891$\times$ & 49,529,900  & 1,029$\times$ & 45,789,000  & 951$\times$ & 303,043,000 & 6,296$\times$  \\\hline
(4) xdp\_cpumap\_kthread  & 24  & 7,414 & 23,387,700  & 3,155$\times$ & 24,583,300  & 3,316$\times$ & 26,223,200  & 3,537$\times$ & 165,283,000 & 22,294 $\times$  \\\hline
(5) xdp\_cpumap\_enqueue  & 26  & 73,769  & 30,974,000  & 420$\times$ & 36,360,800  & 493$\times$ & 26,968,500  & 366$\times$ & 123,350,000 & 1,672$\times$  \\\hline
(14)  xdp\_pktcntr  & 22  & 9,181 & 1,030,280 & 112$\times$ & 43,656,800  & 4,755$\times$ & 80,963,000  & 8,819$\times$ & 23,390,200  & 2,548$\times$  \\\hline
(17)  from-network  & 39  & 9,804 & 4,791,680 & 489$\times$ & 33,758,000  & 3,443$\times$ & 93,730,700  & 9,560$\times$ & 7,836,910,000 & 799,343$\times$  \\\hline
(18)  recvmsg4  & 94  & 6,719 & 58,299,300  & 8,676$\times$ & 1,533,220,000 & 228,181$\times$ & 10,701,400,000  & 1,592,631$\times$ & > 86,400,000,000  & > 12,858,444 $\times$  \\\hline
Avg. of all benchmarks & 31  & 26,447  & 20,336,402  & 1,724$\times$ & 216,252,040 & 30,192$\times$ & 1,372,351,206 & 202,000$\times$ & > 11,860,525,175  & > 1,711,472$\times$  \\\hline
\end{tabular}
\caption{\mycaptionsize Reductions in equivalence-checking time
    (\Sec{optimizing-equivalence-checking},
    \Sec{benefits-optimizations-equivalence-checking}). We study the following
    optimizations: (I) memory type, (II) map type, and (III) memory offset
    concretizations, and (IV) modular verification. All optimizations are turned
    on (I, II, III, IV) as the baseline. Slowdowns relative to this baseline are
    reported as optimizations are turned off progressively. }
\label{tab:opti-bpf-Optimization-impact}
\end{table*}

\Para{Benefits of equivalence-checking optimizations.}
\label{sec:benefits-optimizations-equivalence-checking}
We show the benefits of the optimizations (in
\Sec{optimizing-equivalence-checking}) to reducing equivalence-checking time and
also the number of calls to the solver.  \Tab{opti-bpf-Optimization-impact}
shows the benefits of optimizations I--IV (memory type, map type, and memory
offset concretization, and modular verification) by starting with all
optimizations turned on (I, II, III, IV) as the baseline. We progressively turn
off each optimization, and show absolute verification times and slowdown
relative to the baseline. We find that, across benchmarks, the collective
benefits of the optimizations range between \MinEqChkTimeReduction--\MaxEqChkTimeReduction
orders of magnitude, with a mean improvement of 
\AvgEqChkTimeReduction orders of magnitude across programs. The larger
the program, the more pronounced the impact of the optimizations. Among all the
optimizations we apply, modular verification produces the most consistent and
significant gains across programs.

\Tab{caching-efficacy} in \App{additional-eval} shows the impact of reductions
in the number of queries to the logic solver by caching canonical versions of
programs (optimization V, \Sec{optimizing-equivalence-checking}). We find
caching to be very effective: 93\% or more queries otherwise discharged to the
solver can be eliminated by caching the equivalence-checking outcomes of
syntactically-similar programs checked earlier.

\Para{Impact of parameter choices on stochastic search.}
\label{sec:parameter-choices-in-stochastic-search}
\TheCompiler's stochastic search proceeds in parallel with 16 different sets of
parameters.
These parameters correspond to variants of the cost
functions, with different
coefficients used to combine error and performance, as well
as different program rewrite rules
(\Sec{search-objective-function}). The full set of values parameterizing each
set is described in \App{parameter-set}. Across 13 programs, we show the
efficacy of each set of parameters in optimizing instruction size. Despite
\TheCompiler's current use of 16 parameter sets, some of those sets are much
more likely to produce optimal results than others. Hence,
it is possible to obtain \TheCompiler's gains with
much less parallelism. More generally, exploring the identification of hyperparameters
that provide the best results given a limited compute budget is an interesting
problem deserving further exploration~\cite{hypersched-socc19}.

\Para{Impact of domain-specific rewrite rules.}
\label{sec:benefits-of-domain-specific-rules}
We evaluate the benefits imparted by \TheCompiler's
domain-specific program rewrite rules
(\Sec{proposal-generation}) to the quality of generated
programs.  \Tab{eval-move-types} in
\Sec{benefits-of-domain-specific-rules} shows the results
from optimizing the instruction count of programs with
different settings where we selectively turn the
domain-specific rules on or off. Each domain-specific rule
is necessary to find the best program for each benchmark.
Disabling any one of the rules entirely results in the
quality of the output programs dropping by as much as 12\%
relative to the best outcome.

\section{Optimizations Discovered by \TheCompiler}
\label{sec:optimizations-discovered-by-k2}
We present two classes of optimizations that \TheCompiler discovered
while reducing the number of instructions in the program. Several more
examples from these classes and others are in
\App{detailed-optimization-case-studies}.

\Para{Example 1. Coalescing multiple memory operations.} In the
program {\ct xdp\_pktcntr}~\cite{xdp-pktcntr} developed by Facebook,
\TheCompiler transformed
\begin{lstlisting}
  bpf_mov r1 0          // r1 = 0
  bpf_stx_32 r10 -4 r1  // *(u32*)(r10-4) = r1
  bpf_stx_32 r10 -8 r1  // *(u32*)(r10-8) = r1
\end{lstlisting}
into the single instruction
\begin{lstlisting}
  bpf_st_imm64 r10 -8 0 // *(u64*)(r10-8) = 0
\end{lstlisting}
by coalescing a register assignment and two 32-bit register stores
into a single 64-bit store that writes an immediate value. The
original instruction sequence comes from two assignments in the C
code: {\ct u32 ctl\_flag\_pos = 0; u32 cntr\_pos =\ 0.} This example
is one of the simplest of this class of optimizations that
\TheCompiler found. In a couple of cases, \TheCompiler
shrunk sequences of 12 instructions containing complex swaps
of memory contents into 4--8 instructions.

\Para{Example 2. Context-dependent optimizations.} \TheCompiler
discovered rewrites that depend on the specific context (\eg current
register values) of instructions within a program. For example, in the
{\ct balancer\_kern} program~\cite{balancer-kern} developed by
Facebook, \TheCompiler transformed the sequence
\begin{lstlisting}
  bpf_mov64 r0 r2      // r0 = r2
  bpf_and64 r0 r3      // r0 = r0 & r3
  bpf_rsh64 r0 21      // r0 = r0 >> 21
\end{lstlisting}
into the sequence
\begin{lstlisting}
  bpf_mov32 r0 r2      // r0 = lower32(r2)
  bpf_arsh64 r0 21     // r0 = r0 >> 21
\end{lstlisting}
This transformation does not generally hold under all values of {\ct
  r3}. \TheCompiler used the precondition that the value of
{\ct r3} prior to this sequence was {\ct 0x00000000ffe00000}.
More generally, we found optimizations where
\TheCompiler leveraged {\em both} 
preconditions and postconditions on the values and liveness of
registers and memory addresses.

We believe that the specificity of the optimizations described above
(and in \App{detailed-optimization-case-studies}) may well be
challenging to match with a rule-based optimizing compiler. Beyond the
categories described above, \TheCompiler derived optimizations using
complex opcodes (\eg {\ct bpf\_xadd64 rX off rY} $\Leftrightarrow$
{\ct *(u64*)(rX + off) += rY}) and non-trivial dead code elimination
that leverages the liveness of memory addresses. More
examples are available in \App{detailed-optimization-case-studies}.

\section{Related Work}

Data plane code
optimization has been a topic of recent interest~\cite{chipmunk-sigcomm20,
  morpheus-preprint21,
  lowlatency-kth-thesis19}. Chipmunk~\cite{chipmunk-sigcomm20} generates
code for high-speed switches, where programs must fit within
the available hardware resources or they won't run at all.
In contrast, \TheCompiler starts with a program
that can run; the goal 
is to improve its performance safely. In concurrent work,
Morpheus~\cite{morpheus-preprint21} explores
dynamic recompilation of data plane code based on workload characteristics,
reminiscent of conditionally-correct
optimization~\cite{conditionally-correct-superopt-pldi15}.  \TheCompiler's
approach is orthogonal: it is purely compile-time and the optimizations are
valid across all run-time configurations. 
Farshin's thesis~\cite{lowlatency-kth-thesis19} suggests, but stops short of
applying stochastic search to NFVs, due to performance
variability. hXDP~\cite{hxdp-osdi20} executes BPF code on FPGAs;
\TheCompiler's goal is to optimize BPF over ISA-based
processors.

There is a rich literature on synthesizing data plane rules and control
plane policies for high-speed routers~\cite{propane-sigcomm16,
  propane-at-pldi17, zeppelin-routersynthesis18, netcomplete-nsdi18,
  network-wide-config-synthesis-cav17, netgen-sosr15,
  genesis-popl17}. \TheCompiler must synthesize
BPF instructions, which are more expressive than router data
plane rules and control policies.

\TheCompiler builds significantly on the literature on program
synthesis~\cite{massalin-sigarch87, lens-asplos16, peephole-asplos06,
  denali-pldi02, equalitysaturation-pldi09,
  equality-saturation-tensor-superopt-arxiv21, sygus-cacm18, loopfree-pldi11,
  souper17, dataflow-pruning-oopsla20, dryadsynth,
  oracle-component-synthesis-icse10, stoke-asplos13, loopsuperopt-pldi17} and
accelerating formal verification~\cite{bounded-model-checking-biere99,
  symbolic-exec-king76, symbolic-exec-clarke76, horn-clause-solvers,
  fixing-code-vmcai20, finding-code-oopsla18}. Below, we summarize
three key technical differences from this literature.

First, \TheCompiler makes several domain-specific contributions in formalizing BPF
programs relative to prior work. Sound BPF JITs~\cite{jitterbug-osdi20,
  jitsynth-cav20, serval-sosp19, jitk-osdi14} assert the equivalence between BPF
bytecode instructions and lower-level machine instructions on a per-instruction
basis. \TheCompiler solves a fundamentally different problem: synthesizing new
BPF bytecode and checking the equivalence of synthesized and source BPF
bytecode. Unlike sound JIT compilers, \TheCompiler requires modeling control
flow and pointer aliasing which are not concerns for per-instruction
verification tasks. Prevail~\cite{untrusted-extensions-pldi19} implements a fast
abstract interpretation of BPF programs to prove in-bound memory access safety
and control flow safety. In contrast to Prevail, \TheCompiler performs
synthesis, and considers several additional kernel-checker-specific safety
properties to generate kernel-executable BPF bytecode.
To our knowledge, none of the prior works formalize BPF maps and helpers in
sufficient detail to support equivalence-checking, which requires modeling two
levels of aliasing (\Sec{map-formalization}).
Further, \TheCompiler contributes several domain-specific
techniques to accelerate equivalence-checking by \AvgEqChkTimeReduction orders of
magnitude. 

Second, most prior x86 code synthesizers do not handle program safety
considerations~\cite{stoke-asplos13, peephole-asplos06, lens-asplos16,
  ddec-oopsla13, conditionally-correct-superopt-pldi15, souper17,
  dataflow-pruning-oopsla20}. To our knowledge, the only
prior approach
to synthesize safe code is the NaCl loop
superoptimizer~\cite{loopsuperopt-pldi17}, which considers only access
alignment (\Sec{safety}).

Finally, \TheCompiler includes several domain-specific
program rewrites (\Sec{stochastic-search}) that
accelerate convergence to better programs.

\nop{
  \TheCompiler's formalization of memory and
BPF maps is related to the theory of arrays~\cite{calculus-of-computation-book}
and the axioms of partial maps~\cite{software-foundations-book}.  However, we
encode these axioms directly within the theory of bit vectors, similar to prior
work~\cite{cryptographic-verif-cgo19, stoke-asplos13, peephole-asplos06} to
avoid the inefficiencies of combining theories within solvers.
}

\section{Conclusion}
\label{sec:discussion}

We presented \TheCompiler, a compiler for BPF based on
program synthesis technology. \TheCompiler can produce safe
and optimized drop-in replacements for existing BPF
bytecode.

\TheCompiler naturally leads to several avenues for
follow-up research. (1) {\em Scaling to larger programs:}
Currently, \TheCompiler cannot optimize large programs (200+
instructions) within a short time (\eg a minute). Developing
techniques to optimize large programs quickly is a direction
ripe for further research. (2) {\em Designing better cost
  functions:} \TheCompiler's latency cost function is a weak
predictor of actual latency. The design of high-fidelity
cost functions to statically estimate program performance
metrics such as tail latency and maximum per-core throughput
will help boost the throughput and latency gains available
from synthesis. (3) {\em Addressing engineering challenges:}
The active evolution of the BPF
ecosystem~\cite{bpf-mailing-list} makes it challenging to
keep \TheCompiler's safety checks in sync with that of the
kernel checker and to develop support for emerging BPF hooks
and helper functions.

We hope that the community will build on our compiler and
the techniques in this paper. \TheCompiler's source code,
including all of our experimental scripts, is available at
\url{https://k2.cs.rutgers.edu/}.

\Para{This work does not raise any ethical issues.}

\label{lastpage}
\begin{acks}
This work was funded by the National Science Foundation
grants CNS-1910796 and CNS-2008048.  We thank the SIGCOMM
anonymous reviewers, our shepherd Nate Foster, Akshay
Narayan, Paul Chaignon, and Vibhaalakshmi Sivaraman for their
thoughtful feedback and discussions. We are grateful to
Nikolaj Bj{\o}rner for improvements to the Z3 solver that
helped support \TheCompiler's requirements.
\end{acks}

\bibliographystyle{ACM-Reference-Format}
\bibliography{superopt}


\begin{thebibliography}{139}

%%% the Web address.

\ifx \showCODEN    \undefined \def \showCODEN     #1{\unskip}     \fi
\ifx \showDOI      \undefined \def \showDOI       #1{#1}\fi
\ifx \showISBNx    \undefined \def \showISBNx     #1{\unskip}     \fi
\ifx \showISBNxiii \undefined \def \showISBNxiii  #1{\unskip}     \fi
\ifx \showISSN     \undefined \def \showISSN      #1{\unskip}     \fi
\ifx \showLCCN     \undefined \def \showLCCN      #1{\unskip}     \fi
\ifx \shownote     \undefined \def \shownote      #1{#1}          \fi
\ifx \showarticletitle \undefined \def \showarticletitle #1{#1}   \fi
\ifx \showURL      \undefined \def \showURL       {\relax}        \fi
% invisible to TeX
\providecommand\bibfield[2]{#2}
\providecommand\bibinfo[2]{#2}
\providecommand\natexlab[1]{#1}
\providecommand\showeprint[2][]{arXiv:#2}

\bibitem[\protect\citeauthoryear{??}{rfc}{1999}]%
        {rfc2544}
 \bibinfo{year}{1999}\natexlab{}.
\newblock \bibinfo{title}{{RFC 2544: Benchmarking Methodology for Network
  Interconnect Devices}}.
\newblock \bibinfo{howpublished}{[Online. Retrieved Jan 27, 2021.]
  \url{https://tools.ietf.org/html/rfc2544}}.
\newblock


\bibitem[\protect\citeauthoryear{??}{bpf}{2014}]%
        {bpf-kernel-interpreter-source}
 \bibinfo{year}{2014}\natexlab{}.
\newblock \bibinfo{title}{{BPF kernel interpreter}}.
\newblock \bibinfo{howpublished}{[Online. Retrieved Jan 21, 2021.]
  \url{https://github.com/torvalds/linux/blob/master/kernel/bpf/core.c\#L1356}}.
\newblock


\bibitem[\protect\citeauthoryear{??}{doc}{2016}]%
        {documentation-on-bpf-complexity-2}
 \bibinfo{year}{2016}\natexlab{}.
\newblock \bibinfo{title}{{Some notes on verifier complexity}}.
\newblock \bibinfo{howpublished}{[Online. Retrieved Jul 12, 2021.]
  \url{https://github.com/cilium/cilium/commit/ff7c6767180a9923fb1c0646945f29709da6fb6e}}.
\newblock


\bibitem[\protect\citeauthoryear{??}{bpf}{2017a}]%
        {bpf-instruction-set}
 \bibinfo{year}{2017}\natexlab{a}.
\newblock \bibinfo{title}{{BPF instruction set}}.
\newblock \bibinfo{howpublished}{[Online, Retrieved Jan 20, 2021.]
  \url{https://github.com/iovisor/bpf-docs/blob/master/eBPF.md}}.
\newblock


\bibitem[\protect\citeauthoryear{??}{bpf}{2017b}]%
        {bpf-verifier-selftests}
 \bibinfo{year}{2017}\natexlab{b}.
\newblock \bibinfo{title}{{Linux BPF verifier selftests}}.
\newblock \bibinfo{howpublished}{[Online. Retrieved Jan 21, 2021.]
  \url{https://github.com/torvalds/linux/tree/master/tools/testing/selftests/bpf/verifier}}.
\newblock


\bibitem[\protect\citeauthoryear{??}{bpf}{2017c}]%
        {bpf-verifier-source-code}
 \bibinfo{year}{2017}\natexlab{c}.
\newblock \bibinfo{title}{{The Linux kernel BPF static checker}}.
\newblock \bibinfo{howpublished}{[Online. Retrieved Jan 20, 2021.]
  \url{https://github.com/torvalds/linux/blob/master/kernel/bpf/verifier.c}}.
\newblock


\bibitem[\protect\citeauthoryear{??}{tre}{2017}]%
        {trex-trafficgen}
 \bibinfo{year}{2017}\natexlab{}.
\newblock \bibinfo{title}{{TRex traffic generator}}.
\newblock \bibinfo{howpublished}{[Online. Retrieved Jan 27, 2021.]
  \url{https://trex-tgn.cisco.com/trex/doc/trex_manual.html}}.
\newblock


\bibitem[\protect\citeauthoryear{??}{sta}{2017}]%
        {stack-aligned-restriction-verifier}
 \bibinfo{year}{2017}\natexlab{}.
\newblock \bibinfo{title}{{Verifier restriction on stack accesses being
  aligned}}.
\newblock \bibinfo{howpublished}{[Online. Retrieved Jan 23, 2021.]
  \url{https://github.com/torvalds/linux/blob/v4.18/kernel/bpf/verifier.c\#L1515}}.
\newblock


\bibitem[\protect\citeauthoryear{??}{ctx}{2017}]%
        {ctx-store-restriction-verifier}
 \bibinfo{year}{2017}\natexlab{}.
\newblock \bibinfo{title}{{Verifier restriction on stores into PTR\_TO\_CTX
  pointers}}.
\newblock \bibinfo{howpublished}{[Online. Retrieved Jan 23, 2021.]
  \url{https://github.com/torvalds/linux/blob/v4.18/kernel/bpf/verifier.c\#L4888}}.
\newblock


\bibitem[\protect\citeauthoryear{??}{doc}{2018}]%
        {documentation-on-bpf-complexity-1}
 \bibinfo{year}{2018}\natexlab{}.
\newblock \bibinfo{title}{{Document navigating BPF verifier complexity}}.
\newblock \bibinfo{howpublished}{[Online. Retrieved Jul 12, 2021.]
  \url{https://github.com/cilium/cilium/issues/5130}}.
\newblock


\bibitem[\protect\citeauthoryear{??}{bal}{2018}]%
        {balancer-kern}
 \bibinfo{year}{2018}\natexlab{}.
\newblock \bibinfo{title}{{Facebook XDP load balancer benchmark}}.
\newblock \bibinfo{howpublished}{[Online, Retrieved Jun 15, 2021.]
  \url{https://github.com/facebookincubator/katran/blob/master/katran/lib/bpf/balancer_kern.c}}.
\newblock


\bibitem[\protect\citeauthoryear{??}{xdp}{2018}]%
        {xdp-pktcntr}
 \bibinfo{year}{2018}\natexlab{}.
\newblock \bibinfo{title}{{Facebook XDP packet counter benchmark}}.
\newblock \bibinfo{howpublished}{[Online, Retrieved Jun 15, 2021.]
  \url{https://github.com/facebookincubator/katran/blob/6f86aa82c5b3422313e0a63d195b35e7e2f7539a/katran/lib/bpf/xdp_pktcntr.c\#L52-L53}}.
\newblock


\bibitem[\protect\citeauthoryear{??}{bpf}{2019}]%
        {bpf-increase-complexity-limit}
 \bibinfo{year}{2019}\natexlab{}.
\newblock \bibinfo{title}{{BPF: Increase complexity limit and maximum program
  size}}.
\newblock \bibinfo{howpublished}{[Online. Retrieved Jul 12, 2021.]
  \url{https://git.kernel.org/pub/scm/linux/kernel/git/torvalds/linux.git/commit/?id=c04c0d2b968ac45d6ef020316808ef6c82325a82}}.
\newblock


\bibitem[\protect\citeauthoryear{??}{cla}{2019}]%
        {clang-verifier-3}
 \bibinfo{year}{2019}\natexlab{}.
\newblock \bibinfo{title}{{[DebugInfo] Support to emit debugInfo for extern
  variables}}.
\newblock \bibinfo{howpublished}{[Online. Retrieved Jan 20, 2021.]
  \url{https://github.com/llvm/llvm-project-staging/commit/d77ae1552fc21a9f3877f3ed7e13d631f517c825}}.
\newblock


\bibitem[\protect\citeauthoryear{??}{mel}{2019}]%
        {mellanox-bluefield}
 \bibinfo{year}{2019}\natexlab{}.
\newblock \bibinfo{title}{{Mellanox BlueField SmartNIC for Ethernet}}.
\newblock \bibinfo{howpublished}{[Online. Retrieved Jan 20, 2021.]
  \url{https://www.mellanox.com/sites/default/files/related-docs/prod_adapter_cards/PB_BlueField_Smart_NIC.pdf}}.
\newblock


\bibitem[\protect\citeauthoryear{??}{cla}{2020a}]%
        {clang-verifier-1}
 \bibinfo{year}{2020}\natexlab{a}.
\newblock \bibinfo{title}{{BPF: add a SimplifyCFG IR pass during generic
  Scalar/IPO optimization}}.
\newblock \bibinfo{howpublished}{[Online. Retrieved Jan 20, 2021.]
  \url{https://github.com/llvm/llvm-project-staging/commit/87cba434027bf6ad370629f5b924ebd4543ddabc}}.
\newblock


\bibitem[\protect\citeauthoryear{??}{cla}{2020b}]%
        {clang-verifier-4}
 \bibinfo{year}{2020}\natexlab{b}.
\newblock \bibinfo{title}{{[BPF] disable ReduceLoadWidth during SelectionDag
  phase}}.
\newblock \bibinfo{howpublished}{[Online. Retrieved Jan 20, 2021.]
  \url{https://github.com/llvm/llvm-project-staging/commit/d96c1bbaa03574daf759e5e9a6c75047c5e3af64}}.
\newblock


\bibitem[\protect\citeauthoryear{??}{cla}{2020c}]%
        {clang-verifier-2}
 \bibinfo{year}{2020}\natexlab{c}.
\newblock \bibinfo{title}{{[BPF] fix a bug in BPFMISimplifyPatchable pass with
  -O0}}.
\newblock \bibinfo{howpublished}{[Online. Retrieved Jan 20, 2021.]
  \url{https://github.com/llvm/llvm-project-staging/commit/795bbb366266e83d2bea8dc04c19919b52ab3a2a}}.
\newblock


\bibitem[\protect\citeauthoryear{??}{cla}{2020d}]%
        {clang-verifier-5}
 \bibinfo{year}{2020}\natexlab{d}.
\newblock \bibinfo{title}{{[BPF] simplify zero extension with MOV\_32\_64}}.
\newblock \bibinfo{howpublished}{[Online. Retrieved Jan 20, 2021.]
  \url{https://github.com/llvm/llvm-project-staging/commit/13f6c81c5d9a7a34a684363bcaad8eb7c65356fd}}.
\newblock


\bibitem[\protect\citeauthoryear{??}{btf}{2020}]%
        {btf-kernel-documentation}
 \bibinfo{year}{2020}\natexlab{}.
\newblock \bibinfo{title}{{BPF Type Format (BTF)}}.
\newblock \bibinfo{howpublished}{[Online. Retrieved Jan 20, 2021.]
  \url{https://www.kernel.org/doc/html/latest/bpf/btf.html}}.
\newblock


\bibitem[\protect\citeauthoryear{??}{cal}{2020}]%
        {calico-ebpf-dataplane}
 \bibinfo{year}{2020}\natexlab{}.
\newblock \bibinfo{title}{{Calico's eBPF dataplane}}.
\newblock \bibinfo{howpublished}{[Online, Retrieved Jan 20, 2021.]
  \url{https://docs.projectcalico.org/about/about-ebpf}}.
\newblock


\bibitem[\protect\citeauthoryear{??}{fun}{2020}]%
        {fungible-dpu}
 \bibinfo{year}{2020}\natexlab{}.
\newblock \bibinfo{title}{{Fungible F-1 Data Processing Unit}}.
\newblock \bibinfo{howpublished}{[Online. Retrieved Jan 20, 2021.]
  \url{https://www.fungible.com/wp-content/uploads/2020/08/PB0028.01.02020820-Fungible-F1-Data-Processing-Unit.pdf}}.
\newblock


\bibitem[\protect\citeauthoryear{??}{cpu}{2020}]%
        {cpu-savings-xdp-load-balancer-kubeproxy-replacement-cilium}
 \bibinfo{year}{2020}\natexlab{}.
\newblock \bibinfo{title}{{Kube-proxy replacement at the XDP layer}}.
\newblock \bibinfo{howpublished}{[Online. Retrieved Jan 20, 2021.]
  \url{https://cilium.io/blog/2020/06/22/cilium-18\#kubeproxy-removal}}.
\newblock


\bibitem[\protect\citeauthoryear{??}{mar}{2020}]%
        {marvell-octeon-tx}
 \bibinfo{year}{2020}\natexlab{}.
\newblock \bibinfo{title}{{Marvell Octeon TX-2 product brief}}.
\newblock \bibinfo{howpublished}{[Online. Retrieved Jan 20, 2021.]
  \url{https://www.marvell.com/content/dam/marvell/en/public-collateral/embedded-processors/marvell-infrastructure-processors-octeon-tx2-cn913x-product-brief-2020-02.pdf}}.
\newblock


\bibitem[\protect\citeauthoryear{??}{mel}{2020}]%
        {mellanox-bluefield2}
 \bibinfo{year}{2020}\natexlab{}.
\newblock \bibinfo{title}{{Nvidia Mellanox BlueField 2}}.
\newblock \bibinfo{howpublished}{[Online, Retrived Jan 20, 2021.]
  \url{https://www.mellanox.com/files/doc-2020/pb-bluefield-2-dpu.pdf}}.
\newblock


\bibitem[\protect\citeauthoryear{??}{bpf}{2021a}]%
        {bpf-mailing-list}
 \bibinfo{year}{2021}\natexlab{a}.
\newblock \bibinfo{title}{{BPF archive on lore.kernel.org}}.
\newblock \bibinfo{howpublished}{[Online. Retrieved Jun 09, 2021.]
  \url{https://lore.kernel.org/bpf/}}.
\newblock


\bibitem[\protect\citeauthoryear{??}{bpf}{2021b}]%
        {bpf-design-qa}
 \bibinfo{year}{2021}\natexlab{b}.
\newblock \bibinfo{title}{{BPF design Q \& A}}.
\newblock \bibinfo{howpublished}{[Online. Retrieved Jan 20, 2021.]
  \url{https://www.kernel.org/doc/html/v5.6/bpf/bpf_design_QA.html}}.
\newblock


\bibitem[\protect\citeauthoryear{??}{fri}{2021}]%
        {fristonio-bpf-size-issue}
 \bibinfo{year}{2021}\natexlab{}.
\newblock \bibinfo{title}{{BPF size issue in bpf\_lxc's IPv6 egress path}}.
\newblock \bibinfo{howpublished}{[Online. Retrieved Jan 20, 2021.]
  \url{https://cilium.slack.com/archives/CDKG8NNHK/p1605601543139700}}.
\newblock


\bibitem[\protect\citeauthoryear{??}{cil}{2021a}]%
        {cilium-complexity-issues-list}
 \bibinfo{year}{2021}\natexlab{a}.
\newblock \bibinfo{title}{Cilium complexity issues}.
\newblock \bibinfo{howpublished}{[Online. Retrieved Jul 1, 2021.]
  \url{https://github.com/cilium/cilium/issues?q=is\%3Aissue+is\%3Aopen+label\%3Akind\%2Fcomplexity-issue}}.
\newblock


\bibitem[\protect\citeauthoryear{??}{cil}{2021b}]%
        {cilium-complexity-issue-2}
 \bibinfo{year}{2021}\natexlab{b}.
\newblock \bibinfo{title}{{Complexity issue on 5.10+ with
  kubeProxyReplacement=disabled}}.
\newblock \bibinfo{howpublished}{[Online. Retrieved Jul 12, 2021.]
  \url{https://github.com/cilium/cilium/issues/14726}}.
\newblock


\bibitem[\protect\citeauthoryear{??}{cil}{2021c}]%
        {cilium-complexity-issue-3}
 \bibinfo{year}{2021}\natexlab{c}.
\newblock \bibinfo{title}{{Complexity issue on 5.4+ using
  kubeProxyReplacement=disabled + IPSec}}.
\newblock \bibinfo{howpublished}{[Online. Retrieved Jul 12, 2021.]
  \url{https://github.com/cilium/cilium/issues/14784}}.
\newblock


\bibitem[\protect\citeauthoryear{??}{cil}{2021d}]%
        {cilium-complexity-issue-4}
 \bibinfo{year}{2021}\natexlab{d}.
\newblock \bibinfo{title}{{Complexity Issue with cilium v1.9.5 when
  enable-endpoint-routes=true}}.
\newblock \bibinfo{howpublished}{[Online. Retrieved Jul 12, 2021.]
  \url{https://github.com/cilium/cilium/issues/16144}}.
\newblock


\bibitem[\protect\citeauthoryear{??}{cil}{2021e}]%
        {cilium-complexity-issue-6}
 \bibinfo{year}{2021}\natexlab{e}.
\newblock \bibinfo{title}{{Complexity issue with socket-level LB disabled on
  Linux 5.10 and Cilium 1.8.7}}.
\newblock \bibinfo{howpublished}{[Online. Retrieved Jul 12, 2021.]
  \url{https://github.com/cilium/cilium/issues/15249}}.
\newblock


\bibitem[\protect\citeauthoryear{??}{bpf}{2021c}]%
        {bpf-program-size-limit}
 \bibinfo{year}{2021}\natexlab{c}.
\newblock \bibinfo{title}{{Did you know? BPF program size limit}}.
\newblock \bibinfo{howpublished}{[Online. Retrieved Jul 12, 2021.]
  \url{https://ebpf.io/blog/ebpf-updates-2021-02\#did-you-know-program-size-limit}}.
\newblock


\bibitem[\protect\citeauthoryear{??}{bpf}{2021d}]%
        {bpf-nonprivileged-program-types}
 \bibinfo{year}{2021}\natexlab{d}.
\newblock \bibinfo{title}{{System-call check for BPF non-privileged program
  types}}.
\newblock \bibinfo{howpublished}{[Online. Retrieved Jul 12, 2021.]
  \url{https://elixir.bootlin.com/linux/v5.13/source/kernel/bpf/syscall.c\#L2115}}.
\newblock


\bibitem[\protect\citeauthoryear{??}{cil}{2021f}]%
        {cilium-complexity-issue-5}
 \bibinfo{year}{2021}\natexlab{f}.
\newblock \bibinfo{title}{{v1.9: CI: K8sVerifier Runs the kernel verifier
  against Cilium's BPF datapath on 5.4 }}.
\newblock \bibinfo{howpublished}{[Online. Retrieved Jul 12, 2021.]
  \url{https://github.com/cilium/cilium/issues/16050}}.
\newblock


\bibitem[\protect\citeauthoryear{Ahern}{Ahern}{2020}]%
        {cpu-cost-of-networking}
\bibfield{author}{\bibinfo{person}{David Ahern}.}
  \bibinfo{year}{2020}\natexlab{}.
\newblock \bibinfo{title}{The CPU cost of networking on a host}.
\newblock \bibinfo{howpublished}{[Online. Retrieved Jan 25, 2021.]
  \url{https://people.kernel.org/dsahern/the-cpu-cost-of-networking-on-a-host}}.
\newblock


\bibitem[\protect\citeauthoryear{Alur, Singh, Fisman, and Solar-Lezama}{Alur
  et~al\mbox{.}}{2018}]%
        {sygus-cacm18}
\bibfield{author}{\bibinfo{person}{Rajeev Alur}, \bibinfo{person}{Rishabh
  Singh}, \bibinfo{person}{Dana Fisman}, {and} \bibinfo{person}{Armando
  Solar-Lezama}.} \bibinfo{year}{2018}\natexlab{}.
\newblock \showarticletitle{Search-based program synthesis}.
\newblock \bibinfo{journal}{\emph{Commun. ACM}} \bibinfo{volume}{61},
  \bibinfo{number}{12} (\bibinfo{year}{2018}), \bibinfo{pages}{84--93}.
\newblock


\bibitem[\protect\citeauthoryear{Appel}{Appel}{2004}]%
        {compiler-implementation-appel04}
\bibfield{author}{\bibinfo{person}{Andrew~W Appel}.}
  \bibinfo{year}{2004}\natexlab{}.
\newblock \bibinfo{booktitle}{\emph{Modern compiler implementation in C}}.
\newblock \bibinfo{publisher}{Cambridge university press}.
\newblock


\bibitem[\protect\citeauthoryear{Bansal and Aiken}{Bansal and Aiken}{2006}]%
        {peephole-asplos06}
\bibfield{author}{\bibinfo{person}{Sorav Bansal} {and} \bibinfo{person}{Alex
  Aiken}.} \bibinfo{year}{2006}\natexlab{}.
\newblock \showarticletitle{Automatic generation of peephole superoptimizers}.
  In \bibinfo{booktitle}{\emph{ASPLOS}}.
\newblock


\bibitem[\protect\citeauthoryear{Bansal and Aiken}{Bansal and Aiken}{2008}]%
        {binary-translation-osdi08}
\bibfield{author}{\bibinfo{person}{Sorav Bansal} {and} \bibinfo{person}{Alex
  Aiken}.} \bibinfo{year}{2008}\natexlab{}.
\newblock \showarticletitle{Binary translation using peephole superoptimizers}.
  In \bibinfo{booktitle}{\emph{Proceedings of the 8th USENIX conference on
  Operating systems design and implementation}}. USENIX Association,
  \bibinfo{pages}{177--192}.
\newblock


\bibitem[\protect\citeauthoryear{Baubeau}{Baubeau}{2020}]%
        {datadog-file-integrity-monitoring}
\bibfield{author}{\bibinfo{person}{Sylvain Baubeau}.}
  \bibinfo{year}{2020}\natexlab{}.
\newblock \bibinfo{title}{{File integrity monitoring using eBPF}}.
\newblock \bibinfo{howpublished}{[Online. Retrieved Jan 23, 2021.]
  \url{https://www.devseccon.com/file-integrity-monitoring-using-ebpf-secadvent-day-19/}}.
\newblock


\bibitem[\protect\citeauthoryear{Beckett, Mahajan, Millstein, Padhye, and
  Walker}{Beckett et~al\mbox{.}}{2016}]%
        {propane-sigcomm16}
\bibfield{author}{\bibinfo{person}{Ryan Beckett}, \bibinfo{person}{Ratul
  Mahajan}, \bibinfo{person}{Todd Millstein}, \bibinfo{person}{Jitendra
  Padhye}, {and} \bibinfo{person}{David Walker}.}
  \bibinfo{year}{2016}\natexlab{}.
\newblock \showarticletitle{Don't Mind the Gap: Bridging Network-Wide
  Objectives and Device-Level Configurations}. In
  \bibinfo{booktitle}{\emph{Proceedings of the 2016 ACM SIGCOMM Conference}}
  (Florianopolis, Brazil) \emph{(\bibinfo{series}{SIGCOMM '16})}.
  \bibinfo{publisher}{Association for Computing Machinery},
  \bibinfo{address}{New York, NY, USA}, \bibinfo{pages}{328–341}.
\newblock
\showISBNx{9781450341936}
\urldef\tempurl%
\url{https://doi.org/10.1145/2934872.2934909}
\showDOI{\tempurl}


\bibitem[\protect\citeauthoryear{Beckett, Mahajan, Millstein, Padhye, and
  Walker}{Beckett et~al\mbox{.}}{2017}]%
        {propane-at-pldi17}
\bibfield{author}{\bibinfo{person}{Ryan Beckett}, \bibinfo{person}{Ratul
  Mahajan}, \bibinfo{person}{Todd Millstein}, \bibinfo{person}{Jitendra
  Padhye}, {and} \bibinfo{person}{David Walker}.}
  \bibinfo{year}{2017}\natexlab{}.
\newblock \showarticletitle{Network Configuration Synthesis with Abstract
  Topologies}.
\newblock \bibinfo{journal}{\emph{SIGPLAN Not.}} \bibinfo{volume}{52},
  \bibinfo{number}{6} (\bibinfo{date}{June} \bibinfo{year}{2017}),
  \bibinfo{pages}{437–451}.
\newblock
\showISSN{0362-1340}
\urldef\tempurl%
\url{https://doi.org/10.1145/3140587.3062367}
\showDOI{\tempurl}


\bibitem[\protect\citeauthoryear{Belay, Prekas, Klimovic, Grossman, Kozyrakis,
  and Bugnion}{Belay et~al\mbox{.}}{2014}]%
        {ix}
\bibfield{author}{\bibinfo{person}{Adam Belay}, \bibinfo{person}{George
  Prekas}, \bibinfo{person}{Ana Klimovic}, \bibinfo{person}{Samuel Grossman},
  \bibinfo{person}{Christos Kozyrakis}, {and} \bibinfo{person}{Edouard
  Bugnion}.} \bibinfo{year}{2014}\natexlab{}.
\newblock \showarticletitle{{IX}: A Protected Dataplane Operating System for
  High Throughput and Low Latency}. In \bibinfo{booktitle}{\emph{11th {USENIX}
  Symposium on Operating Systems Design and Implementation ({OSDI} 14)}}.
\newblock


\bibitem[\protect\citeauthoryear{Biere, Cimatti, Clarke, and Zhu}{Biere
  et~al\mbox{.}}{1999}]%
        {bounded-model-checking-biere99}
\bibfield{author}{\bibinfo{person}{Armin Biere}, \bibinfo{person}{Alessandro
  Cimatti}, \bibinfo{person}{Edmund Clarke}, {and} \bibinfo{person}{Yunshan
  Zhu}.} \bibinfo{year}{1999}\natexlab{}.
\newblock \showarticletitle{Symbolic model checking without BDDs}. In
  \bibinfo{booktitle}{\emph{International conference on tools and algorithms
  for the construction and analysis of systems}}. Springer,
  \bibinfo{pages}{193--207}.
\newblock


\bibitem[\protect\citeauthoryear{{Bjorn Topel et al}}{{Bjorn Topel et
  al}}{2018}]%
        {af-xdp}
\bibfield{author}{\bibinfo{person}{{Bjorn Topel et al}}.}
  \bibinfo{year}{2018}\natexlab{}.
\newblock \bibinfo{title}{AF\_XDP}.
\newblock \bibinfo{howpublished}{[Online, Retrieved Jan 20, 2021.]
  \url{https://www.kernel.org/doc/html/latest/networking/af_xdp.html}}.
\newblock


\bibitem[\protect\citeauthoryear{Bj{\o}rner, Gurfinkel, McMillan, and
  Rybalchenko}{Bj{\o}rner et~al\mbox{.}}{2015}]%
        {horn-clause-solvers}
\bibfield{author}{\bibinfo{person}{Nikolaj Bj{\o}rner}, \bibinfo{person}{Arie
  Gurfinkel}, \bibinfo{person}{Ken McMillan}, {and} \bibinfo{person}{Andrey
  Rybalchenko}.} \bibinfo{year}{2015}\natexlab{}.
\newblock \bibinfo{booktitle}{\emph{Horn Clause Solvers for Program
  Verification}}.
\newblock \bibinfo{publisher}{Springer International Publishing},
  \bibinfo{address}{Cham}, \bibinfo{pages}{24--51}.
\newblock
\showISBNx{978-3-319-23534-9}
\urldef\tempurl%
\url{https://doi.org/10.1007/978-3-319-23534-9_2}
\showDOI{\tempurl}


\bibitem[\protect\citeauthoryear{Borkmann and Pumputis}{Borkmann and
  Pumputis}{2020}]%
        {kubernetes-load-balancing-with-xdp}
\bibfield{author}{\bibinfo{person}{Daniel Borkmann} {and}
  \bibinfo{person}{Martynas Pumputis}.} \bibinfo{year}{2020}\natexlab{}.
\newblock \bibinfo{title}{{K8s Service Load Balancing with BPF \& XDP}}.
\newblock \bibinfo{howpublished}{[Online. Retrieved Jan 23, 2021.]
  \url{https://linuxplumbersconf.org/event/7/contributions/674/attachments/568/1002/plumbers_2020_cilium_load_balancer.pdf}}.
\newblock


\bibitem[\protect\citeauthoryear{Bornholt and Torlak}{Bornholt and
  Torlak}{2018}]%
        {finding-code-oopsla18}
\bibfield{author}{\bibinfo{person}{James Bornholt} {and} \bibinfo{person}{Emina
  Torlak}.} \bibinfo{year}{2018}\natexlab{}.
\newblock \showarticletitle{Finding code that explodes under symbolic
  evaluation}.
\newblock \bibinfo{journal}{\emph{Proceedings of the ACM on Programming
  Languages}} \bibinfo{volume}{2}, \bibinfo{number}{OOPSLA}
  (\bibinfo{year}{2018}), \bibinfo{pages}{1--26}.
\newblock


\bibitem[\protect\citeauthoryear{Bradley and Manna}{Bradley and Manna}{2007}]%
        {calculus-of-computation-book}
\bibfield{author}{\bibinfo{person}{Aaron~R Bradley} {and}
  \bibinfo{person}{Zohar Manna}.} \bibinfo{year}{2007}\natexlab{}.
\newblock \bibinfo{booktitle}{\emph{The calculus of computation: decision
  procedures with applications to verification}}.
\newblock \bibinfo{publisher}{Springer Science \& Business Media}.
\newblock


\bibitem[\protect\citeauthoryear{Brunella, Belocchi, Bonola, Pontarelli,
  Siracusano, Bianchi, Cammarano, Palumbo, Petrucci, and Bifulco}{Brunella
  et~al\mbox{.}}{2020}]%
        {hxdp-osdi20}
\bibfield{author}{\bibinfo{person}{Marco~Spaziani Brunella},
  \bibinfo{person}{Giacomo Belocchi}, \bibinfo{person}{Marco Bonola},
  \bibinfo{person}{Salvatore Pontarelli}, \bibinfo{person}{Giuseppe
  Siracusano}, \bibinfo{person}{Giuseppe Bianchi}, \bibinfo{person}{Aniello
  Cammarano}, \bibinfo{person}{Alessandro Palumbo}, \bibinfo{person}{Luca
  Petrucci}, {and} \bibinfo{person}{Roberto Bifulco}.}
  \bibinfo{year}{2020}\natexlab{}.
\newblock \showarticletitle{hXDP: Efficient Software Packet Processing on
  {FPGA} NICs}. In \bibinfo{booktitle}{\emph{14th {USENIX} Symposium on
  Operating Systems Design and Implementation ({OSDI} 20)}}.
  \bibinfo{publisher}{{USENIX} Association}, \bibinfo{pages}{973--990}.
\newblock
\showISBNx{978-1-939133-19-9}
\urldef\tempurl%
\url{https://www.usenix.org/conference/osdi20/presentation/brunella}
\showURL{%
\tempurl}


\bibitem[\protect\citeauthoryear{{Chonggang Li, Craig Gallek, Eddie Hao, Kevin
  Athey, Maciej Żenczykowski, Vlad Dumitrescu, Willem de Bruijn, Xiaotian
  Pei}}{{Chonggang Li, Craig Gallek, Eddie Hao, Kevin Athey, Maciej
  Żenczykowski, Vlad Dumitrescu, Willem de Bruijn, Xiaotian Pei}}{2018}]%
        {scaling-linux-networking-stack}
\bibfield{author}{\bibinfo{person}{{Chonggang Li, Craig Gallek, Eddie Hao,
  Kevin Athey, Maciej Żenczykowski, Vlad Dumitrescu, Willem de Bruijn,
  Xiaotian Pei}}.} \bibinfo{year}{2018}\natexlab{}.
\newblock \bibinfo{title}{{Scaling in the Linux Networking Stack}}.
\newblock \bibinfo{howpublished}{[Online, Retrieved Jan 20, 2021.]
  \url{https://www.kernel.org/doc/html/v5.8/networking/scaling.html}}.
\newblock


\bibitem[\protect\citeauthoryear{{Chonggang Li, Craig Gallek, Eddie Hao,Kevin
  Athey, Maciej Żenczykowski,Vlad Dumitrescu, Willem de Bruijn,Xiaotian
  Pei}}{{Chonggang Li, Craig Gallek, Eddie Hao,Kevin Athey, Maciej
  Żenczykowski,Vlad Dumitrescu, Willem de Bruijn,Xiaotian Pei}}{2018}]%
        {bpf-traffic-control}
\bibfield{author}{\bibinfo{person}{{Chonggang Li, Craig Gallek, Eddie Hao,Kevin
  Athey, Maciej Żenczykowski,Vlad Dumitrescu, Willem de Bruijn,Xiaotian
  Pei}}.} \bibinfo{year}{2018}\natexlab{}.
\newblock \bibinfo{title}{{Scaling Linux Traffic Shaping with BPF}}.
\newblock \bibinfo{howpublished}{[Online. Retrieved Jan 25, 2021.]
  \url{http://vger.kernel.org/lpc_bpf2018_talks/lpc-bpf-2018-shaping.pdf}}.
\newblock


\bibitem[\protect\citeauthoryear{Churchill, Sharma, Bastien, and
  Aiken}{Churchill et~al\mbox{.}}{2017}]%
        {loopsuperopt-pldi17}
\bibfield{author}{\bibinfo{person}{Berkeley Churchill}, \bibinfo{person}{Rahul
  Sharma}, \bibinfo{person}{JF Bastien}, {and} \bibinfo{person}{Alex Aiken}.}
  \bibinfo{year}{2017}\natexlab{}.
\newblock \showarticletitle{Sound loop superoptimization for google native
  client}.
\newblock \bibinfo{journal}{\emph{ACM SIGPLAN Notices}} \bibinfo{volume}{52},
  \bibinfo{number}{4} (\bibinfo{year}{2017}), \bibinfo{pages}{313--326}.
\newblock


\bibitem[\protect\citeauthoryear{{Cilium}}{{Cilium}}{2017}]%
        {kubeproxy-replacement}
\bibfield{author}{\bibinfo{person}{{Cilium}}.}
  \bibinfo{year}{{2017}}\natexlab{}.
\newblock \bibinfo{title}{{Kubernetes Without kube-proxy}}.
\newblock \bibinfo{howpublished}{[Online, Retrieved Jan 20, 2021.]
  \url{https://docs.cilium.io/en/v1.9/gettingstarted/kubeproxy-free/}}.
\newblock


\bibitem[\protect\citeauthoryear{Clarke}{Clarke}{1976}]%
        {symbolic-exec-clarke76}
\bibfield{author}{\bibinfo{person}{Lori~A. Clarke}.}
  \bibinfo{year}{1976}\natexlab{}.
\newblock \showarticletitle{A system to generate test data and symbolically
  execute programs}.
\newblock \bibinfo{journal}{\emph{IEEE Transactions on software engineering}}
  \bibinfo{number}{3} (\bibinfo{year}{1976}), \bibinfo{pages}{215--222}.
\newblock


\bibitem[\protect\citeauthoryear{Corbet}{Corbet}{2002}]%
        {tso}
\bibfield{author}{\bibinfo{person}{Jonathan Corbet}.}
  \bibinfo{year}{2002}\natexlab{}.
\newblock \bibinfo{title}{{TCP segmentation offloading (TSO)}}.
\newblock \bibinfo{howpublished}{[Online, Retrieved Jan 20, 2021.]
  https://lwn.net/Articles/9129/}.
\newblock


\bibitem[\protect\citeauthoryear{Corbet}{Corbet}{2014}]%
        {bpf-universal-kernel-vm}
\bibfield{author}{\bibinfo{person}{Jonathan Corbet}.}
  \bibinfo{year}{2014}\natexlab{}.
\newblock \bibinfo{title}{{BPF: the universal in-kernel virtual machine}}.
\newblock \bibinfo{howpublished}{[Online, Retrieved Jan 20, 2021.]
  \url{https://lwn.net/Articles/599755/}}.
\newblock


\bibitem[\protect\citeauthoryear{Corbet}{Corbet}{2019a}]%
        {compiling-to-bpf-with-gcc}
\bibfield{author}{\bibinfo{person}{Jonathan Corbet}.}
  \bibinfo{year}{2019}\natexlab{a}.
\newblock \bibinfo{title}{{Compiling to BPF with gcc}}.
\newblock \bibinfo{howpublished}{[Online. Retrieved Jan 23, 2021.]
  \url{https://lwn.net/Articles/800606/}}.
\newblock


\bibitem[\protect\citeauthoryear{Corbet}{Corbet}{2019b}]%
        {bpf-concurrency-management}
\bibfield{author}{\bibinfo{person}{Jonathan Corbet}.}
  \bibinfo{year}{2019}\natexlab{b}.
\newblock \bibinfo{title}{{Concurrency management in BPF}}.
\newblock \bibinfo{howpublished}{{[Online, Retrieved Jun 19, 2021.]
  \url{https://lwn.net/Articles/779120/}}}.
\newblock


\bibitem[\protect\citeauthoryear{Corbet}{Corbet}{2021}]%
        {bpf-calling-kernel-functions}
\bibfield{author}{\bibinfo{person}{Jonathan Corbet}.}
  \bibinfo{year}{2021}\natexlab{}.
\newblock \bibinfo{title}{{Calling kernel functions from BPF}}.
\newblock \bibinfo{howpublished}{{[Online, Retrieved Jun 19, 2021.]
  \url{https://lwn.net/Articles/856005/}}}.
\newblock


\bibitem[\protect\citeauthoryear{Cytron, Ferrante, Rosen, Wegman, and
  Zadeck}{Cytron et~al\mbox{.}}{1991}]%
        {ssa-toplas91}
\bibfield{author}{\bibinfo{person}{Ron Cytron}, \bibinfo{person}{Jeanne
  Ferrante}, \bibinfo{person}{Barry~K Rosen}, \bibinfo{person}{Mark~N Wegman},
  {and} \bibinfo{person}{F~Kenneth Zadeck}.} \bibinfo{year}{1991}\natexlab{}.
\newblock \showarticletitle{Efficiently computing static single assignment form
  and the control dependence graph}.
\newblock \bibinfo{journal}{\emph{ACM Transactions on Programming Languages and
  Systems (TOPLAS)}} \bibinfo{volume}{13}, \bibinfo{number}{4}
  (\bibinfo{year}{1991}), \bibinfo{pages}{451--490}.
\newblock


\bibitem[\protect\citeauthoryear{Dalton, Schultz, Adriaens, Arefin, Gupta,
  Fahs, Rubinstein, Zermeno, Rubow, Docauer, et~al\mbox{.}}{Dalton
  et~al\mbox{.}}{2018}]%
        {andromeda-nsdi18}
\bibfield{author}{\bibinfo{person}{Michael Dalton}, \bibinfo{person}{David
  Schultz}, \bibinfo{person}{Jacob Adriaens}, \bibinfo{person}{Ahsan Arefin},
  \bibinfo{person}{Anshuman Gupta}, \bibinfo{person}{Brian Fahs},
  \bibinfo{person}{Dima Rubinstein}, \bibinfo{person}{Enrique~Cauich Zermeno},
  \bibinfo{person}{Erik Rubow}, \bibinfo{person}{James~Alexander Docauer},
  {et~al\mbox{.}}} \bibinfo{year}{2018}\natexlab{}.
\newblock \showarticletitle{Andromeda: Performance, isolation, and velocity at
  scale in cloud network virtualization}. In \bibinfo{booktitle}{\emph{15th
  $\{$USENIX$\}$ Symposium on Networked Systems Design and Implementation
  ($\{$NSDI$\}$ 18)}}. \bibinfo{pages}{373--387}.
\newblock


\bibitem[\protect\citeauthoryear{De~Moura and Bj{\o}rner}{De~Moura and
  Bj{\o}rner}{2008}]%
        {z3}
\bibfield{author}{\bibinfo{person}{Leonardo De~Moura} {and}
  \bibinfo{person}{Nikolaj Bj{\o}rner}.} \bibinfo{year}{2008}\natexlab{}.
\newblock \showarticletitle{Z3: An efficient SMT solver}. In
  \bibinfo{booktitle}{\emph{International conference on Tools and Algorithms
  for the Construction and Analysis of Systems}}. Springer,
  \bibinfo{pages}{337--340}.
\newblock


\bibitem[\protect\citeauthoryear{Duplyakin, Ricci, Maricq, Wong, Duerig, Eide,
  Stoller, Hibler, Johnson, Webb, Akella, Wang, Ricart, Landweber, Elliott,
  Zink, Cecchet, Kar, and Mishra}{Duplyakin et~al\mbox{.}}{2019}]%
        {cloudlab-atc19}
\bibfield{author}{\bibinfo{person}{Dmitry Duplyakin}, \bibinfo{person}{Robert
  Ricci}, \bibinfo{person}{Aleksander Maricq}, \bibinfo{person}{Gary Wong},
  \bibinfo{person}{Jonathon Duerig}, \bibinfo{person}{Eric Eide},
  \bibinfo{person}{Leigh Stoller}, \bibinfo{person}{Mike Hibler},
  \bibinfo{person}{David Johnson}, \bibinfo{person}{Kirk Webb},
  \bibinfo{person}{Aditya Akella}, \bibinfo{person}{Kuangching Wang},
  \bibinfo{person}{Glenn Ricart}, \bibinfo{person}{Larry Landweber},
  \bibinfo{person}{Chip Elliott}, \bibinfo{person}{Michael Zink},
  \bibinfo{person}{Emmanuel Cecchet}, \bibinfo{person}{Snigdhaswin Kar}, {and}
  \bibinfo{person}{Prabodh Mishra}.} \bibinfo{year}{2019}\natexlab{}.
\newblock \showarticletitle{The Design and Operation of {CloudLab}}. In
  \bibinfo{booktitle}{\emph{Proceedings of the {USENIX} Annual Technical
  Conference (ATC)}}. \bibinfo{pages}{1--14}.
\newblock
\urldef\tempurl%
\url{https://www.flux.utah.edu/paper/duplyakin-atc19}
\showURL{%
\tempurl}


\bibitem[\protect\citeauthoryear{Edge}{Edge}{2020}]%
        {bpf-in-gcc}
\bibfield{author}{\bibinfo{person}{Jake Edge}.}
  \bibinfo{year}{2020}\natexlab{}.
\newblock \bibinfo{title}{{BPF in GCC}}.
\newblock \bibinfo{howpublished}{[Online. Retrieved Jan 23, 2021.]
  \url{https://lwn.net/Articles/831402/}}.
\newblock


\bibitem[\protect\citeauthoryear{El-Hassany, Tsankov, Vanbever, and
  Vechev}{El-Hassany et~al\mbox{.}}{2017}]%
        {network-wide-config-synthesis-cav17}
\bibfield{author}{\bibinfo{person}{Ahmed El-Hassany}, \bibinfo{person}{Petar
  Tsankov}, \bibinfo{person}{Laurent Vanbever}, {and} \bibinfo{person}{Martin
  Vechev}.} \bibinfo{year}{2017}\natexlab{}.
\newblock \showarticletitle{Network-Wide Configuration Synthesis}. In
  \bibinfo{booktitle}{\emph{Computer Aided Verification}},
  \bibfield{editor}{\bibinfo{person}{Rupak Majumdar} {and}
  \bibinfo{person}{Viktor Kun{\v{c}}ak}} (Eds.). \bibinfo{publisher}{Springer
  International Publishing}, \bibinfo{address}{Cham},
  \bibinfo{pages}{261--281}.
\newblock
\showISBNx{978-3-319-63390-9}


\bibitem[\protect\citeauthoryear{El-Hassany, Tsankov, Vanbever, and
  Vechev}{El-Hassany et~al\mbox{.}}{2018}]%
        {netcomplete-nsdi18}
\bibfield{author}{\bibinfo{person}{Ahmed El-Hassany}, \bibinfo{person}{Petar
  Tsankov}, \bibinfo{person}{Laurent Vanbever}, {and} \bibinfo{person}{Martin
  Vechev}.} \bibinfo{year}{2018}\natexlab{}.
\newblock \showarticletitle{NetComplete: Practical Network-Wide Configuration
  Synthesis with Autocompletion}. In \bibinfo{booktitle}{\emph{15th {USENIX}
  Symposium on Networked Systems Design and Implementation ({NSDI} 18)}}.
  \bibinfo{publisher}{{USENIX} Association}, \bibinfo{address}{Renton, WA},
  \bibinfo{pages}{579--594}.
\newblock
\showISBNx{978-1-939133-01-4}
\urldef\tempurl%
\url{https://www.usenix.org/conference/nsdi18/presentation/el-hassany}
\showURL{%
\tempurl}


\bibitem[\protect\citeauthoryear{{Eric Leblond}}{{Eric Leblond}}{2016}]%
        {suricata-xdp}
\bibfield{author}{\bibinfo{person}{{Eric Leblond}}.}
  \bibinfo{year}{2016}\natexlab{}.
\newblock \bibinfo{title}{Suricata bypass feature}.
\newblock \bibinfo{howpublished}{[Online, Retrieved Jan 20, 2021.]
  \url{https://www.stamus-networks.com/blog/2016/09/28/suricata-bypass-feature
  }}.
\newblock


\bibitem[\protect\citeauthoryear{Fabre}{Fabre}{2018}]%
        {cloudflare-l4drop}
\bibfield{author}{\bibinfo{person}{Arthur Fabre}.}
  \bibinfo{year}{2018}\natexlab{}.
\newblock \bibinfo{title}{L4Drop: XDP DDoS Mitigations}.
\newblock \bibinfo{howpublished}{[Online, Retrieved Jan 20, 2021.]
  \url{https://blog.cloudflare.com/l4drop-xdp-ebpf-based-ddos-mitigations/}}.
\newblock


\bibitem[\protect\citeauthoryear{Farshin}{Farshin}{2019}]%
        {lowlatency-kth-thesis19}
\bibfield{author}{\bibinfo{person}{Alireza Farshin}.}
  \bibinfo{year}{2019}\natexlab{}.
\newblock \emph{\bibinfo{title}{Realizing Low-Latency Internet Services via
  Low-Level Optimization of NFV Service Chains}}.
\newblock \bibinfo{thesistype}{Ph.D. Dissertation}. \bibinfo{school}{KTH,
  Stockholm}.
\newblock
\urldef\tempurl%
\url{https://doi.org/10.13140/RG.2.2.22044.95361}
\showDOI{\tempurl}


\bibitem[\protect\citeauthoryear{Fetaya}{Fetaya}{2016}]%
        {ethanfetaya-mcmc-lecture-notes}
\bibfield{author}{\bibinfo{person}{Ethan Fetaya}.}
  \bibinfo{year}{2016}\natexlab{}.
\newblock \bibinfo{title}{Stochastic Optimization with MCMC}.
\newblock \bibinfo{howpublished}{[Online, Retrieved Jan 17, 2021.]
  \url{http://www.wisdom.weizmann.ac.il/~ethanf/MCMC/stochastic\%20optimization.pdf}}.
\newblock


\bibitem[\protect\citeauthoryear{Firestone, Putnam, Mundkur, Chiou, Dabagh,
  Andrewartha, Angepat, Bhanu, Caulfield, Chung, Chandrappa, Chaturmohta,
  Humphrey, Lavier, Lam, Liu, Ovtcharov, Padhye, Popuri, Raindel, Sapre, Shaw,
  Silva, Sivakumar, Srivastava, Verma, Zuhair, Bansal, Burger, Vaid, Maltz, and
  Greenberg}{Firestone et~al\mbox{.}}{2018}]%
        {azurenic-nsdi18}
\bibfield{author}{\bibinfo{person}{Daniel Firestone}, \bibinfo{person}{Andrew
  Putnam}, \bibinfo{person}{Sambhrama Mundkur}, \bibinfo{person}{Derek Chiou},
  \bibinfo{person}{Alireza Dabagh}, \bibinfo{person}{Mike Andrewartha},
  \bibinfo{person}{Hari Angepat}, \bibinfo{person}{Vivek Bhanu},
  \bibinfo{person}{Adrian Caulfield}, \bibinfo{person}{Eric Chung},
  \bibinfo{person}{Harish~Kumar Chandrappa}, \bibinfo{person}{Somesh
  Chaturmohta}, \bibinfo{person}{Matt Humphrey}, \bibinfo{person}{Jack Lavier},
  \bibinfo{person}{Norman Lam}, \bibinfo{person}{Fengfen Liu},
  \bibinfo{person}{Kalin Ovtcharov}, \bibinfo{person}{Jitu Padhye},
  \bibinfo{person}{Gautham Popuri}, \bibinfo{person}{Shachar Raindel},
  \bibinfo{person}{Tejas Sapre}, \bibinfo{person}{Mark Shaw},
  \bibinfo{person}{Gabriel Silva}, \bibinfo{person}{Madhan Sivakumar},
  \bibinfo{person}{Nisheeth Srivastava}, \bibinfo{person}{Anshuman Verma},
  \bibinfo{person}{Qasim Zuhair}, \bibinfo{person}{Deepak Bansal},
  \bibinfo{person}{Doug Burger}, \bibinfo{person}{Kushagra Vaid},
  \bibinfo{person}{David~A. Maltz}, {and} \bibinfo{person}{Albert Greenberg}.}
  \bibinfo{year}{2018}\natexlab{}.
\newblock \showarticletitle{{Azure Accelerated Networking: SmartNICs in the
  Public Cloud}}. In \bibinfo{booktitle}{\emph{{USENIX} Symposium on Networked
  Systems Design and Implementation ({NSDI} 18)}}.
\newblock


\bibitem[\protect\citeauthoryear{Gao, Kim, Wong, Raghunathan, Varma, Kannan,
  Sivaraman, Narayana, and Gupta}{Gao et~al\mbox{.}}{2020}]%
        {chipmunk-sigcomm20}
\bibfield{author}{\bibinfo{person}{Xiangyu Gao}, \bibinfo{person}{Taegyun Kim},
  \bibinfo{person}{Michael~D Wong}, \bibinfo{person}{Divya Raghunathan},
  \bibinfo{person}{Aatish~Kishan Varma}, \bibinfo{person}{Pravein~Govindan
  Kannan}, \bibinfo{person}{Anirudh Sivaraman}, \bibinfo{person}{Srinivas
  Narayana}, {and} \bibinfo{person}{Aarti Gupta}.}
  \bibinfo{year}{2020}\natexlab{}.
\newblock \showarticletitle{Switch code generation using program synthesis}. In
  \bibinfo{booktitle}{\emph{Proceedings of the Annual conference of the ACM
  Special Interest Group on Data Communication on the applications,
  technologies, architectures, and protocols for computer communication}}.
  \bibinfo{pages}{44--61}.
\newblock


\bibitem[\protect\citeauthoryear{{Gavin Stark and Sakin Sezer}}{{Gavin Stark
  and Sakin Sezer}}{2020}]%
        {netronome-hotchips-talk}
\bibfield{author}{\bibinfo{person}{{Gavin Stark and Sakin Sezer}}.}
  \bibinfo{year}{2020}\natexlab{}.
\newblock \bibinfo{title}{{A 22nm High-Performance Flow Processor for 200Gb/s
  Software Defined Networking}}.
\newblock \bibinfo{howpublished}{[Online, Retrieved July 1, 2021.]
  \url{https://old.hotchips.org/wp-content/uploads/hc_archives/hc25/HC25.60-Networking-epub/HC25.27.620-22nm-Flow-Proc-Stark-Netronome.pdf}}.
\newblock


\bibitem[\protect\citeauthoryear{Gershuni, Amit, Gurfinkel, Narodytska, Navas,
  Rinetzky, Ryzhyk, and Sagiv}{Gershuni et~al\mbox{.}}{2019}]%
        {untrusted-extensions-pldi19}
\bibfield{author}{\bibinfo{person}{Elazar Gershuni}, \bibinfo{person}{Nadav
  Amit}, \bibinfo{person}{Arie Gurfinkel}, \bibinfo{person}{Nina Narodytska},
  \bibinfo{person}{Jorge~A Navas}, \bibinfo{person}{Noam Rinetzky},
  \bibinfo{person}{Leonid Ryzhyk}, {and} \bibinfo{person}{Mooly Sagiv}.}
  \bibinfo{year}{2019}\natexlab{}.
\newblock \showarticletitle{Simple and precise static analysis of untrusted
  linux kernel extensions}. In \bibinfo{booktitle}{\emph{Proceedings of the
  40th ACM SIGPLAN Conference on Programming Language Design and
  Implementation}}. \bibinfo{pages}{1069--1084}.
\newblock


\bibitem[\protect\citeauthoryear{Gilks, Richardson, and Spiegelhalter}{Gilks
  et~al\mbox{.}}{1996}]%
        {mcmc-in-practice-text96}
\bibfield{author}{\bibinfo{person}{WR Gilks}, \bibinfo{person}{S Richardson},
  {and} \bibinfo{person}{DJ Spiegelhalter}.} \bibinfo{year}{1996}\natexlab{}.
\newblock \bibinfo{booktitle}{\emph{Markov Chain Monte Carlo in Practice}}.
\newblock \bibinfo{publisher}{Chapman \& Hall}, \bibinfo{address}{London}.
\newblock


\bibitem[\protect\citeauthoryear{Gregg}{Gregg}{2019a}]%
        {bpf-microkernel}
\bibfield{author}{\bibinfo{person}{Brendan Gregg}.}
  \bibinfo{year}{2019}\natexlab{a}.
\newblock \bibinfo{title}{BPF: a new type of software}.
\newblock \bibinfo{howpublished}{[Online, Retrieved Jan 19, 2020.]
  \url{http://www.brendangregg.com/blog/2019-12-02/bpf-a-new-type-of-software.html}}.
\newblock


\bibitem[\protect\citeauthoryear{Gregg}{Gregg}{2019b}]%
        {brendan-gregg-reinvent-talk-2019}
\bibfield{author}{\bibinfo{person}{Brendan Gregg}.}
  \bibinfo{year}{2019}\natexlab{b}.
\newblock \bibinfo{title}{BPF Performance Analysis at Netflix}.
\newblock \bibinfo{howpublished}{[Online, Retrieved Jan 19, 2020.]
  \url{https://www.slideshare.net/brendangregg/reinvent-2019-bpf-performance-analysis-at-netflix}}.
\newblock


\bibitem[\protect\citeauthoryear{gro}{gro}{2009}]%
        {gro}
gro \bibinfo{year}{2009}\natexlab{}.
\newblock \bibinfo{title}{{Generic Receive Offload (GRO)}}.
\newblock \bibinfo{howpublished}{[Online, Retrieved Nov 15, 2018.]
  https://lwn.net/Articles/358910/}.
\newblock


\bibitem[\protect\citeauthoryear{Gulwani, Jha, Tiwari, and Venkatesan}{Gulwani
  et~al\mbox{.}}{2011}]%
        {loopfree-pldi11}
\bibfield{author}{\bibinfo{person}{Sumit Gulwani}, \bibinfo{person}{Susmit
  Jha}, \bibinfo{person}{Ashish Tiwari}, {and} \bibinfo{person}{Ramarathnam
  Venkatesan}.} \bibinfo{year}{2011}\natexlab{}.
\newblock \showarticletitle{Synthesis of loop-free programs}.
\newblock \bibinfo{journal}{\emph{ACM SIGPLAN Notices}} \bibinfo{volume}{46},
  \bibinfo{number}{6} (\bibinfo{year}{2011}), \bibinfo{pages}{62--73}.
\newblock


\bibitem[\protect\citeauthoryear{H\o{}iland-J\o{}rgensen, Brouer, Borkmann,
  Fastabend, Herbert, Ahern, and Miller}{H\o{}iland-J\o{}rgensen
  et~al\mbox{.}}{2018}]%
        {xdp-conext18}
\bibfield{author}{\bibinfo{person}{Toke H\o{}iland-J\o{}rgensen},
  \bibinfo{person}{Jesper~Dangaard Brouer}, \bibinfo{person}{Daniel Borkmann},
  \bibinfo{person}{John Fastabend}, \bibinfo{person}{Tom Herbert},
  \bibinfo{person}{David Ahern}, {and} \bibinfo{person}{David Miller}.}
  \bibinfo{year}{2018}\natexlab{}.
\newblock \showarticletitle{The EXpress Data Path: Fast Programmable Packet
  Processing in the Operating System Kernel}. In
  \bibinfo{booktitle}{\emph{Proceedings of the 14th International Conference on
  Emerging Networking EXperiments and Technologies}} (Heraklion, Greece)
  \emph{(\bibinfo{series}{CoNEXT ’18})}. \bibinfo{publisher}{Association for
  Computing Machinery}, \bibinfo{address}{New York, NY, USA},
  \bibinfo{pages}{54–66}.
\newblock
\showISBNx{9781450360807}
\urldef\tempurl%
\url{https://doi.org/10.1145/3281411.3281443}
\showDOI{\tempurl}


\bibitem[\protect\citeauthoryear{Huang, Qiu, Shen, and Wang}{Huang
  et~al\mbox{.}}{2020}]%
        {dryadsynth-pldi20}
\bibfield{author}{\bibinfo{person}{Kangjing Huang}, \bibinfo{person}{Xiaokang
  Qiu}, \bibinfo{person}{Peiyuan Shen}, {and} \bibinfo{person}{Yanjun Wang}.}
  \bibinfo{year}{2020}\natexlab{}.
\newblock \showarticletitle{Reconciling Enumerative and Deductive Program
  Synthesis}. In \bibinfo{booktitle}{\emph{Proceedings of the 41st ACM SIGPLAN
  Conference on Programming Language Design and Implementation}} (London, UK)
  \emph{(\bibinfo{series}{PLDI 2020})}. \bibinfo{publisher}{Association for
  Computing Machinery}, \bibinfo{address}{New York, NY, USA},
  \bibinfo{pages}{1159–1174}.
\newblock
\showISBNx{9781450376136}
\urldef\tempurl%
\url{https://doi.org/10.1145/3385412.3386027}
\showDOI{\tempurl}


\bibitem[\protect\citeauthoryear{Huang, Qiu, and Wang}{Huang
  et~al\mbox{.}}{2017}]%
        {dryadsynth}
\bibfield{author}{\bibinfo{person}{Kangjing Huang}, \bibinfo{person}{Xiaokang
  Qiu}, {and} \bibinfo{person}{Yanjun Wang}.} \bibinfo{year}{2017}\natexlab{}.
\newblock \showarticletitle{DRYADSYNTH: A Concolic SyGuS Solver}.
\newblock  (\bibinfo{year}{2017}).
\newblock


\bibitem[\protect\citeauthoryear{{Ingo Molnar and Max Krasnyansky}}{{Ingo
  Molnar and Max Krasnyansky}}{2021}]%
        {smp-irq-affinity}
\bibfield{author}{\bibinfo{person}{{Ingo Molnar and Max Krasnyansky}}.}
  \bibinfo{year}{2021}\natexlab{}.
\newblock \bibinfo{title}{{SMP IRQ affinity}}.
\newblock \bibinfo{howpublished}{[Online. Retrieved Jan 27, 2021.]
  \url{https://www.kernel.org/doc/html/latest/core-api/irq/irq-affinity.html}}.
\newblock


\bibitem[\protect\citeauthoryear{Intel}{Intel}{2010}]%
        {dpdk}
\bibfield{author}{\bibinfo{person}{Intel}.} \bibinfo{year}{2010}\natexlab{}.
\newblock \bibinfo{title}{{Data Plane Development Kit (DPDK)}}.
\newblock \bibinfo{howpublished}{[Online, Retrieved Nov 15, 2018.]
  https://www.dpdk.org/}.
\newblock


\bibitem[\protect\citeauthoryear{{Jay Schulist, Daniel Borkmann, Alexei
  Starovoitov}}{{Jay Schulist, Daniel Borkmann, Alexei Starovoitov}}{[n.d.]}]%
        {bpf-kernel-documentation}
\bibfield{author}{\bibinfo{person}{{Jay Schulist, Daniel Borkmann, Alexei
  Starovoitov}}.} \bibinfo{year}{[n.d.]}\natexlab{}.
\newblock \bibinfo{title}{{Linux Socket Filtering aka Berkeley Packet Filter
  (BPF)}}.
\newblock \bibinfo{howpublished}{[Online, Retrieved Oct 29, 2020.]
  \url{https://www.kernel.org/doc/Documentation/networking/filter.txt}}.
\newblock


\bibitem[\protect\citeauthoryear{Jha, Gulwani, Seshia, and Tiwari}{Jha
  et~al\mbox{.}}{2010}]%
        {oracle-component-synthesis-icse10}
\bibfield{author}{\bibinfo{person}{Susmit Jha}, \bibinfo{person}{Sumit
  Gulwani}, \bibinfo{person}{Sanjit~A Seshia}, {and} \bibinfo{person}{Ashish
  Tiwari}.} \bibinfo{year}{2010}\natexlab{}.
\newblock \showarticletitle{Oracle-guided component-based program synthesis}.
  In \bibinfo{booktitle}{\emph{2010 ACM/IEEE 32nd International Conference on
  Software Engineering}}, Vol.~\bibinfo{volume}{1}. IEEE,
  \bibinfo{pages}{215--224}.
\newblock


\bibitem[\protect\citeauthoryear{{Johar, Gobind and Marupadi, Varun}}{{Johar,
  Gobind and Marupadi, Varun}}{2020}]%
        {cilium-google-kubernetes-engine}
\bibfield{author}{\bibinfo{person}{{Johar, Gobind and Marupadi, Varun}}.}
  \bibinfo{year}{2020}\natexlab{}.
\newblock \bibinfo{title}{{New GKE Dataplane V2 increases security and
  visibility for containers}}.
\newblock \bibinfo{howpublished}{[Online, Retrieved Jan 20, 2021.]
  \url{https://cloud.google.com/blog/products/containers-kubernetes/bringing-ebpf-and-cilium-to-google-kubernetes-engine}}.
\newblock


\bibitem[\protect\citeauthoryear{{Jonathan Corbet}}{{Jonathan Corbet}}{2018}]%
        {accelerate-with-af-xdp}
\bibfield{author}{\bibinfo{person}{{Jonathan Corbet}}.}
  \bibinfo{year}{2018}\natexlab{}.
\newblock \bibinfo{title}{{Accelerating networking with AF\_XDP}}.
\newblock \bibinfo{howpublished}{[Online. Retrieved Jan 20, 2021.]
  \url{https://lwn.net/Articles/750845/}}.
\newblock


\bibitem[\protect\citeauthoryear{{Jonathan Corbet}}{{Jonathan Corbet}}{2020}]%
        {bpf-tcp-cc-struct-ops}
\bibfield{author}{\bibinfo{person}{{Jonathan Corbet}}.}
  \bibinfo{year}{2020}\natexlab{}.
\newblock \bibinfo{title}{{Kernel operations structures in BPF}}.
\newblock \bibinfo{howpublished}{[Online, Retrieved Jan 20, 2021.]
  \url{https://lwn.net/Articles/811631/}}.
\newblock


\bibitem[\protect\citeauthoryear{Joshi, Nelson, and Randall}{Joshi
  et~al\mbox{.}}{2002}]%
        {denali-pldi02}
\bibfield{author}{\bibinfo{person}{Rajeev Joshi}, \bibinfo{person}{Greg
  Nelson}, {and} \bibinfo{person}{Keith Randall}.}
  \bibinfo{year}{2002}\natexlab{}.
\newblock \showarticletitle{Denali: a goal-directed superoptimizer}.
\newblock \bibinfo{journal}{\emph{ACM SIGPLAN Notices}} \bibinfo{volume}{37},
  \bibinfo{number}{5} (\bibinfo{year}{2002}), \bibinfo{pages}{304--314}.
\newblock


\bibitem[\protect\citeauthoryear{Kerrisk}{Kerrisk}{2021a}]%
        {tc-bpf-man}
\bibfield{author}{\bibinfo{person}{Michael Kerrisk}.}
  \bibinfo{year}{2021}\natexlab{a}.
\newblock \bibinfo{title}{BPF classifier and actions in tc}.
\newblock \bibinfo{howpublished}{[Online, Retrieved Jan 20, 2021.]
  \url{https://www.man7.org/linux/man-pages/man8/tc-bpf.8.html}}.
\newblock


\bibitem[\protect\citeauthoryear{Kerrisk}{Kerrisk}{2021b}]%
        {bpf-helper-calls}
\bibfield{author}{\bibinfo{person}{Michael Kerrisk}.}
  \bibinfo{year}{2021}\natexlab{b}.
\newblock \bibinfo{title}{BPF-helpers: a list of eBPF helper functions}.
\newblock \bibinfo{howpublished}{[Online, Retrieved Oct 29, 2020.]
  \url{https://www.man7.org/linux/man-pages/man7/bpf-helpers.7.html}}.
\newblock


\bibitem[\protect\citeauthoryear{Khalid, Rozner, Felter, Xu, Rajamani,
  Ferreira, and Akella}{Khalid et~al\mbox{.}}{2018}]%
        {iron-nsdi18}
\bibfield{author}{\bibinfo{person}{Junaid Khalid}, \bibinfo{person}{Eric
  Rozner}, \bibinfo{person}{Wesley Felter}, \bibinfo{person}{Cong Xu},
  \bibinfo{person}{Karthick Rajamani}, \bibinfo{person}{Alexandre Ferreira},
  {and} \bibinfo{person}{Aditya Akella}.} \bibinfo{year}{2018}\natexlab{}.
\newblock \showarticletitle{Iron: Isolating Network-based $\{$CPU$\}$ in
  Container Environments}. In \bibinfo{booktitle}{\emph{15th $\{$USENIX$\}$
  Symposium on Networked Systems Design and Implementation ($\{$NSDI$\}$ 18)}}.
  \bibinfo{pages}{313--328}.
\newblock


\bibitem[\protect\citeauthoryear{King}{King}{1976}]%
        {symbolic-exec-king76}
\bibfield{author}{\bibinfo{person}{James~C King}.}
  \bibinfo{year}{1976}\natexlab{}.
\newblock \showarticletitle{Symbolic execution and program testing}.
\newblock \bibinfo{journal}{\emph{Commun. ACM}} \bibinfo{volume}{19},
  \bibinfo{number}{7} (\bibinfo{year}{1976}), \bibinfo{pages}{385--394}.
\newblock


\bibitem[\protect\citeauthoryear{Kroening and Strichman}{Kroening and
  Strichman}{2008}]%
        {decision-procedures-book}
\bibfield{author}{\bibinfo{person}{Daniel Kroening} {and} \bibinfo{person}{Ofer
  Strichman}.} \bibinfo{year}{2008}\natexlab{}.
\newblock \bibinfo{booktitle}{\emph{Decision procedures: an algorithmic point
  of view}}.
\newblock \bibinfo{publisher}{Springer}.
\newblock


\bibitem[\protect\citeauthoryear{Levine}{Levine}{1999}]%
        {linkers-and-loaders-levine1999}
\bibfield{author}{\bibinfo{person}{John~R Levine}.}
  \bibinfo{year}{1999}\natexlab{}.
\newblock \bibinfo{booktitle}{\emph{Linkers \& loaders}}.
\newblock \bibinfo{publisher}{Morgan-Kaufmann}.
\newblock


\bibitem[\protect\citeauthoryear{Liaw, Bhardwaj, Dunlap, Zou, Gonzalez, Stoica,
  and Tumanov}{Liaw et~al\mbox{.}}{2019}]%
        {hypersched-socc19}
\bibfield{author}{\bibinfo{person}{Richard Liaw}, \bibinfo{person}{Romil
  Bhardwaj}, \bibinfo{person}{Lisa Dunlap}, \bibinfo{person}{Yitian Zou},
  \bibinfo{person}{Joseph~E Gonzalez}, \bibinfo{person}{Ion Stoica}, {and}
  \bibinfo{person}{Alexey Tumanov}.} \bibinfo{year}{2019}\natexlab{}.
\newblock \showarticletitle{Hypersched: Dynamic resource reallocation for model
  development on a deadline}. In \bibinfo{booktitle}{\emph{Proceedings of the
  ACM Symposium on Cloud Computing}}. \bibinfo{pages}{61--73}.
\newblock


\bibitem[\protect\citeauthoryear{Lim and Nagarakatte}{Lim and
  Nagarakatte}{2019}]%
        {cryptographic-verif-cgo19}
\bibfield{author}{\bibinfo{person}{Jay~P. Lim} {and} \bibinfo{person}{Santosh
  Nagarakatte}.} \bibinfo{year}{2019}\natexlab{}.
\newblock \showarticletitle{{Automatic Equivalence Checking for Assembly
  Implementations of Cryptography Libraries}}. In
  \bibinfo{booktitle}{\emph{{Proceedings of the 17th International Symposium on
  Code Generation and Optimization}}}. \bibinfo{publisher}{{IEEE}},
  \bibinfo{pages}{37--49}.
\newblock
\showISBNx{978-1-7281-1436-1}
\urldef\tempurl%
\url{https://doi.org/10.1109/CGO.2019.8661180}
\showDOI{\tempurl}


\bibitem[\protect\citeauthoryear{Lopes, Menendez, Nagarakatte, and
  Regehr}{Lopes et~al\mbox{.}}{2015}]%
        {peephole-alive-pldi15}
\bibfield{author}{\bibinfo{person}{Nuno~P Lopes}, \bibinfo{person}{David
  Menendez}, \bibinfo{person}{Santosh Nagarakatte}, {and} \bibinfo{person}{John
  Regehr}.} \bibinfo{year}{2015}\natexlab{}.
\newblock \showarticletitle{Provably correct peephole optimizations with
  alive}. In \bibinfo{booktitle}{\emph{Proceedings of the 36th ACM SIGPLAN
  Conference on Programming Language Design and Implementation}}.
  \bibinfo{pages}{22--32}.
\newblock


\bibitem[\protect\citeauthoryear{Marsden}{Marsden}{2019}]%
        {bpf-program-types}
\bibfield{author}{\bibinfo{person}{Greg Marsden}.}
  \bibinfo{year}{2019}\natexlab{}.
\newblock \bibinfo{title}{BPF: A Tour of Program Types}.
\newblock \bibinfo{howpublished}{[Online, Retrieved Oct 29, 2020.]
  \url{https://blogs.oracle.com/linux/notes-on-bpf-1}}.
\newblock


\bibitem[\protect\citeauthoryear{Marty, de~Kruijf, Adriaens, Alfeld, Bauer,
  Contavalli, Dalton, Dukkipati, Evans, Gribble, et~al\mbox{.}}{Marty
  et~al\mbox{.}}{2019}]%
        {snap-sigcomm19}
\bibfield{author}{\bibinfo{person}{Michael Marty}, \bibinfo{person}{Marc de
  Kruijf}, \bibinfo{person}{Jacob Adriaens}, \bibinfo{person}{Christopher
  Alfeld}, \bibinfo{person}{Sean Bauer}, \bibinfo{person}{Carlo Contavalli},
  \bibinfo{person}{Michael Dalton}, \bibinfo{person}{Nandita Dukkipati},
  \bibinfo{person}{William~C Evans}, \bibinfo{person}{Steve Gribble},
  {et~al\mbox{.}}} \bibinfo{year}{2019}\natexlab{}.
\newblock \showarticletitle{Snap: a microkernel approach to host networking}.
  In \bibinfo{booktitle}{\emph{Proceedings of the 27th ACM Symposium on
  Operating Systems Principles}}. \bibinfo{pages}{399--413}.
\newblock


\bibitem[\protect\citeauthoryear{Massalin}{Massalin}{1987}]%
        {massalin-sigarch87}
\bibfield{author}{\bibinfo{person}{Henry Massalin}.}
  \bibinfo{year}{1987}\natexlab{}.
\newblock \showarticletitle{Superoptimizer: a look at the smallest program}.
\newblock \bibinfo{journal}{\emph{ACM SIGARCH Computer Architecture News}}
  \bibinfo{volume}{15}, \bibinfo{number}{5} (\bibinfo{year}{1987}),
  \bibinfo{pages}{122--126}.
\newblock


\bibitem[\protect\citeauthoryear{McCanne and Jacobson}{McCanne and
  Jacobson}{1993}]%
        {bpf-packet-capture-usenix93}
\bibfield{author}{\bibinfo{person}{Steven McCanne} {and} \bibinfo{person}{Van
  Jacobson}.} \bibinfo{year}{1993}\natexlab{}.
\newblock \showarticletitle{The BSD Packet Filter: A New Architecture for
  User-level Packet Capture.}. In \bibinfo{booktitle}{\emph{USENIX winter}},
  Vol.~\bibinfo{volume}{46}.
\newblock


\bibitem[\protect\citeauthoryear{McKeeman}{McKeeman}{1965}]%
        {peephole-cacm65}
\bibfield{author}{\bibinfo{person}{W.~M. McKeeman}.}
  \bibinfo{year}{1965}\natexlab{}.
\newblock \showarticletitle{Peephole Optimization}.
\newblock \bibinfo{journal}{\emph{Commun. ACM}} \bibinfo{volume}{8},
  \bibinfo{number}{7} (\bibinfo{date}{July} \bibinfo{year}{1965}),
  \bibinfo{pages}{443–444}.
\newblock
\showISSN{0001-0782}
\urldef\tempurl%
\url{https://doi.org/10.1145/364995.365000}
\showDOI{\tempurl}


\bibitem[\protect\citeauthoryear{Menendez and Nagarakatte}{Menendez and
  Nagarakatte}{2017}]%
        {alive-infer-pldi17}
\bibfield{author}{\bibinfo{person}{David Menendez} {and}
  \bibinfo{person}{Santosh Nagarakatte}.} \bibinfo{year}{2017}\natexlab{}.
\newblock \showarticletitle{Alive-infer: Data-driven precondition inference for
  peephole optimizations in llvm}. In \bibinfo{booktitle}{\emph{Proceedings of
  the 38th ACM SIGPLAN Conference on Programming Language Design and
  Implementation}}. \bibinfo{pages}{49--63}.
\newblock


\bibitem[\protect\citeauthoryear{Miano, Sanaee, Risso, Rétvári, and
  Antichi}{Miano et~al\mbox{.}}{2021}]%
        {morpheus-preprint21}
\bibfield{author}{\bibinfo{person}{Sebastiano Miano}, \bibinfo{person}{Alireza
  Sanaee}, \bibinfo{person}{Fulvio Risso}, \bibinfo{person}{Gábor Rétvári},
  {and} \bibinfo{person}{Gianni Antichi}.} \bibinfo{year}{2021}\natexlab{}.
\newblock \bibinfo{title}{Dynamic Recompilation of Software Network Services
  with Morpheus}.
\newblock
\newblock
\showeprint[arxiv]{2106.08833}~[cs.NI]


\bibitem[\protect\citeauthoryear{M\o{}ller and Schwartzbach}{M\o{}ller and
  Schwartzbach}{2018}]%
        {static-analysis-book}
\bibfield{author}{\bibinfo{person}{Anders M\o{}ller} {and}
  \bibinfo{person}{Michael~I. Schwartzbach}.} \bibinfo{year}{2018}\natexlab{}.
\newblock \bibinfo{title}{Static Program Analysis}.
\newblock
\newblock
\newblock
\shownote{Department of Computer Science, Aarhus University,
  \texttt{http://cs.au.dk/\~{}amoeller/spa/}.}


\bibitem[\protect\citeauthoryear{Morrisett, Tan, Tassarotti, Tristan, and
  Gan}{Morrisett et~al\mbox{.}}{2012}]%
        {rocksalt-pldi12}
\bibfield{author}{\bibinfo{person}{Greg Morrisett}, \bibinfo{person}{Gang Tan},
  \bibinfo{person}{Joseph Tassarotti}, \bibinfo{person}{Jean-Baptiste Tristan},
  {and} \bibinfo{person}{Edward Gan}.} \bibinfo{year}{2012}\natexlab{}.
\newblock \showarticletitle{RockSalt: better, faster, stronger SFI for the
  x86}. In \bibinfo{booktitle}{\emph{Proceedings of the 33rd ACM SIGPLAN
  conference on Programming Language Design and Implementation}}.
  \bibinfo{pages}{395--404}.
\newblock


\bibitem[\protect\citeauthoryear{Mukherjee, Kant, Liu, and Regehr}{Mukherjee
  et~al\mbox{.}}{2020}]%
        {dataflow-pruning-oopsla20}
\bibfield{author}{\bibinfo{person}{Manasij Mukherjee}, \bibinfo{person}{Pranav
  Kant}, \bibinfo{person}{Zhengyang Liu}, {and} \bibinfo{person}{John Regehr}.}
  \bibinfo{year}{2020}\natexlab{}.
\newblock \showarticletitle{Dataflow-based Pruning for Speeding up
  Superoptimization}. In \bibinfo{booktitle}{\emph{ACM SIGPLAN Conference on
  Object-Oriented Programming, Systems, Languages, and Applications}}.
\newblock


\bibitem[\protect\citeauthoryear{Nakryiko}{Nakryiko}{2020}]%
        {bpf-portability-core}
\bibfield{author}{\bibinfo{person}{Andrii Nakryiko}.}
  \bibinfo{year}{2020}\natexlab{}.
\newblock \bibinfo{title}{{BPF Portability and CO-RE}}.
\newblock \bibinfo{howpublished}{[Online. Retrieved Jan 20, 2021.]
  \url{https://facebookmicrosites.github.io/bpf/blog/2020/02/19/bpf-portability-and-co-re.html}}.
\newblock


\bibitem[\protect\citeauthoryear{Nelson, Bornholt, Gu, Baumann, Torlak, and
  Wang}{Nelson et~al\mbox{.}}{2019}]%
        {serval-sosp19}
\bibfield{author}{\bibinfo{person}{Luke Nelson}, \bibinfo{person}{James
  Bornholt}, \bibinfo{person}{Ronghui Gu}, \bibinfo{person}{Andrew Baumann},
  \bibinfo{person}{Emina Torlak}, {and} \bibinfo{person}{Xi Wang}.}
  \bibinfo{year}{2019}\natexlab{}.
\newblock \showarticletitle{Scaling symbolic evaluation for automated
  verification of systems code with Serval}. In
  \bibinfo{booktitle}{\emph{Proceedings of the 27th ACM Symposium on Operating
  Systems Principles}}. \bibinfo{pages}{225--242}.
\newblock


\bibitem[\protect\citeauthoryear{Nelson, Van~Geffen, Torlak, and Wang}{Nelson
  et~al\mbox{.}}{2020}]%
        {jitterbug-osdi20}
\bibfield{author}{\bibinfo{person}{Luke Nelson}, \bibinfo{person}{Jacob
  Van~Geffen}, \bibinfo{person}{Emina Torlak}, {and} \bibinfo{person}{Xi
  Wang}.} \bibinfo{year}{2020}\natexlab{}.
\newblock \showarticletitle{Specification and verification in the field:
  Applying formal methods to BPF just-in-time compilers in the Linux kernel}.
  In \bibinfo{booktitle}{\emph{Usenix Operating Systems Design and
  Implementation (OSDI)}}.
\newblock


\bibitem[\protect\citeauthoryear{Peter, Li, Zhang, Ports, Woos, Krishnamurthy,
  Anderson, and Roscoe}{Peter et~al\mbox{.}}{2014}]%
        {arrakis}
\bibfield{author}{\bibinfo{person}{Simon Peter}, \bibinfo{person}{Jialin Li},
  \bibinfo{person}{Irene Zhang}, \bibinfo{person}{Dan R.~K. Ports},
  \bibinfo{person}{Doug Woos}, \bibinfo{person}{Arvind Krishnamurthy},
  \bibinfo{person}{Thomas Anderson}, {and} \bibinfo{person}{Timothy Roscoe}.}
  \bibinfo{year}{2014}\natexlab{}.
\newblock \showarticletitle{{Arrakis: The Operating System is the Control
  Plane}}. In \bibinfo{booktitle}{\emph{{USENIX} Symposium on Operating Systems
  Design and Implementation ({OSDI} 14)}}.
\newblock


\bibitem[\protect\citeauthoryear{Phothilimthana, Thakur, Bodik, and
  Dhurjati}{Phothilimthana et~al\mbox{.}}{2016}]%
        {lens-asplos16}
\bibfield{author}{\bibinfo{person}{Phitchaya~Mangpo Phothilimthana},
  \bibinfo{person}{Aditya Thakur}, \bibinfo{person}{Rastislav Bodik}, {and}
  \bibinfo{person}{Dinakar Dhurjati}.} \bibinfo{year}{2016}\natexlab{}.
\newblock \showarticletitle{Scaling up superoptimization}. In
  \bibinfo{booktitle}{\emph{Proceedings of the Twenty-First International
  Conference on Architectural Support for Programming Languages and Operating
  Systems}}. \bibinfo{pages}{297--310}.
\newblock


\bibitem[\protect\citeauthoryear{Pierce, Casinghino, Gaboardi, Greenberg,
  Hri{\c{t}}cu, Sj{\"o}berg, and Yorgey}{Pierce et~al\mbox{.}}{2010}]%
        {software-foundations-book}
\bibfield{author}{\bibinfo{person}{Benjamin~C Pierce}, \bibinfo{person}{Chris
  Casinghino}, \bibinfo{person}{Marco Gaboardi}, \bibinfo{person}{Michael
  Greenberg}, \bibinfo{person}{C{\u{a}}t{\u{a}}lin Hri{\c{t}}cu},
  \bibinfo{person}{Vilhelm Sj{\"o}berg}, {and} \bibinfo{person}{Brent Yorgey}.}
  \bibinfo{year}{2010}\natexlab{}.
\newblock \bibinfo{title}{Software foundations}.
\newblock \bibinfo{howpublished}{[Online, Retrieved Oct 29, 2020.]
  \url{http://www.cis.upenn.edu/bcpierce/sf/current/index.html}}.
\newblock


\bibitem[\protect\citeauthoryear{Porncharoenwase, Bornholt, and
  Torlak}{Porncharoenwase et~al\mbox{.}}{2020}]%
        {fixing-code-vmcai20}
\bibfield{author}{\bibinfo{person}{Sorawee Porncharoenwase},
  \bibinfo{person}{James Bornholt}, {and} \bibinfo{person}{Emina Torlak}.}
  \bibinfo{year}{2020}\natexlab{}.
\newblock \showarticletitle{Fixing Code that Explodes Under Symbolic
  Evaluation}. In \bibinfo{booktitle}{\emph{International Conference on
  Verification, Model Checking, and Abstract Interpretation}}. Springer,
  \bibinfo{pages}{44--67}.
\newblock


\bibitem[\protect\citeauthoryear{{Qiongwen Xu, Michael D. Wong, Tanvi Wagle,
  Srinivas Narayana, Anirudh Sivaraman}}{{Qiongwen Xu, Michael D. Wong, Tanvi
  Wagle, Srinivas Narayana, Anirudh Sivaraman}}{2021}]%
        {project-web-page}
\bibfield{author}{\bibinfo{person}{{Qiongwen Xu, Michael D. Wong, Tanvi Wagle,
  Srinivas Narayana, Anirudh Sivaraman}}.} \bibinfo{year}{2021}\natexlab{}.
\newblock \bibinfo{title}{The K2 compiler}.
\newblock \bibinfo{howpublished}{[Online, Retrieved Jun 30, 2021.]
  \url{https://k2.cs.rutgers.edu}}.
\newblock


\bibitem[\protect\citeauthoryear{{Quentin Monnet}}{{Quentin Monnet}}{2020}]%
        {netronome-smaller-programs-greater-performance}
\bibfield{author}{\bibinfo{person}{{Quentin Monnet}}.}
  \bibinfo{year}{2020}\natexlab{}.
\newblock \bibinfo{title}{{Optimizing BPF: Smaller Programs for Greater
  Performance}}.
\newblock \bibinfo{howpublished}{[Online. Retrieved Jan 20, 2021.]
  \url{https://www.netronome.com/blog/optimizing-bpf-smaller-programs-greater-performance/}}.
\newblock


\bibitem[\protect\citeauthoryear{Rizzo}{Rizzo}{2012}]%
        {netmap}
\bibfield{author}{\bibinfo{person}{Luigi Rizzo}.}
  \bibinfo{year}{2012}\natexlab{}.
\newblock \showarticletitle{netmap: A Novel Framework for Fast Packet I/O}. In
  \bibinfo{booktitle}{\emph{2012 {USENIX} Annual Technical Conference ({USENIX}
  {ATC} 12)}}.
\newblock


\bibitem[\protect\citeauthoryear{Sadok, Zhao, Choung, Atre, Berger, Hoe, Panda,
  and Sherry}{Sadok et~al\mbox{.}}{2021}]%
        {kopi-hotos21}
\bibfield{author}{\bibinfo{person}{Hugo Sadok}, \bibinfo{person}{Zhipeng Zhao},
  \bibinfo{person}{Valerie Choung}, \bibinfo{person}{Nirav Atre},
  \bibinfo{person}{Daniel~S Berger}, \bibinfo{person}{James~C Hoe},
  \bibinfo{person}{Aurojit Panda}, {and} \bibinfo{person}{Justine Sherry}.}
  \bibinfo{year}{2021}\natexlab{}.
\newblock \showarticletitle{We need kernel interposition over the network
  dataplane}. In \bibinfo{booktitle}{\emph{Proceedings of the Workshop on Hot
  Topics in Operating Systems}}. \bibinfo{pages}{152--158}.
\newblock


\bibitem[\protect\citeauthoryear{Saha, Prabhu, and Madhusudan}{Saha
  et~al\mbox{.}}{2015}]%
        {netgen-sosr15}
\bibfield{author}{\bibinfo{person}{Shambwaditya Saha},
  \bibinfo{person}{Santhosh Prabhu}, {and} \bibinfo{person}{P Madhusudan}.}
  \bibinfo{year}{2015}\natexlab{}.
\newblock \showarticletitle{NetGen: Synthesizing data-plane configurations for
  network policies}. In \bibinfo{booktitle}{\emph{Proceedings of the 1st ACM
  SIGCOMM Symposium on Software Defined Networking Research}}.
  \bibinfo{pages}{1--6}.
\newblock


\bibitem[\protect\citeauthoryear{Salim, Olsson, and Kuznetsov}{Salim
  et~al\mbox{.}}{2001}]%
        {napi-linuxconf01}
\bibfield{author}{\bibinfo{person}{Jamal~Hadi Salim}, \bibinfo{person}{Robert
  Olsson}, {and} \bibinfo{person}{Alexey Kuznetsov}.}
  \bibinfo{year}{2001}\natexlab{}.
\newblock \showarticletitle{Beyond Softnet.}. In
  \bibinfo{booktitle}{\emph{Annual Linux Showcase \& Conference}},
  Vol.~\bibinfo{volume}{5}. \bibinfo{pages}{18--18}.
\newblock


\bibitem[\protect\citeauthoryear{Sasnauskas, Chen, Collingbourne, Ketema,
  Taneja, and Regehr}{Sasnauskas et~al\mbox{.}}{2017}]%
        {souper17}
\bibfield{author}{\bibinfo{person}{Raimondas Sasnauskas}, \bibinfo{person}{Yang
  Chen}, \bibinfo{person}{Peter Collingbourne}, \bibinfo{person}{Jeroen
  Ketema}, \bibinfo{person}{Jubi Taneja}, {and} \bibinfo{person}{John Regehr}.}
  \bibinfo{year}{2017}\natexlab{}.
\newblock \showarticletitle{Souper: {A} Synthesizing Superoptimizer}.
\newblock \bibinfo{journal}{\emph{CoRR}}  \bibinfo{volume}{abs/1711.04422}
  (\bibinfo{year}{2017}).
\newblock
\showeprint[arxiv]{1711.04422}
\urldef\tempurl%
\url{http://arxiv.org/abs/1711.04422}
\showURL{%
\tempurl}


\bibitem[\protect\citeauthoryear{Schkufza, Sharma, and Aiken}{Schkufza
  et~al\mbox{.}}{2013}]%
        {stoke-asplos13}
\bibfield{author}{\bibinfo{person}{Eric Schkufza}, \bibinfo{person}{Rahul
  Sharma}, {and} \bibinfo{person}{Alex Aiken}.}
  \bibinfo{year}{2013}\natexlab{}.
\newblock \showarticletitle{Stochastic superoptimization}.
\newblock \bibinfo{journal}{\emph{ACM SIGARCH Computer Architecture News}}
  \bibinfo{volume}{41}, \bibinfo{number}{1} (\bibinfo{year}{2013}),
  \bibinfo{pages}{305--316}.
\newblock


\bibitem[\protect\citeauthoryear{Sharma, Schkufza, Churchill, and Aiken}{Sharma
  et~al\mbox{.}}{2013}]%
        {ddec-oopsla13}
\bibfield{author}{\bibinfo{person}{Rahul Sharma}, \bibinfo{person}{Eric
  Schkufza}, \bibinfo{person}{Berkeley Churchill}, {and} \bibinfo{person}{Alex
  Aiken}.} \bibinfo{year}{2013}\natexlab{}.
\newblock \showarticletitle{Data-driven equivalence checking}. In
  \bibinfo{booktitle}{\emph{Proceedings of the 2013 ACM SIGPLAN international
  conference on Object oriented programming systems languages \&
  applications}}. \bibinfo{pages}{391--406}.
\newblock


\bibitem[\protect\citeauthoryear{Sharma, Schkufza, Churchill, and Aiken}{Sharma
  et~al\mbox{.}}{2015}]%
        {conditionally-correct-superopt-pldi15}
\bibfield{author}{\bibinfo{person}{Rahul Sharma}, \bibinfo{person}{Eric
  Schkufza}, \bibinfo{person}{Berkeley Churchill}, {and} \bibinfo{person}{Alex
  Aiken}.} \bibinfo{year}{2015}\natexlab{}.
\newblock \showarticletitle{Conditionally correct superoptimization}.
\newblock \bibinfo{journal}{\emph{ACM SIGPLAN Notices}} \bibinfo{volume}{50},
  \bibinfo{number}{10} (\bibinfo{year}{2015}), \bibinfo{pages}{147--162}.
\newblock


\bibitem[\protect\citeauthoryear{Shirokov}{Shirokov}{2018}]%
        {katran-facebook-talk}
\bibfield{author}{\bibinfo{person}{Nikita~V. Shirokov}.}
  \bibinfo{year}{2018}\natexlab{}.
\newblock \showarticletitle{{XDP: 1.5 years in production. Evolution and
  lessons learned.}}.
  \bibinfo{howpublished}{\url{http://vger.kernel.org/lpc_net2018_talks/LPC_XDP_Shirokov_v2.pdf}}.
  In \bibinfo{booktitle}{\emph{{Linux Plumbers Conference}}}.
\newblock


\bibitem[\protect\citeauthoryear{Solar-Lezama, Rabbah, Bod{\'\i}k, and
  Ebcio{\u{g}}lu}{Solar-Lezama et~al\mbox{.}}{2005}]%
        {sketch-pldi05}
\bibfield{author}{\bibinfo{person}{Armando Solar-Lezama},
  \bibinfo{person}{Rodric Rabbah}, \bibinfo{person}{Rastislav Bod{\'\i}k},
  {and} \bibinfo{person}{Kemal Ebcio{\u{g}}lu}.}
  \bibinfo{year}{2005}\natexlab{}.
\newblock \showarticletitle{Programming by sketching for bit-streaming
  programs}. In \bibinfo{booktitle}{\emph{Proceedings of the 2005 ACM SIGPLAN
  conference on Programming language design and implementation}}.
  \bibinfo{pages}{281--294}.
\newblock


\bibitem[\protect\citeauthoryear{Solar-Lezama, Tancau, Bodik, Seshia, and
  Saraswat}{Solar-Lezama et~al\mbox{.}}{2006}]%
        {sketch-asplos06}
\bibfield{author}{\bibinfo{person}{Armando Solar-Lezama},
  \bibinfo{person}{Liviu Tancau}, \bibinfo{person}{Rastislav Bodik},
  \bibinfo{person}{Sanjit Seshia}, {and} \bibinfo{person}{Vijay Saraswat}.}
  \bibinfo{year}{2006}\natexlab{}.
\newblock \showarticletitle{Combinatorial sketching for finite programs}. In
  \bibinfo{booktitle}{\emph{Proceedings of the 12th international conference on
  Architectural support for programming languages and operating systems}}.
  \bibinfo{pages}{404--415}.
\newblock


\bibitem[\protect\citeauthoryear{Subramanian, D'Antoni, and Akella}{Subramanian
  et~al\mbox{.}}{2017}]%
        {genesis-popl17}
\bibfield{author}{\bibinfo{person}{Kausik Subramanian}, \bibinfo{person}{Loris
  D'Antoni}, {and} \bibinfo{person}{Aditya Akella}.}
  \bibinfo{year}{2017}\natexlab{}.
\newblock \showarticletitle{Genesis: Synthesizing forwarding tables in
  multi-tenant networks}. In \bibinfo{booktitle}{\emph{Proceedings of the 44th
  ACM SIGPLAN Symposium on Principles of Programming Languages}}.
  \bibinfo{pages}{572--585}.
\newblock


\bibitem[\protect\citeauthoryear{Subramanian, D'Antoni, and Akella}{Subramanian
  et~al\mbox{.}}{2018}]%
        {zeppelin-routersynthesis18}
\bibfield{author}{\bibinfo{person}{Kausik Subramanian}, \bibinfo{person}{Loris
  D'Antoni}, {and} \bibinfo{person}{Aditya Akella}.}
  \bibinfo{year}{2018}\natexlab{}.
\newblock \showarticletitle{Synthesis of Fault-Tolerant Distributed Router
  Configurations}.
\newblock \bibinfo{journal}{\emph{Proc. ACM Meas. Anal. Comput. Syst.}}
  \bibinfo{volume}{2}, \bibinfo{number}{1}, Article \bibinfo{articleno}{22}
  (\bibinfo{date}{April} \bibinfo{year}{2018}), \bibinfo{numpages}{26}~pages.
\newblock
\urldef\tempurl%
\url{https://doi.org/10.1145/3179425}
\showDOI{\tempurl}


\bibitem[\protect\citeauthoryear{Tate, Stepp, Tatlock, and Lerner}{Tate
  et~al\mbox{.}}{2009}]%
        {equalitysaturation-pldi09}
\bibfield{author}{\bibinfo{person}{Ross Tate}, \bibinfo{person}{Michael Stepp},
  \bibinfo{person}{Zachary Tatlock}, {and} \bibinfo{person}{Sorin Lerner}.}
  \bibinfo{year}{2009}\natexlab{}.
\newblock \showarticletitle{Equality saturation: a new approach to
  optimization}. In \bibinfo{booktitle}{\emph{Proceedings of the 36th annual
  ACM SIGPLAN-SIGACT symposium on Principles of programming languages}}.
  \bibinfo{pages}{264--276}.
\newblock


\bibitem[\protect\citeauthoryear{Torlak and Bodik}{Torlak and Bodik}{2013}]%
        {rosette-onward13}
\bibfield{author}{\bibinfo{person}{Emina Torlak} {and}
  \bibinfo{person}{Rastislav Bodik}.} \bibinfo{year}{2013}\natexlab{}.
\newblock \showarticletitle{Growing solver-aided languages with Rosette}. In
  \bibinfo{booktitle}{\emph{Proceedings of the 2013 ACM international symposium
  on New ideas, new paradigms, and reflections on programming \& software}}.
  \bibinfo{pages}{135--152}.
\newblock


\bibitem[\protect\citeauthoryear{Van~Geffen, Nelson, Dillig, Wang, and
  Torlak}{Van~Geffen et~al\mbox{.}}{2020}]%
        {jitsynth-cav20}
\bibfield{author}{\bibinfo{person}{Jacob Van~Geffen}, \bibinfo{person}{Luke
  Nelson}, \bibinfo{person}{Isil Dillig}, \bibinfo{person}{Xi Wang}, {and}
  \bibinfo{person}{Emina Torlak}.} \bibinfo{year}{2020}\natexlab{}.
\newblock \showarticletitle{Synthesizing JIT Compilers for In-Kernel DSLs}. In
  \bibinfo{booktitle}{\emph{International Conference on Computer Aided
  Verification}}. Springer, \bibinfo{pages}{564--586}.
\newblock


\bibitem[\protect\citeauthoryear{Wang, Lazar, Zeldovich, Chlipala, and
  Tatlock}{Wang et~al\mbox{.}}{2014}]%
        {jitk-osdi14}
\bibfield{author}{\bibinfo{person}{Xi Wang}, \bibinfo{person}{David Lazar},
  \bibinfo{person}{Nickolai Zeldovich}, \bibinfo{person}{Adam Chlipala}, {and}
  \bibinfo{person}{Zachary Tatlock}.} \bibinfo{year}{2014}\natexlab{}.
\newblock \showarticletitle{Jitk: A trustworthy in-kernel interpreter
  infrastructure}. In \bibinfo{booktitle}{\emph{11th $\{$USENIX$\}$ Symposium
  on Operating Systems Design and Implementation ($\{$OSDI$\}$ 14)}}.
  \bibinfo{pages}{33--47}.
\newblock


\bibitem[\protect\citeauthoryear{Yang, Phothilimtha, Wang, Willsey, Roy, and
  Pienaar}{Yang et~al\mbox{.}}{2021}]%
        {equality-saturation-tensor-superopt-arxiv21}
\bibfield{author}{\bibinfo{person}{Yichen Yang},
  \bibinfo{person}{Phitchaya~Mangpo Phothilimtha}, \bibinfo{person}{Yisu~Remy
  Wang}, \bibinfo{person}{Max Willsey}, \bibinfo{person}{Sudip Roy}, {and}
  \bibinfo{person}{Jacques Pienaar}.} \bibinfo{year}{2021}\natexlab{}.
\newblock \bibinfo{title}{Equality Saturation for Tensor Graph
  Superoptimization}.
\newblock
\newblock
\showeprint[arxiv]{2101.01332}~[cs.AI]


\end{thebibliography}
\balance

\clearpage
\appendix
Appendices are supporting material that has not been peer-reviewed.
\section{Approaches to Program Synthesis}
\label{app:program-synthesis-background}

In this paper, we are given a sequence of instructions in a fixed instruction
set, \ie a BPF bytecode source program. We are interested in generating an
alternative sequence of instructions, \ie a synthesized program, that satisfies
the specification that (i) the synthesized program is equivalent to the source
program in its input-output behaviors, (ii) the synthesized program is safe, and
(iii) the synthesized program is more efficient than the source program. The
precise meanings of efficiency and safety in the BPF context are described in
\Sec{compiler-overview} and \Sec{safety}.

To simplify the following discussion, suppose the program specification is
simply (i) above, \ie $spec := p_{synth}(x) == p_{src}(x)$
for source program $p_{src}$ and synthesized program
$p_{synth}$ for all program inputs $x$. At a
high level, the program synthesis problem we are interested in can be formulated
as the logical query
\begin{equation}
  \label{eq:general-synthesis}
  \exists p . \forall x . p(x)\ ==\ p_{src}(x)
\end{equation}
where $p$ is any program composed of instructions from the BPF instruction
set. As written down, this problem is a quantified boolean
formula (QBF) with alternating quantifiers, which
does not permit efficient decision procedures for problems that arise in
synthesis~\cite{sketch-pldi05}. Hence, program synthesizers take the approach of
{\em counterexample guided inductive synthesis} (CEGIS)~\cite{sketch-pldi05,
  sketch-asplos06}. First, a candidate program $p_{cand}$ is determined through a search
procedure. Then, the synthesizer checks whether the candidate satisfies the
specification, by asking
\begin{equation}
  \label{eq:cegis-synthesis}
  \exists x . p_{cand}(x)\ !=\ p_{src}(x)
\end{equation}
for the {\em fixed} program $p_{cand}$.

Typically, synthesis algorithms use test cases to
quickly prune candidates that do not satisfy the specification.
If test cases do not eliminate $p_{cand}$ as a candidate satisfying the specification,
the query above can usually be formulated in a first-order logic
theory~\cite{calculus-of-computation-book} which permits efficient decision
procedures. The query above is an example of {\em equivalence-checking}, which
determines whether two programs produce the same output on all inputs. If the
query (\ref{eq:cegis-synthesis}) is satisfiable, we get a {\em counterexample},
which can be added to the set of test cases to prune 
the same or similar programs in the future without
discharging computationally-expensive logic queries. If
(\ref{eq:cegis-synthesis}) is unsatisfiable, we have found a
program $p_{cand}$ that produces the same output as
$p_{src}$ on all inputs, and hence meets the specification.

The synthesis approaches in the literature differ mainly in the search
procedures they use to propose candidate programs $p$.  There are broadly four
search approaches. Enumerative search (\eg~\cite{massalin-sigarch87,
  lens-asplos16, peephole-asplos06}) searches the space in order from smallest
to largest programs, terminating the search with the smallest program that
satisfies the specification. Rule-based search~\cite{denali-pldi02,
  equalitysaturation-pldi09, equality-saturation-tensor-superopt-arxiv21} uses
targeted rewrite rules to transform the source program into another program that
satisfies the specification yet is more optimal.  Deductive
search~\cite{sygus-cacm18, loopfree-pldi11, souper17, dataflow-pruning-oopsla20,
  dryadsynth, oracle-component-synthesis-icse10} encodes the search into a
deductive query discharged to a solver whose solution implies a program within a
specified grammar.  Stochastic search~\cite{stoke-asplos13, loopsuperopt-pldi17}
searches the space guided by cost functions that enable sampling new random
programs.  Further, there are search algorithms that combine the best of
multiple approaches~\cite{lens-asplos16, dryadsynth-pldi20}.

Among the synthesis approaches above, in our view,
stochastic synthesis is the easiest to generalize to new and
diverse contexts, due to its ability to support very
expressive cost functions and constraints. We adopt and
adapt stochastic synthesis in this paper. 

\section{Verification Conditions for BPF Maps and Helper Functions}
\label{app:map-formalization}

BPF maps are a special kind of memory. Like memory, a BPF program could read a map with a specific key ({\em look-up}) and write a value corresponding to an existing key ({\em update}). However, three aspects make BPF maps very different from memory from the formalization perspective.  First, the BPF map API~\cite{bpf-helper-calls} requires input keys in the form of {\em pointers} to memory, and further, returns values that are themselves pointers to memory. Second, keys can be {\em  deleted} in a BPF map, unlike addresses in memory. Third, the keys and values in BPF maps are {\em persistent} structures that exist before and after program execution. Together, these aspects of BPF maps prevent us from directly applying the same formalization methods as memory access (\Sec{memory-instruction-formalization}) or other existing axiomatizations (\eg partial maps~\cite{software-foundations-book}) to model BPF maps in first-order logic.  None of the prior works on formalizing BPF~\cite{untrusted-extensions-pldi19, serval-sosp19, jitterbug-osdi20, jitsynth-cav20} handle any aspect of BPF maps.

In the rest of this subsection, we show how our compiler handles pointer inputs/outputs as well as deletions. We show how we address the last aspect (map persistence) while performing program equivalence checking (\Sec{program-equivalence-checking}).

\subsection{Handling pointer access to map memory}

Supplying an input key to a map operation (\eg look-up) as a pointer creates {\em two} levels of aliasing behaviors: First, two distinct registers may point to the same memory address (same as regular pointer aliasing). Second, a map needs to return the same value given the same (value of) key, {\em even if the keys are supplied from distinct memory addresses.} 

We formalize map access in bit vector theory by decomposing the two levels of aliasing as follows. First, we create a new symbolic variable for each key supplied to the map API and each pointer value it returns.  We call these variables the {\em valuations} of the (input) key pointer and the (output) value pointer. We apply the formulas in \Sec{memory-instruction-formalization} to the key pointer, where the data that is read is the key's valuation.

To handle aliasing among the valuations of the keys themselves, we write down formulas analogous to memory aliasing (\Sec{memory-instruction-formalization}) over the valuations of the key and the value. This entails maintaining {\em map write and read tables} analogous to memory read and write tables.

Addressing aliasing among key valuations has the added benefit of encoding the partial map axioms~\cite{software-foundations-book}, \eg a look-up of a key following an update of that same key returns the value from that update. No guarantees are made over the returned value {\em pointer} itself (this is a reference to the kernel's memory) except that it is non-zero.

\Para{Handling map manipulations through memory-access instructions.}  BPF map
values can be manipulated directly through memory-access instructions, \eg to
increment a counter stored as a map value using a {\ct bpf\_xadd} instruction,
which has the impact of performing {\ct *ptr = *ptr + reg}.  \TheCompiler uses
the static analysis described in \Sec{optimizing-equivalence-checking} to
determine the memory region corresponding to the pointer being loaded or stored.
The memory read and write tables (\Sec{memory-instruction-formalization}) ensure
that \TheCompiler puts down the right formulas for subsequent operations over
these map value memory regions.

\subsection{Handling map deletions}

Keys in a BPF map can be deleted, unlike memory locations. If a deleted key is subsequently looked up, the result is a null pointer. We model this by the simple trick of setting the value address corresponding to the valuation of a deleted key to {\ct 0} in the map write table. When another key with the same valuation is looked up, the returned value address is {\ct 0}, indicating to the program that the key does not exist in the map. Luckily, this return value coincides with BPF's semantics for the return value of map lookup, which is {\ct 0} whenever the key does not exist in the map. We also handle setting the right return value for the BPF map delete API call, which provides distinct return values depending on whether the key currently exists in the map. 

\subsection{Handling multiple maps}

The discussion so far assumes the existence of a single map with all operations
over this map. BPF programs may look-up several maps in one program by loading
map descriptors. However, the set of all possible maps accessed by the program
is known at compile time. Hence, we can handle the general case by prefixing the
precondition {\ct map\_id == K} as the head of an implication, with the body of
the implication being the formula generated using the methods described above. 

\subsection{Limitation of map model}

The current modeling of maps ignores concurrency among
threads (kernel or user) accessing the same
map. \TheCompiler does not make formal guarantees about the
equivalence of its output programs to input programs when
running in multi-threaded contexts where data races are
permissible. However, BPF provides mutual exclusion for map
contents through BPF
spinlocks~\cite{bpf-concurrency-management}, and for code
that enforces mutual exclusion to map access, the semantics
of \TheCompiler's output programs are indeed equivalent to
those of the input program under concurrent access.

\subsection{Other kernel helper functions}
\label{sec:helper-call-formalization}
In addition to map helper functions, we also modeled a number of other BPF kernel helper functions, including functions to get random numbers, obtain a Unix timestamp, adjust the headroom in a memory buffer containing a packet, get the processor ID on which the program is executing, among others. The list of all BPF helper functions numbers over a hundred (and growing~\cite{bpf-helper-calls}). Out of engineering expediency, we only formalized the actions of helper calls as we needed them, rather than support all of them. 

\section{Details on Optimizations to Equivalence-Checking}
\label{app:optimizing-equivalence-checking}

The techniques described in \Sec{formalization} are sufficient to produce a
working version of a BPF bytecode compiler. However, we found that the time to
compile a single program of even modest size ($\sim$100 BPF instructions) is
intractably large. The underlying reason is the significant time required to
perform equivalence checking, which runs to several hours on even just a single
pair of BPF programs on a modern server
(\Sec{benefits-optimizations-equivalence-checking}).

This section presents domain-specific optimizations that bring down this
verification time by several orders of magnitude to a few milliseconds. We show
the impact of each optimization in \Sec{evaluation}. Several features of the BPF
instruction set and packet-processing programs contribute to increased
equivalence checking time: aliasing in memory and map access, multiple types of
memories (stack, packet, metadata, and so on), usage of multiple maps, encoding
the semantics of helper functions, and the existence of control flow within
programs. \TheCompiler incorporates two broad approaches to reduce verification
time: {\em concretization} of symbolic terms in the formula
(\Sec{best-effort-concretization}) and {\em modular verification} of smaller
parts of the program (\Sec{modular-verification}) to verify the larger formula
from \Sec{formalization}.

\subsection{Concretizations of Symbolic Terms}
\label{sec:best-effort-concretization}

It is well known that an explosion in the size and solving difficulty of a
formula under symbolic evaluation can be mitigated by assigning specific values
to, \ie {\em concretizing,} terms in the
formula~\cite{bounded-model-checking-biere99, symbolic-exec-clarke76,
  symbolic-exec-king76, finding-code-oopsla18, fixing-code-vmcai20}.

We leverage the unique characteristics and simplicity of packet-processing BPF
programs, as well as the feasibility of constructing simple static analyses (\eg
sound inference of types) within the \TheCompiler compiler, to infer concrete
values of several terms within the equivalence-checking formulas dispatched to
the solver. These concretizations are ``best effort'' in the sense that we do
not need the analysis within \TheCompiler to be complete: we apply the
simplifications where possible, falling back to the general versions in
\Sec{formalization} where they are not applicable. We eschewed complicated alias
relationship mining techniques~\cite{cryptographic-verif-cgo19,
  loopsuperopt-pldi17} in favor of simpler domain-specific ones.

To simplify the discussion, we describe our techniques for straight-line
programs (no branches) first, and then show how they generalize to loop-free
programs with control flow.

\Para{I. Memory type concretization.}  BPF programs use multiple memory regions:
the stack, packet, maps, and various kernel data structures like sockets.
The handling of pointer-based aliasing of memory access discussed in
\Sec{memory-instruction-formalization} uses a single write table and read table
for all of memory.

Instead, \TheCompiler leverages multiple read/write tables, one corresponding to
each distinct memory region. This is feasible to do since any reference to a
specific memory originates from pre-defined inputs to the BPF
program~\cite{bpf-kernel-documentation}, such as the {\tt R10} register (stack),
an input register like R1 (packet or packet metadata), or return values from
specific helper calls (map memory, kernel socket structure, \etc). The type of
each pointer in a program can then be easily inferred through a simple
information-flow static analysis. For example, given a code sequence {\ct
  bpf\_mov r1 r10; bpf\_mov r2 r1}, it is straightforward to infer that {\ct r2}
is pointing into the stack memory. \TheCompiler tracks types for all registers
across each instruction in this manner.

The benefit of maintaining separate tables for separate memories is that the
size of aliasing formulas in a memory read reduces from $O(N_{all\ mem\ types}^2)$
to $\sum_{t \in mem\ types} O(N_{t}^2)$, where $N_{(.)}$ refers to the number of
accesses to memory of a specific type (or all types).

\Para{II. Memory offset concretization.}  The exact memory address contained in
a register during a BPF program's execution is in general hard to determine at
compile time, since the program stack or the packet contents may be located
anywhere within the kernel's memory. However, in many packet-processing
programs, the {\em offsets} relative to the base address of the memory region,
\eg such as the offset (from the beginning of the packet) of a header field, are
known at program compile time. Further, BPF stack locations can only be read
after they are written, and it is feasible to track which concrete offsets are
being written to in the first place. \TheCompiler attempts to maintain a
concrete offset into the memory region for each register known to be a
pointer. Each subsequent operation on the register is associated with the
corresponding operation on the concrete offset. For example, the instruction
{\ct bpf\_mov r1 r10; bpf\_sub r1 2 /* r1 := r10 - 2 */} associates the offset
{\ct -2} with the register {\ct r1.} A subsequent operation {\ct bpf\_mov r2 r1;
  bpf\_sub r2 4} would also update the corresponding offset of {\ct r2} to {\ct
  -6}. (Both registers have a STACK memory type.)

Offset concretization has the effect of turning a clause like {\ct addr\_i ==
  addr\_j} (\Sec{memory-instruction-formalization}) into a clause like {\ct
  offset\_i == offset\_j}, where {\ct offset\_i} and {\ct offset\_j} are
concrete values, \eg -2 == -6. When a load instruction results in comparing a
register offset to other offsets stored in the write table, any comparison of a
concrete pair of offsets can be evaluated at compile time without even a solver
call. If all pairs of offsets compared are concrete, this can simplify an entire
memory read into a single clause of the form {\ct value\_i == value\_j}. Even if
only one of the offsets being compared is known concretely, it simplifies the
clause overall.

In the limit, this optimization has the effect of turning the entire memory into
nameable locations akin to registers. Similar ideas have been applied in prior
efforts to scale verification time~\cite{peephole-asplos06,
  stoke-asplos13}. \TheCompiler applies this idea in a best-effort manner,
falling back to general aliasing when concrete offsets cannot be soundly
inferred.

\Para{III. Map concretization.} BPF programs access several maps and frequently
perform operations such as lookups, which are among the most expensive BPF
operations in terms of verification. Map lookups generate large verification
conditions because of two levels of aliasing
(\Sec{map-formalization}). Accessing multiple maps in a single BPF program
further explodes the formula sizes, since one set of clauses is generated for
each possible map the program may have accessed.  Similar to the idea of
concretizing memory types for each access, \TheCompiler statically determines
the map to which a given lookup or update occurs. This is feasible due to a
property of BPF's map-related instructions: there is a unique opcode {\ct
  LD\_MAP\_ID} which loads a unique and concrete {\ct map\_id} (obtained from a
file descriptor), which is then supplied as a parameter to each BPF map call.
\qwx{add the case of using map-in-map map to get the map id or just regard it 
as a future work?}

We did not generalize our approach to concretize the keys that are looked up or
values that are presented to the map update function call. The BPF map API
(\Sec{background}) mandates map function parameters and return values to be {\em
  pointers} to keys and values, rather than registers containing the keys and
values themselves.
Map keys are pointers to the stack, and the keys on the stack are often from 
input memory regions, such as packet fields. As
noted above (cf. memory offset concretization), we concretize memory addresses
that are accessed, but not the bits stored at concrete addresses in memory. If
the keys are symbolic, then the values are symbolic, too. We leave leveraging
opportunities to concretize map keys and values to future work.

\Para{Incorporating control flow into optimizations.}  So far, the discussion of
the static analysis in \TheCompiler to determine concrete terms has focused on
straight-line code. In the presence of branches, this analysis is generalized as
follows. First, statically known information (\eg pointer type, offset) is
annotated with the basic block and path condition under which it is
determined. Next, for a subsequent read, to determine the concrete information
(\eg offset) to apply, we use the following procedure iteratively for each prior
write to the pointer, starting from the latest to the earliest in order of
program execution:
\begin{itemize}
\item If the last write is from a basic block that {\em
  dominates}~\cite{ssa-toplas91} the basic block of the reading instruction, we
  use the corresponding entry's concrete information and stop. SSA dominance
  analysis is performed just once per program.
\item If the last write is from a basic block from which control never {\em
  reaches} the basic block of the reading instruction, we skip the corresponding
  entry's concrete information and move to the next entry. Reachability analysis
  is performed just once per program.
\item If neither of the above is true, there is a path through the basic block
  of the write to the current instruction, but there are also other
  paths. \TheCompiler conjoins the clause {\ct path condition of the entry
    $\Rightarrow$ read offset == entry offset}. The process continues to the
  next entry.
\end{itemize}

For example, after processing the instruction sequence
\begin{lstlisting}
jneq r5 r6 1 /* if r5 == r6 */
b1: bpf_mov r1 r10 -2 /* r1 = r10 - 2 */
b2: bpf_mov r1 r10 -4 /* else r1 = r10 - 4 */
b3: bpf_load_32 r3 r1 /* r3 = (uint32) *r1 */
\end{lstlisting}

the compiler stores the following information:

\begin{lstlisting}
(block b1, cond r5 == r6, r1 STACK, offset -2)
(block b2, cond !(r5 == r6), r1 STACK, offset -4)
\end{lstlisting}

To determine the offset read by block b3 ({\ct r3 = *r1}), the compiler
determines that neither b1 nor b2 dominates b3, but they can both reach b3,
hence producing the verification conditions

\begin{lstlisting}
(r5 == r6) (*$\Rightarrow$*) read_offset == -2 (*$\land$*)
 ! (r5 == r6) (*$\Rightarrow$*) read_offset == -4
\end{lstlisting}

It is not always possible to make the {\ct read\_offset} concrete, as seen in
this example.

\Para{IV. Caching} We also cache the verification outcomes of {\em
  canonicalized} versions of a given program to quickly determine if a
structurally similar program was equivalence-checked earlier.
We canonicalize the program by removing dead code. Then, the canonicalized
program is hashed into a key that is used to look-up a program cache of
verification outcomes. Equivalence-checking is only performed when there is a
cache miss; the outcome of the equivalence checking is inserted into the cache
along with the program that was checked.

\subsection{Modular Verification}
\label{sec:modular-verification}

Synthesis and optimization typically scale to large programs by operating over a
sequence of smaller {\em windows}, where a window is a contiguous sequence of
instructions as they appear in the text of the program~\cite{lens-asplos16,
  peephole-alive-pldi15, alive-infer-pldi17,
  conditionally-correct-superopt-pldi15}. Small windows (rather than the full
programs) lead to smaller verification conditions, which are much faster to
solve.

Verifying the correctness of synthesized windows be thought of in one of two
ways. A peephole optimizer~\cite{peephole-cacm65, peephole-asplos06} would only
produce instruction sequences that exhibit equivalent behavior under {\em any}
values that the window consumes (precondition) and ensure {\em any} values it
produces are equal to that produced by the original program
(postcondition). This is similar to the verification conditions from
\Sec{program-equivalence-checking}, written just for a window in the program.

An alternative possibility is to consider stronger preconditions and weaker
postconditions to verify the window~\cite{lens-asplos16,
  conditionally-correct-superopt-pldi15}. This enables stronger optimizations:
for example, the instruction {\ct bpf\_mul r2 r1} can be optimized into {\ct
  bpf\_lshift r2 2} {\em provided} a compiler can infer that {\ct r1 == 4}
during the multiplication, but not in general. Similarly, other optimizations
become possible with weaker postconditions.

We call the part of the program before the window the {\em prefix} and the part
after the window the {\em postfix.} Automatically inferring strong preconditions
(that hold after the prefix) and weak postconditions (that must hold before the
postfix) is a challenging problem~\cite{alive-infer-pldi17, peephole-alive-pldi15,
  conditionally-correct-superopt-pldi15}. Recent work~\cite{lens-asplos16} takes
the approach of using the strongest possible precondition and weakest possible
postcondition, by constructing the entire first-order
logical representation of the prefix and
postfix programs. This approach, while enabling strong optimizations, reduces to
using the verification conditions for the full program.

Instead, we build on an earlier approach~\cite{peephole-asplos06} that uses the
{\em live variables}~\cite{static-analysis-book} to generate stronger
preconditions and weaker postconditions than peephole
optimizers. Additionally, \TheCompiler also infers sets of
concrete values for variables, if they exist, and pose those
as preconditions for the optimization.
\TheCompiler chooses windows among basic blocks of a certain maximum instruction
count and uses the following window-based verification condition:

\begin{lstlisting}
variables live into window 1
== variables live into window 2
(* $\land$ *) inferred concrete valuations of variables
(* $\land$ *) input-output behavior of window 1 
(* $\land$ *) input-output behavior of window 2
(* $\Rightarrow$ *) variables live out of window 1
    != variables live out of window 2
\end{lstlisting}

Here, variable set ``live into'' the window contains all
variables written in the prefix that are readable in the
window, and the set ``live out of'' the window contains all
variables that are written inside the window {\em and} read
by the postfix. The concrete valuations of variables are
inferred through static analysis that determines the set of
concrete values that a register might take on, along with
the corresponding path conditions. For example, in the
code sequence
\begin{lstlisting}
if r0 > 4:
  r1 = 6
else if r2 < 6:
  r1 = 10
else:
  r1 = 8    
---- window begins ----
/* use r1 somewhere here */
---- window  ends  ----
\end{lstlisting}
the clause of the precondition with inferred values is
\begin{lstlisting}
p1 (* $\Rightarrow$ *) r1 == 6
(* $\land$ *) p2 (* $\Rightarrow$ *) r1 == 10
(* $\land$ *) p3 (* $\Rightarrow$ *) r1 == 8
(* $\land$ *) exactly one of p1, p2, p3 is true
\end{lstlisting}
where {\ct p1, p2,} and {\ct p3} are fresh boolean
variables corresponding to the path conditions where {\ct
  r1} takes on distinct concrete values.

If the formula is satisfiable, the solver produces an input
state at the beginning of the window that results in the two windows producing
different output states at the end of the two window executions. If not, the two
windows are conditionally-equivalent under the specific
precondition and postcondition above. 

We specialize the liveness analysis to the BPF context by handling BPF registers
as well as BPF memory (stack, packet, map values). We build on optimizations
(I--III) in \Sec{best-effort-concretization} and track liveness information
along with concrete types and offsets. In particular, with map values, we treat
each map value as a distinct type of memory, with its own concretizations of
each access within the window, since we do not have concrete offsets into the
kernel's (value) memory.

Window-based verification condition generation and equivalence checking are
particularly useful in the BPF context for many reasons. First, it becomes
unnecessary to track aliasing for map access, since each window typically just
contains one live map value pointer, and we only care about aliasing within the
value memory itself.  In fact, we don't even use any information about the key
in the verification condition. Second, it becomes possible to handle several new
helpers in a much more expedient way with window verification, since we can
optimize ``around'' the function by simply tracking the inputs and outputs of
the helper, rather than ``across'' the function, which requires capturing its
input-output semantics. Third, window verification allows us to guide the
stochastic search better for programs that 
output a small number of output bits---programs known to be hard to optimize with
typical error functions~\cite{stoke-asplos13}. This is
because \TheCompiler's window
postcondition ensures that the window obtains other intermediate values in the
program correctly, rather than (just) the output.

The fundamental disadvantage of window-based verification relative to
full-program verification is its inability to discover some strong optimizations
that would be found with equivalence-checking full
programs~\cite{lens-asplos16}. Unlike window-based equivalence checking, full
program checking can detect aliasing conditions over the entire program,
optimize across helper functions (with full semantics), and can optimize
instructions under the strongest possible preconditions and weakest possible
postconditions.

\section{More Details on the Implementation of \TheCompiler}
\label{app:implementation}

Please see \Sec{implementation} for an overview of \TheCompiler's
implementation. 

\TheCompiler uses the same internal BPF instruction structure as the kernel, so
that \TheCompiler can consume BPF instructions directly from the binary ELF
format of the instructions (\ie the {\ct .o} files) without explicitly decoding
from and encoding into another format. Binary encode/decode is known to be a
significant source of compiler bugs~\cite{rocksalt-pldi12}.

The compiler consumes inputs from pre-compiled BPF bytecode
object files as well as instruction sequences encoded
through the kernel {\ct bpf\_insn} data structure.  The
extraction of the inputs from ELF binaries uses {\ct
  libbpf}, the kernel's standard library to work with BPF
bytecode. A tricky aspect of BPF bytecode is that the text
section of the ELF file is not directly executable. The text
section must be processed using {\em
  relocations}~\cite{linkers-and-loaders-levine1999} during
the loading process~\cite{bpf-portability-core} to ensure
that run-time references to map file descriptors and other
symbols are updated before execution (the text section is
independent of these run-time parameters).  \TheCompiler
consumes instructions from relocated ELF.

The output of the compiler can take on two forms: a sequence of binary
instructions, which is useful to test simple programs quickly, or a patched ELF
object file that contains the sections of the original ELF input with the
optimized instruction sequence patched in lieu of the original instructions in
the text section. \TheCompiler uses {\ct pyelftools} to perform the patching.

\TheCompiler uses relocation metadata, specifically the set of instructions that
are touched by the {\ct libbpf} loader, to ensure that the linkages between the
text section and the remaining ELF sections are unmodified. Hence, outputs from
\TheCompiler can serve as drop-in replacements to existing BPF object
files.

\TheCompiler includes a high-performance BPF interpreter that runs BPF bytecode
instructions using an optimized jumptable implementation akin to the kernel's
internal BPF interpreter~\cite{bpf-kernel-interpreter-source}. Using C
preprocessor directives, we share code for most instructions between the
interpreter and the equivalence checker. Before adopting this design, we found
that the mismatch between the behaviors of the interpreter and the semantics of
the first-order logic formulation of the same program was a significant and
vexing source of bugs. We found that using a shared semantic description of each
instruction that is common to the interpreter and the first-order logic
verification condition generator was helpful in ensuring that the input-output
behaviors of the interpreter and the program's formalization were compatible.

We construct first-order logic representations using {\ct Z3}~\cite{z3} and also
use it as our solver. To reduce the impact of large memory-bound queries on the
overall compile time, we use two separate Z3 solver processes, with the query
discharged to the solvers using a serialized {\ct smtlib2} format. We pick the
solver that returns a response first and respawn server processes after a fixed
number of discharged queries.

\TheCompiler's interpreter and the equivalence-checking query formulation can
distinguish different BPF program types and fix the inputs and outputs
appropriately.

\begin{table}
\resizebox{0.45\textwidth}{!}{%
\begin{tabular}{ |c|c|c|c|}
\hline
Benchmark & \# Variants & \# accepted by & Cause(s) of  \\
& produced & kernel checker & checker failure \\
\hline
xdp1 & 5 & 5 & - \\ \hline
xdp2 & 5 & 5 & - \\ \hline
xdp\_redirect & 5 & 5 & -\\ \hline
xdp\_map\_access & 5 & 5 & - \\ \hline
xdp\_router\_ipv4 & 5 & 5 & - \\ \hline
xdp\_pktcntr & 3 & 3 & - \\ \hline
xdp\_fwd & 5 & 5 & - \\ \hline
xdp\_fw & 5 & 5 & - \\ \hline
\end{tabular}%
}
\caption{\mycaptionsize We loaded 38 \TheCompiler-optimized program variants (known to
    be equivalent to the corresponding source programs) into the kernel. All programs were
    successfully accepted by the kernel checker.}
\label{tab:eval-safety}
\end{table}

\begin{table}[!htbp]
\centering
\begin{tabular}{|c|c|c|c|c|}
\hline
\multicolumn{1}{|c|}{ \multirow{2}*{Benchmark} }& \multicolumn{1}{c|}{ \# progs.} &\multicolumn{1}{c|}{ total \# calls} & \multicolumn{1}{c|}{ \multirow{2}*{Hit rate}} & \multicolumn{1}{c|}{ \multirow{2}*{\# iters.}}\\
\multicolumn{1}{|c|}{} & \multicolumn{1}{c|}{hit cache} & \multicolumn{1}{c|}{to the cache} & \multicolumn{1}{c|}{} & \multicolumn{1}{c|}{} \\\hline
(1) & 13,479 & 14,470 & 93\% & 1,000,000 \\\hline
(2) & 32,706 & 35,072 & 93\% & 1,500,000 \\\hline
(3) & 30,557 & 31,833 & 96\% & 4,000,000 \\\hline
(4) & 68,828 & 72,220 & 95\% & 4,000,000 \\\hline
(14) & 15,827 & 16,552 & 96\% & 2,000,000 \\\hline
(17) & 134,505 & 140,686 & 96\% & 5,000,000 \\\hline
(18) & 100,641 & 109,046 & 92\% & 5,000,000 \\\hline
\end{tabular}
\caption{\mycaptionsize The benefit of caching
  (\Sec{optimizing-equivalence-checking}) in reducing the
  number of solver calls. The benchmark numbers correspond
  to those in \Tab{eval-instruction-count}.}
\label{tab:caching-efficacy}
\end{table}

\section{Estimated Performance from \TheCompiler}
\label{app:estimated-runtime-gains}

In this section, we show the performance estimated by \TheCompiler with the
latency goal. We perform latency and throughput measurements of the top-k
programs from these experiments in \Sec{throughput-latency-improvements}.

\begin{table*}[!htbp]
\centering
\begin{tabular}{|l|c|c|c|c|c|c|}
\hline
\multicolumn{1}{|c|}{ \multirow{2}*{Benchmark} }& \multicolumn{4}{c|}{Program Runtime (sec)} & \multicolumn{2}{c|}{When lowest perf cost prog. is found} \\
\cline{2-7}
\multicolumn{1}{|c|}{}& \texttt{-O1} & \texttt{-O2/-O3} & \TheCompiler & Gain & Time (sec) & Iterations \\\hline
(9) xdp\_router\_ipv4 & 57.45 & 50.34 & 47.21 & 6.22\% & 1,455 & 450,502 \\\hline
(10) xdp\_redirect & 17.15 & 16.08 & 14.52 & 9.70\% & 472 & 203,332 \\\hline
(11) xdp1\_kern/xdp1 & 31.33 & 25.57 & 24.55 & 3.99\% & 155 & 240,116 \\\hline
(12) xdp2\_kern/xdp1 & 34.79 & 30.88 & 28.86 & 6.54\% & 145 & 107,089 \\\hline
(13) xdp\_fwd & 57.02 & 51.56 & 43.73 & 15.19\% & 4,316 & 2,850,176 \\\hline
(14) xdp\_pktcntr & 12.32 & 12.32 & 11.85 & 3.81\% & 48 & 201,057 \\\hline
(15) xdp\_fw & 65.02 & 46.61 & 45.01 & 3.43\% & 858 & 345,320 \\\hline
(16) xdp\_map\_access & 27.96 & 27.96 & 27.28 & 2.43\% & 20 & 50,814 \\\hline
(17) from-network & 32.27 & 30.96 & 29.17 & 5.78\% & 713 & 401,730 \\\hline
(18) recvmsg4 & 68.12 & 68.12 & 63.83 & 6.30\% & 3,444 & 907,834 \\\hline
(19) xdp-balancer & DNL & 760.12 & 724.73 & 4.66\% & 170,154 & 7,502,210 \\\hline
Avg. of all benchmarks & & & &  6.19\% & 16,525 & 1,205,471 \\\hline
\end{tabular}
\caption{\mycaptionsize Improvements in {\em estimated}
  performance of programs according to \TheCompiler. }
\end{table*}

\section{Additional Evaluation Results}

\label{app:additional-eval}

\subsection{Parameters of stochastic search}
\label{app:parameter-set}
\Tab{parameter-settings-sets} shows the best-performing
parameter settings of
\TheCompiler. \Tab{perf-under-different-param-settings}
shows \TheCompiler's instruction count reductions from those
parameter settings. \Tab{eval-move-types} shows the benefits
of \TheCompiler's domain-specific rules in stochastic
synthesis.
\begin{table*}[!htbp]
\centering
\begin{tabular}{|l|c|c|c|c|c|}\hline
Setting ID  & 1 & 2 & 3 & 4 & 5 \\\hline
Compute error cost  & ABS & POP & POP & ABS & ABS \\\hline
Avg. total error cost by \# test cases or not & No  & No  & No  & No  & Yes \\\hline
Weight of error cost ($\alpha$) & 0.5 & 0.5 & 0.5 & 0.5 & 0.5 \\\hline
Weight of performance cost ($\beta$)  & 5 & 5 & 5 & 5 & 1.5 \\\hline
Probability of instruction replacement ($prob_{ir}$)  & 0.2 & 0.17  & 0.2 & 0.17  & 0.17  \\\hline
Probability of operand replacement ($prob_{or}$)  & 0.4 & 0.33  & 0.4 & 0.33  & 0.33  \\\hline
Probability of replacement by NOP ($prob_{nr}$) & 0.15  & 0.15  & 0.15  & 0.15  & 0.15  \\\hline
Probability of memory exchange type 1 ($prob_{me1}$)  & 0.2 & 0.17  & 0.2 & 0     & 0  \\\hline
Probability of memory exchange type 2 ($prob_{me2}$)  & 0   & 0     & 0  & 0.17  & 0.17  \\\hline
Probability of replacement of contiguous instructions ($prob_{cir}$)  & 0.05  & 0.18  & 0.05  & 0.18  & 0.18  \\\hline
\end{tabular}
\caption{\mycaptionsize Details of \TheCompiler's five
  best-performing parameter
  settings (\Sec{stochastic-search}). ABS corresponds to the
  absolute error cost function, while POP corresponds to
  population count of the bitwise difference. }
\label{tab:parameter-settings-sets}
\end{table*}

\begin{table*}[!htbp]
\centering
\begin{tabular}{|l|c|c|c|c|c|c|c|c|}
\hline
\multicolumn{1}{|c|}{ \multirow{2}*{Benchmark} }& \multicolumn{5}{c|}{Parameter setting ID} &\multicolumn{1}{c|}{\#  instructions} & \multicolumn{1}{c|}{ \% of paras} & \multicolumn{1}{c|}{ Time to find }\\
\cline{2-6}
\multicolumn{1}{|c|}{} & 1 & 2 & 3 & 4 & 5 & \multicolumn{1}{c|}{of the smallest prog.} & \multicolumn{1}{c|}{get the smallest prog.} & \multicolumn{1}{c|}{the smallest prog. (sec)} \\
\hline
(1) xdp\_exception & 16 & 18 & 16 & 16 & 16 & 16 & 80\% & 323 \\\hline
(2) xdp\_redirect\_err & 16 & 16 & 16 & 16 & 16 & 16 & 100\% & 10 \\\hline
(3) xdp\_devmap\_xmit & 30 & 29 & 30 & 32 & 30 & 29 & 20\% & 1,213 \\\hline
(4) xdp\_cpumap\_kthread & 19 & 18 & 19 & 22 & 18 & 18 & 40\% & 1,170 \\\hline
(5) xdp\_cpumap\_enqueue & 22 & 21 & 22 & 26 & 22 & 21 & 20\% & 1,848 \\\hline
(6) sys\_enter\_open & 21 & 21 & 21 & 20 & 21 & 20 & 20\% & 519 \\\hline
(7) socket/0 & 27 & 27 & 27 & 27 & 27 & 27 & 100\% & 6 \\\hline
(8) socket/1 & 30 & 30 & 30 & 30 & 30 & 30 & 100\% & 9 \\\hline
(9) xdp\_router\_ipv4 & 101 & 99 & 99 & 100 & 100 & 99 & 40\% & 898 \\\hline
(10) xdp\_redirect & 35 & 35 & 35 & 35 & 37 & 35 & 80\% & 523 \\\hline
(11) xdp1\_kern/xdp1 & 56 & 56 & 56 & 57 & 58 & 56 & 60\% & 598 \\\hline
(12) xdp2\_kern/xdp1 & 71 & 72 & 71 & 71 & 72 & 71 & 60\% & 159 \\\hline
(13) xdp\_fwd & 128 & 130 & 130 & 133 & 134 & 128 & 20\% & 6,137 \\\hline
(14) xdp\_pktcntr & 19 & 19 & 19 & 20 & 19 & 19 & 80\% & 288 \\\hline
(15) xdp\_fw & 65 & 65 & 65 & 68 & 69 & 65 & 60\% & 826 \\\hline
(16) xdp\_map\_access & 26 & 26 & 26 & 27 & 26 & 26 & 80\% & 32 \\\hline
(17) from-network & 30 & 30 & 30 & 31 & 29 & 29 & 20\% & 6,871 \\\hline
(18) recvmsg4 & 81 & 81 & 81 & 83 & 84 & 81 & 60\% & 3,350 \\\hline
(19) xdp-balancer & 1,624 & 1,607 & 1,627 & 1,679 & 1,684 & 1,607 & 20\% & 167,428 \\\hline
\end{tabular}
\caption{\mycaptionsize Performance of \TheCompiler under
  different parameter settings.}
\label{tab:perf-under-different-param-settings}
\end{table*}

\begin{table*}[!htbp]
\footnotesize
\centering
\begin{tabular}{|l|c|c|c|c|c|c|c|c|c|c|c|c|}
\hline
\multicolumn{1}{|c|}{ \multirow{2}*{Benchmark} }& \multicolumn{2}{c|}{$MEM_1$ \& $CONT$} & \multicolumn{2}{c|}{$MEM_2$ \& $CONT$} & \multicolumn{2}{c|}{$MEM_1$ only} & \multicolumn{2}{c|}{$MEM_2$ only} & \multicolumn{2}{c|}{$CONT$ only} & \multicolumn{2}{c|}{None}\\

\cline{2-13}
\multicolumn{1}{|c|}{}& \# instr.  & time (sec) & \# instr.  & time (sec) & \# instr. & time (sec) & \# instr. & time (sec) & \# instr.  & time (sec) & \# instr.  & time (sec) \\
\hline
(1) xdp\_exception & 16* & 8 & 16* & 315 & 16* & 90 & 16* & 91 & 16* & 183 & 16* & 93\\\hline
(2) xdp\_redirect\_err & 16* & 10 & 16* & 40 & 16* & 29 & 16* & 160 & 16* & 416 & 16* & 84\\\hline
(3) xdp\_devmap\_xmit & 29* & 1,201 & 30 & 497 & 30 & 914 & 30 & 2,988 & 30 & 791 & 30 & 3,454\\\hline
(4) xdp\_cpumap\_kthread & 18* & 1,848 & 18* & 3,998 & 19 & 344 & 21 & 535 & 19 & 1,364 & 19 & 2,933\\\hline
(5) xdp\_cpumap\_enqueue & 21* & 214 & 21* & 2,647 & 22 & 52 & 22 & 360 & 22 & 1,108 & 22 & 373\\\hline
(6) sys\_enter\_open & 21 & 6 & 20* & 519 & 21 & 55 & 21 & 214 & 21 & 364 & 21 & 31\\\hline
(7) socket/0 & 27* & 9 & 27* & 19 & 27* & 10 & 27* & 72 & 27* & 5 & 27* & 10\\\hline
(8) socket/1 & 30* & 598 & 30* & 10 & 30* & 60 & 30* & 154 & 30* & 5 & 30* & 56\\\hline
(9) xdp\_router\_ipv4 & 99* & 482 & 99* & 2,704 & 99* & 1,073 & 99* & 992 & 105 & 4,949 & 106 & 299\\\hline
(10) xdp\_redirect & 35* & 826 & 35* & 2,681 & 35* & 724 & 35* & 764 & 39 & 507 & 39 & 1,441\\\hline
(11) xdp1\_kern/xdp1 & 56* & 269 & 56* & 472 & 57 & 41 & 57 & 286 & 57 & 62 & 57 & 42\\\hline
(12) xdp2\_kern/xdp1 & 71* & 3,350 & 71* & 334 & 71* & 328 & 71* & 185 & 75 & 6 & 75 & 5\\\hline
(13) xdp\_fwd & 128* & 4,944 & 130 & 3,036 & 128* & 2,847 & 130 & 4,686 & 144 & 4,003 & 145 & 5,091\\\hline
(14) xdp\_pktcntr & 19* & 1,194 & 19* & 260 & 20 & 30 & 20 & 32 & 20 & 44 & 20 & 57\\\hline
(15) xdp\_fw & 65* & 898 & 66 & 1,856 & 65* & 245 & 67 & 46 & 66 & 1,043 & 66 & 2,619\\\hline
(16) xdp\_map\_access & 26* & 523 & 26* & 106 & 26* & 30 & 26* & 256 & 26* & 354 & 26* & 78\\\hline
(17) from-network & 30 & 366 & 29* & 6,871 & 30 & 915 & 30 & 4,449 & 30 & 3,544 & 30 & 3,696\\\hline
(18) recvmsg4 & 81* & 1,170 & 81* & 8,317 & 84 & 4,146 & 85 & 14,829 & 85 & 10,498 & 87 & 16,763\\\hline
\hline
\multicolumn{1}{|c|}{\# benchmarks where this  } & \multicolumn{2}{c|}{\multirow{2}*{16}}& \multicolumn{2}{c|}{\multirow{2}*{15}}&\multicolumn{2}{c|}{\multirow{2}*{10}} & \multicolumn{2}{c|}{\multirow{2}*{8}} & \multicolumn{2}{c|}{\multirow{2}*{5}} & \multicolumn{2}{c|}{\multirow{2}*{5}} \\
\multicolumn{1}{|c|}{setting found the best program} & \multicolumn{2}{c|}{}& \multicolumn{2}{c|}{}&\multicolumn{2}{c|}{} & \multicolumn{2}{c|}{} & \multicolumn{2}{c|}{} & \multicolumn{2}{c|}{} \\\hline
\multicolumn{1}{|c|}{\# benchmarks where {\em only} this} & \multicolumn{2}{c|}{\multirow{2}*{1}}& \multicolumn{2}{c|}{\multirow{2}*{2}}&\multicolumn{2}{c|}{\multirow{2}*{0}} & \multicolumn{2}{c|}{\multirow{2}*{0}} & \multicolumn{2}{c|}{\multirow{2}*{0}} & \multicolumn{2}{c|}{\multirow{2}*{0}} \\
\multicolumn{1}{|c|}{ setting found the best program} & \multicolumn{2}{c|}{}& \multicolumn{2}{c|}{}&\multicolumn{2}{c|}{} & \multicolumn{2}{c|}{} & \multicolumn{2}{c|}{} & \multicolumn{2}{c|}{}\\\hline
\end{tabular}
\caption{\mycaptionsize Improvements in program compactness under
    different proposal generation settings
    (\Sec{proposal-generation}).  We consider turning the
    following proposal-generation rewrite-rules on or off:
    $MEM_1$ implements a type 1 memory exchange, sampling to
    replace all non-pointer operands, $MEM_2$ implements a
    type-2 memory exchange, sampling to replace only the
    memory operation width, and $CONT$ replaces $k = 2$
    contiguous instructions.  Instruction counts with the *
    mark indicate that they are the minimal found among all
    the proposal generation settings tested. }
\label{tab:eval-move-types}
\end{table*}

\subsection{Efficacy of safety checks}
The results of loading the outputs of \TheCompiler using the kernel
checker are shown in \Tab{eval-safety}.

\subsection{Caching efficacy}
The efficacy of caching in reducing the number of solver queries to
perform equivalence checking is illustrated in
\Tab{caching-efficacy}.

\subsection{Benefits of domain-specific rewrite rules}
\label{sec:benefits-of-domain-specific-rewrite-rules}
In \Tab{eval-move-types}, we show the benefits of including
domain-specific program rewrite rules
(\Sec{proposal-generation}) while generating proposals
within the stochastic optimization loop.

\section{More Optimizations Discovered by \TheCompiler}
\label{app:detailed-optimization-case-studies}

Table \ref{tab:detailed-optimization-case-studies-table} lists several optimization
case studies across benchmarks from the Linux kernel, hXDP, and Cilium.

\begin{table*}[!htbp]
    \centering
    \scriptsize
    \begin{tabularx}{\textwidth}{lXXXX}
        \toprule
        \textbf{Benchmark} &  \textbf{BPF program before optimization} 
        & \textbf{BPF program after optimization} & \textbf{C code}  
        & \textbf{Note} \\
        \midrule

        xdp2\_kern/xdp1 & 
        bpf\_load\_16 r1 r7 0  // r1 = *(u16*)(r7+0)\newline
        bpf\_load\_16 r2 r7 6  // r2 = *(u16*)(r7+6)\newline
        bpf\_stx\_16 r7 0 r2   // *(u16*)(r7+0) = r2\newline
        bpf\_load\_16 r2 r7 8  // r2 = *(u16*)(r7+8)\newline
        bpf\_load\_16 r3 r7 2  // r3 = *(u16*)(r7+2)\newline
        bpf\_stx\_16 r7 8 r3   // *(u16*)(r7+8) = r3\newline
        bpf\_stx\_16 r7 2 r2   // *(u16*)(r7+2) = r2\newline
        bpf\_load\_16 r2 r7 10 // r2 = *(u16*)(r7+10)\newline
        bpf\_load\_16 r3 r7 4  // r3 = *(u16*)(r7+4)\newline
        bpf\_stx\_16 r7 10 r3  // *(u16*)(r7+10) = r3\newline
        bpf\_stx\_16 r7 6 r1   // *(u16*)(r7+6) = r1\newline
        bpf\_stx\_16 r7 4 r2   // *(u16*)(r7+4) = r2
        & 
        bpf\_load\_32 r1 r7 0  // r1 = *(u32*)(r7+0)\newline
        bpf\_load\_32 r2 r7 6  // r2 = *(u32*)(r7+6)\newline
        bpf\_stx\_32 r7 0 r2   // *(u32*)(r7+0) = r2\newline
        bpf\_load\_32 r3 r7 4  // r3 = *(u32*)(r7+4)\newline
        bpf\_load\_32 r2 r7 10 // r2 = *(u32*)(r7+10)\newline
        bpf\_stx\_32 r7 10 r3  // *(u16*)(r7+10) = r3\newline
        bpf\_stx\_32 r7 6 r1   // *(u32*)(r7+6) = r1\newline
        bpf\_stx\_16 r7 4 r2   // *(u16*)(r7+4) = r2
        & 
           dst[0] = p[0];\newline
           dst[1] = p[1];\newline
           dst[2] = p[2];\newline
           p[0] = p[3];\newline
           p[1] = p[4];\newline
           p[2] = p[5];\newline
           p[3] = dst[0];\newline
           p[4] = dst[1];\newline
           p[5] = dst[2];\newline
        & This instruction sequence swaps three higher bytes and three lower bytes through six 8-bit loads and stores. K2 coalesced six loads and stores into two 16-bit loads and stores and one 8-bit load and store.
        \\
        \midrule

        xdp\_fwd
        &
        bpf\_load\_16 r1 r10 -2 // r1 = *(u16*)(r10-2)\newline
        bpf\_stx\_8 r7 4 r1     // *(u8*)(r7+4) = r1\newline
        bpf\_rsh64 r1 8        // r1 >>= 8\newline
        bpf\_stx\_8 r7 5 r1     // *(u8*)(r7+5) = r1\newline
        bpf\_load\_16 r1 r10 -4 // r1 = *(u16*)(r10-4)\newline
        bpf\_stx\_8 r7 2 r1     // *(u8*)(r7+2) = r1\newline
        bpf\_rsh64 r1 8        // r1 >>= 8\newline
        bpf\_stx\_8 r7 3 r1     // *(u8*)(r7+3) = r1\newline
        bpf\_load\_16 r1 r10 -6 // r1 = *(u16*)(r10-6)\newline
        bpf\_stx\_8 r7 0 r1     // *(u8*)(r7+0) = r1\newline
        bpf\_rsh64 r1 8        // r1 >>= 8\newline
        bpf\_stx\_8 r7 1 r1     // *(u8*)(r7+1) = r1
        &
        bpf\_load\_32 r1 r10 -4 // r1 = *(u32*)(r10-4)\newline
        bpf\_stx\_32 r7 2 r1    // *(u32*)(r7+2) = r1\newline
        bpf\_load\_16 r1 r10 -6 // r1 = *(u16*)(r10-6)\newline
        bpf\_stx\_16 r7 0 r1    // *(u16*)(r7+0) = r1
        & memcpy(eth->h\_dest, fib\_params.dmac, ETH\_ALEN);
        & This instruction sequence copies 6 bytes from the
        source address (r10-6) to the destination address r7
        by three sets of operations, each involving one 16-bit
        load and two 8-bit
        stores. K2 reduced the instruction count by
        compressing these memory operations into one 32-bit
        load and store, and one 16-bit load and store.
        \\
        \midrule

        sys\_enter\_open
        &
        bpf\_load\_32 r1 r0 0 // r1 = *(u32*)(r0+0)\newline
        bpf\_add64 1 1       // r1 += 1\newline
        bpf\_stx\_32 r0 0 r1  // *(u32*)(r0+0) = r1
        &
        bpf\_mov64 r1 1      // r1 = 1\newline
        bpf\_xadd\_32 r0 0 r1 // *(u32*)(r0+0) += r1
        & 
        & This instruction sequence increases the memory value by 1. It loads the value from the memory and then performs a register addition, finally stores the register value into the memory. K2 utilized the memory addition to reduce one instruction.
        \\
        \midrule

        xdp1\_kern/xdp1
        &
        bpf\_mov64 r1 0       // r1 = 0\newline
        bpf\_stx\_32 r10 -4 r1 // *(u32*)(r10-4) = r1
        & bpf\_st\_imm32 r10 -4 0 // *(u32*)(r10-4) = 0
        & 
        & This transformation coalesces a register assignment and one register store into a store that writes an immediate value.
        \\
        \midrule

        recvmsg4
        &
        bpf\_load\_32 r1 r6 24   // r1 = *(u32*)(r6+24)\newline
        bpf\_stx\_32 r10 -16 r1  // *(u32*)(r10-16)=r1\newline
        bpf\_stx\_16 r10 -26 r7  // *(u16*)(r10-26)=r7\newline
        bpf\_load\_32 r1 r10 -16 // r1 = *(u32*)(r10-16)\newline
        bpf\_load\_16 r10 -28 r1 // *(u16*)(r10-28)=r1
        &
        bpf\_load\_16 r1 r6 24   // r1 = *(u16*)(r6+24)\newline
        bpf\_stx\_32 r10 -28 r1  // *(u32*)(r10-28) = r1
        & 
        & This optimization does not hold under all values of r7. In the prefix program, r7 is assigned as 0. Also, the value written in (r10-16) is not read in the postfix program. K2 found this transformation by leveraging both preconditions and postconditions.
        \\
        \midrule

        xdp\_map\_access
        &
        bpf\_mov64 r3 0       // r3 = 0 \newline
        bpf\_stx\_8 r10 -8 r3  // *(u8*)(r10-8) = r3
        & (no instructions)
        & 
        & K2 removed these two instructions by the postconditions where the values set to the register and the memory were not used in the postfix program.
        \\

        \bottomrule
    \end{tabularx}
    \caption{\mycaptionsize A catalog of optimizations found by \TheCompiler.}
    \label{tab:detailed-optimization-case-studies-table}
\end{table*}

\section{Profiles of program latency \vs offered load}
\label{app:throughput-latency-profile}

These results supplement the numbers in \Sec{evaluation}
regarding the latency, throughput, and drop rates of various
XDP programs as offered load from the traffic generator
increases.

\begin{figure*}
  \centering
  % xdp2
  \subfloat[xdp2: Throughput vs. Offered load]{
    \begin{minipage}[t]{0.32\textwidth}
    \includegraphics[width=\textwidth]{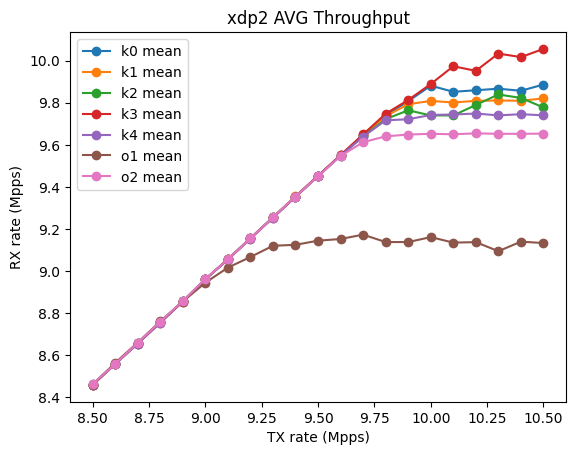}
    \end{minipage}}
  \subfloat[xdp2: Avg. latency vs. Offered load]{
    \begin{minipage}[t]{0.32\textwidth}
    \includegraphics[width=\textwidth]{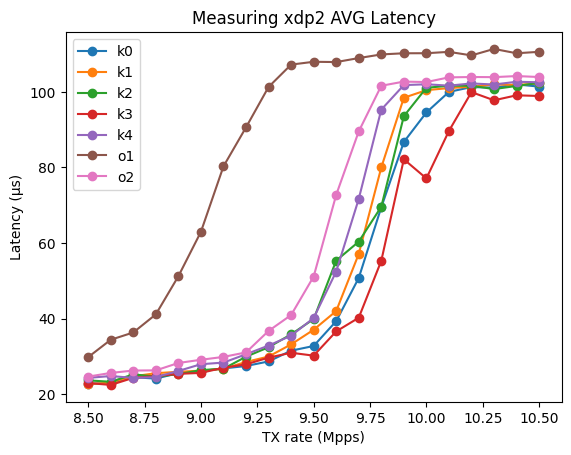}  
    \end{minipage}}
  \subfloat[xdp2: Drop rate vs. Offered load]{
    \begin{minipage}[t]{0.31\textwidth}
    \includegraphics[width=\textwidth]{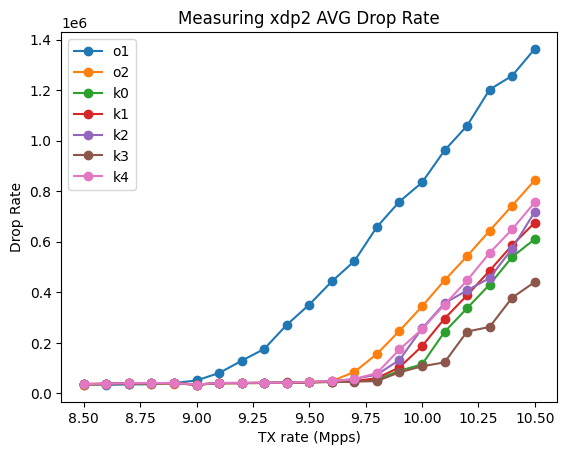}
    \end{minipage}} 

  % xdp_router_ipv4
  \subfloat[xdp\_router\_ipv4: Throughput vs. Offered load]{
    \begin{minipage}[t]{0.32\textwidth}
    \includegraphics[width=\textwidth]{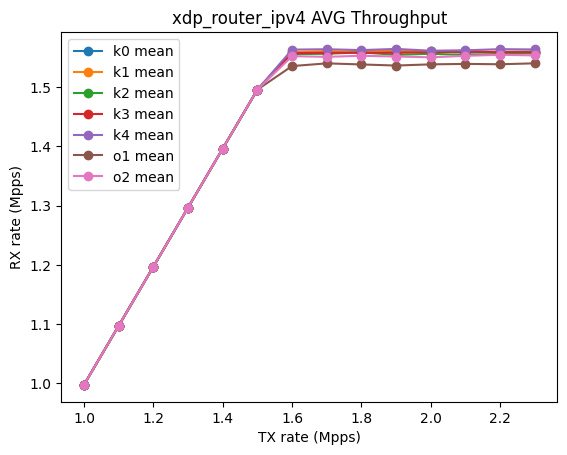}
    \end{minipage}} 
  \subfloat[xdp\_router\_ipv4: Avg. latency vs. Offered load]{
    \begin{minipage}[t]{0.32\textwidth}
    \includegraphics[width=\textwidth]{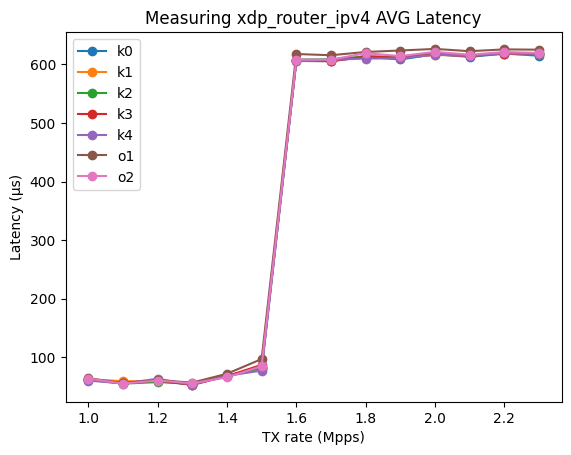}
    \end{minipage}} 
  \subfloat[xdp\_router\_ipv4: Drop rate vs. Offered load]{
    \begin{minipage}[t]{0.33\textwidth}
    \includegraphics[width=\textwidth]{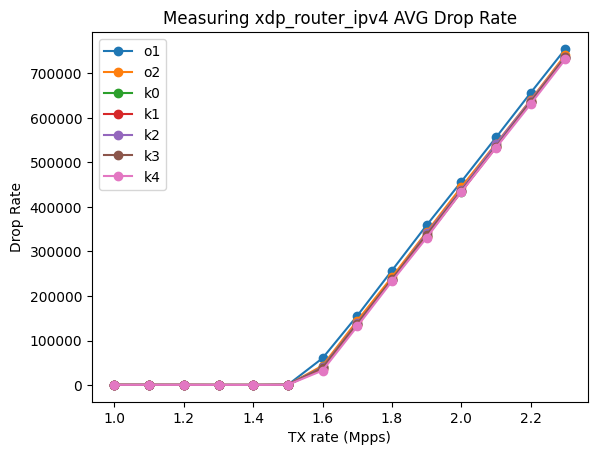}
    \end{minipage}} 

  % xdp_fwd
  \subfloat[xdp\_fwd: Throughput vs. Offered load]{
    \begin{minipage}[t]{0.32\textwidth}
    \includegraphics[width=\textwidth]{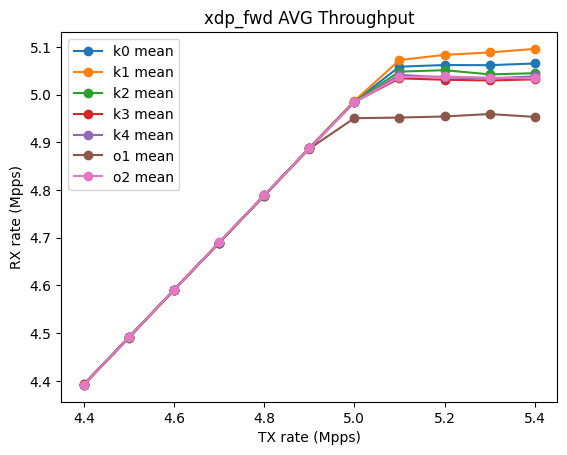}
    \end{minipage}} 
  \subfloat[xdp\_fwd: Avg. latency vs. Offered load]{
    \begin{minipage}[t]{0.32\textwidth}
    \includegraphics[width=\textwidth]{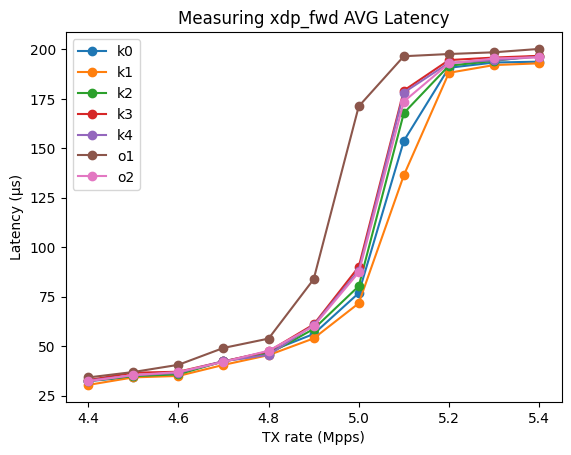}
    \end{minipage}} 
  \subfloat[xdp\_fwd: Drop rate vs. Offered load]{
    \begin{minipage}[t]{0.33\textwidth}
    \includegraphics[width=\textwidth]{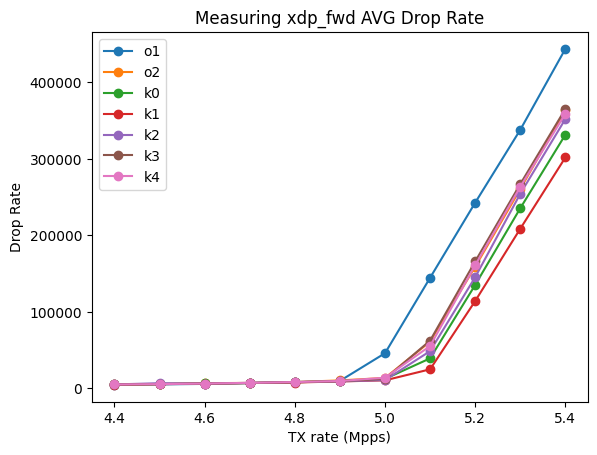}
    \end{minipage}}

  % xdp-balancer
  \subfloat[xdp-balancer: Throughput vs. Offered load]{
    \begin{minipage}[t]{0.32\textwidth}
    \includegraphics[width=\textwidth]{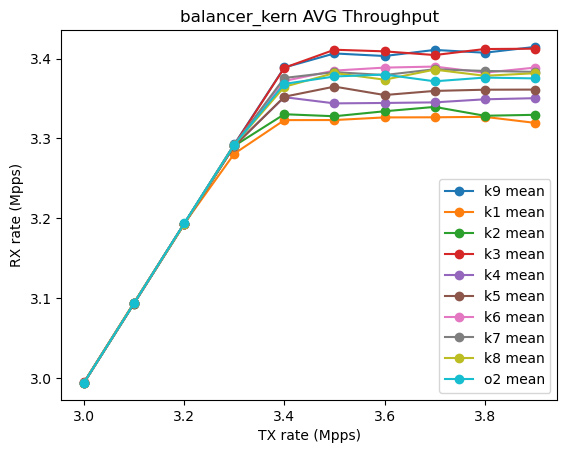}
    \end{minipage}} 
  \subfloat[xdp-balancer: Avg. latency vs. Offered load]{
    \begin{minipage}[t]{0.32\textwidth}
    \includegraphics[width=\textwidth]{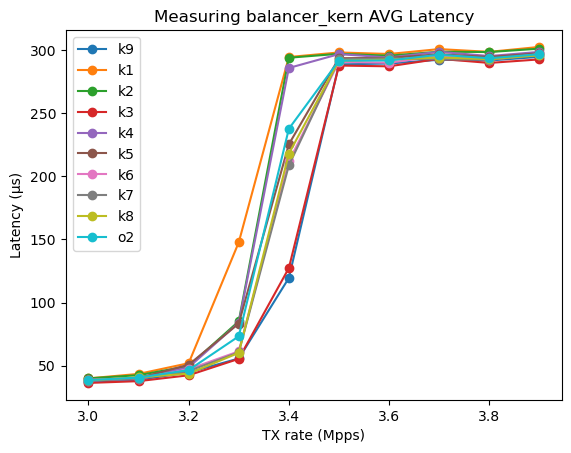}
    \end{minipage}} 
  \subfloat[xdp-balancer: Drop rate vs. Offered load]{
    \begin{minipage}[t]{0.33\textwidth}
    \includegraphics[width=\textwidth]{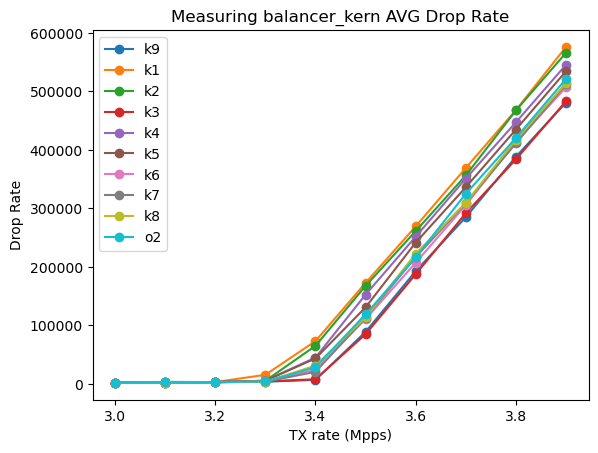}
    \end{minipage}} 
\end{figure*}

\begin{figure*}\ContinuedFloat
  \centering
  % xdp1
  \subfloat[xdp1: Throughput vs. Offered load]{
    \begin{minipage}[t]{0.32\textwidth}
    \includegraphics[width=\textwidth]{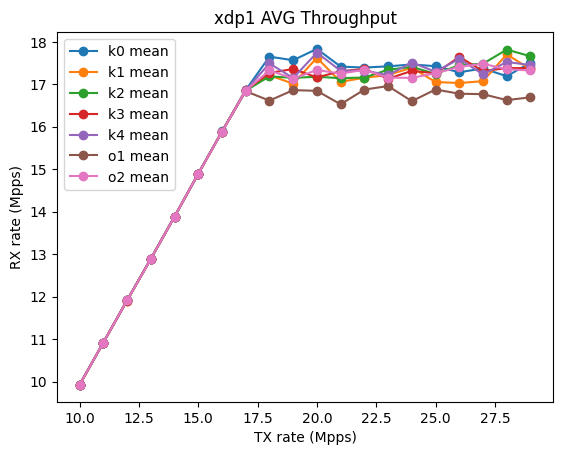}
    \end{minipage}} 
  % xdp_map_access
  \subfloat[xdp\_map\_access: Throughput vs. Offered load]{
    \begin{minipage}[t]{0.33\textwidth}
    \includegraphics[width=\textwidth]{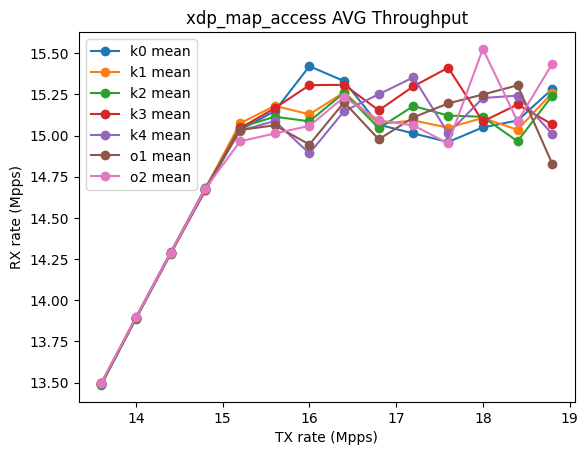}
    \end{minipage}} 

  \caption{\mycaptionsize Throughput, average latency and drop rate under different offered loads. 
           We measured the top-$k$ K2 versions ($k=9$ for xdp-balancer and $k=5$ for other
           benchmarks), and two {\ct clang} versions: {\ct
             -O1} and {\ct -O2}. Note that {\ct -O2} and
           {\ct -O3} are identical for these benchmarks.}
  \label{fig:throughput-latency}
% \vspace{-0.5em}
\end{figure*}

\end{document}